\documentclass[12pt,preprint]{aastex}
\received{}
\revised{}
\accepted{}
\journalid{}{}
\articleid{}{}
\paperid{}
\cpright{}{}
\ccc{}
\shortauthors{Matthews \& Uson}
\shorttitle{{H\,{\sc i}} Imaging of IC~2233 and NGC~2537}
\begin{document}

\newcommand{\ang}{\rm \AA}
\newcommand{\msun}{\mbox{$M_\odot$}}
\newcommand{\lsun}{\mbox{$L_\odot$}}
\newcommand{\kms}{\mbox{km s$^{-1}$}}
\newcommand{\kmsm}{\mbox{km s$^{-1}$ Mpc$^{-1}$}}
\newcommand{\HI}{\mbox{H\,{\sc i}}}
\newcommand{\mhi}{\mbox{$M_{\rm HI}$}}
\newcommand{\HA}{H$\alpha$}
\newcommand{\HII}{\mbox{H\,{\sc ii}}}
\newcommand{\NII}{\mbox{N\,{\sc ii}}}
\newcommand{\OIII}{\mbox{O\,{\sc iii}}}
\newcommand{\lsim}{~\rlap{$<$}{\lower 1.0ex\hbox{$\sim$}}}
\newcommand{\gsim}{~\rlap{$>$}{\lower 1.0ex\hbox{$\sim$}}}
\newcommand{\skp}{\mbox{  }\\}
\newcommand{\bmr}{\mbox{B--R}}
\newcommand{\ad}[2]{$#1^{\circ}\,\hspace{-1.7mm}.\hspace{.0mm}#2$}
\newcommand{\am}[2]{$#1'\,\hspace{-1.7mm}.\hspace{.0mm}#2$}
\newcommand{\as}[2]{$#1''\,\hspace{-1.7mm}.\hspace{.0mm}#2$}
\newcommand{\UV}{$u$-$v$}
\newcommand{\jks}{Jy km s$^{-1}$}

\title{{H\,{\sc i}} Imaging Observations of Superthin Galaxies.
II. IC~2233 and the Blue Compact Dwarf NGC~2537}

\author{Lynn D. Matthews\altaffilmark{1,}\altaffilmark{2}}
\author{Juan M. Uson\altaffilmark{3}}

\altaffiltext{1}{Harvard-Smithsonian Center for Astrophysics,
60 Garden Street, MS-42, Cambridge, MA 02138 USA; lmatthew@cfa.harvard.edu}
\altaffiltext{2}{Visiting Astronomer, Kitt Peak National Observatory.
National Optical Observatories is operated by the Association of
Universities for Research in Astronomy (AURA), Inc., under a
cooperative agreement with the National Science Foundation.}
\altaffiltext{3}{National Radio Astronomy Observatory, 520 Edgemont Road,
Charlottesville, VA 22903 USA; juson@nrao.edu}

\begin{abstract}
We have used the Very Large Array to image the \HI\ 21-cm line emission in the
edge-on Sd galaxy IC~2233 and the blue compact
dwarf NGC~2537.  We also present new optical $B$, $R$, and H$\alpha$ imaging of
IC~2233 obtained with the WIYN telescope. Despite evidence of localized massive star
formation in the form of prominent \HII\ regions and shells, 
supergiant stars, and a blue integrated color, IC~2233 has a 
low surface brightness disk with a very
low global star formation rate 
($\lsim0.05M_{\odot}$~yr$^{-1}$), and 
no significant 21-cm radio continuum emission. 
The \HI\ and ionized gas disks of
IC~2233 are clumpy and vertically distended, with scale heights comparable to
that of the young stellar disk.  Both the stellar and \HI\ disks of IC~2233
appear flared, and we also find a vertically extended,
rotationally anomalous component of \HI\ extending to
$\sim$2.4~$d_{10}$~kpc from
the midplane. The \HI\ disk
exhibits a mild lopsidedness  as well as a global 
corrugation pattern with a period of  $\sim7~d_{10}$~kpc and an 
amplitude of $\sim150~d_{10}$~pc. To our knowledge, this is the first
time corrugations of the gas disk have
been reported in an external galaxy;
these  undulations may be linked to bending instabilities or to 
underlying spiral structure and
suggest that the disk is largely self-gravitating. 

Lying at a projected distance of \am{16}{7} from IC~2233, 
NGC~2537 has an \HI\ disk with a bright, tilted inner ring and 
a flocculent, dynamically cold outer region that extends to 
$\sim$3.5~times the extent of the stellar light ($D_{25}$). Although NGC~2537 is
rotationally-dominated, we measure
\HI\ velocity dispersions as high as $\sigma_{V,\rm HI}\sim25$~\kms\ near its
center, indicative of significant turbulent motions.  The inner
rotation curve rises steeply, implying a strong central
mass concentration. Our data indicate that 
IC~2233 and NGC~2537 do not
constitute a bound pair and most likely lie at different distances.
We also find no compelling evidence of a recent minor
merger in either IC~2233 or NGC~2537, suggesting that both are
examples of small disk galaxies evolving in relative isolation.

\end{abstract}


\section{Introduction\protect\label{intro}}
This is the second in our series of papers presenting \HI\ imaging observations
of edge-on, pure-disk galaxies obtained with the Very Large Array
(VLA)\footnote{The Very Large Array of the National Radio Astronomy Observatory
is a facility of the National Science Foundation, operated under cooperative
agreement by Associated Universities, Inc.}.   We are focusing on `superthin'
Sd galaxies that are characterized by their highly flattened, dynamically cold
stellar disks and the absence of a visible spheroid component
(Goad \& Roberts 1981).  The structural simplicity and edge-on orientation of
superthins makes these galaxies excellent laboratories for exploring how
internal versus external processes regulate many of the key properties of
late-type disk galaxies, including their structures and star formation
histories.  For example, the abundance of thin, bulgeless disks in the local
universe (Karachentsev et al. 1999; Kautsch et al. 2006) poses challenges for
hierarchical galaxy-formation models in which galaxies are built-up
through violent mergers (Abadi et al. 2003; D'Onghia \& Burkert 2004;
Eliche-Moral et al. 2006; Kormendy \& Fisher 2005 and references therein).
The superthin disks that we are studying also inhabit a particularly interesting
galactic mass range ($V_{\rm rot}\sim$100~\kms), where important changes
in the structure and composition of the interstellar medium have been noted
(Dalcanton et al. 2004; Matthews et al. 2005).  Complementary studies of
both stellar and interstellar components are necessary to understand the
physics behind these changes and, in turn, how they affect the regulation
of star formation.  Finally, it is well established that the thinnest
galaxies tend to be chemically unevolved systems of low optical surface
brightness (Goad \& Roberts 1981; Bergvall \& R\"onnback 1995; Matthews
et al. 1999; Gerritsen \& de Blok 1999).  Thus, the vertical structure of superthin
galaxies provides information relevant to understanding the overall disk structure
and mass distribution of low surface brightness (LSB) galaxies in general.  

In a previous paper (Uson \& Matthews 2003; hereafter Paper~I), we presented 
VLA observations of the isolated, LSB superthin spiral
UGC~7321.  Here we turn to IC~2233 (= UGC~4278). In addition to new
VLA \HI\ observations (\S\ref{icHI}), we present optical $B$, $R$ and
H$\alpha$ imaging of IC~2233 obtained 
with the WIYN telescope (\S\ref{optical}). IC~2233 has a comparable
physical size and luminosity to UGC~7321, and both are weak
infrared sources with very low current star formation rates. However,
IC~2233 is slightly less massive and has a slightly
higher optical surface brightness. And while both systems are \HI-rich
and exhibit bulgeless, highly flattened stellar disks with minimal
dust absorption, our new optical
and \HI\ imaging study has uncovered some interesting differences in
the kinematics of these two galaxies and in the
structures of their ionized and neutral gas disks.

One of the goals of our VLA study of superthin, edge-on spiral
galaxies is to explore the role of environment in shaping the
structure, kinematics, and
evolution of these dynamically cold and seemingly ``fragile'' disk systems.
While UGC~7321 appears to be an extremely isolated galaxy (Paper~I), we
targeted IC~2233 because of the presence of a second galaxy at a projected separation of
\am{16}{7} and a radial velocity difference of $\Delta
V\approx114$~\kms: the blue
compact dwarf galaxy (BCD) NGC~2537 (= UGC~4274
= Arp~6 = Mrk~86 = ``The Bear's Paw''). We observed  NGC~2537 simultaneously
with the VLA, and our new observations highlight several interesting
features of this galaxy (\S\ref{bearpawHI}). 
However, we find that IC~2233 and NGC~2537 do not appear to be
physically associated and that a mutual tidal
interaction is unlikely to have shaped the properties of either
galaxy. Indeed, like UGC~7321, both IC~2233 and NGC~2537 appear to be
examples of low-mass disk galaxies evolving in relative isolation
(\S\ref{thefield},\S\ref{discussion},\S\ref{comparison}). 

\section{The IC~2233 and NGC~2537 Field\protect\label{thefield}}
The existence of a true physical association between IC~2233 and NGC~2537
has been a matter of previous debate (see de Vaucouleurs et al. 1976;
Schneider \& Salpeter 1992; Oosterloo 1993; Gil de Paz et al. 2000a,
Wilcots \& Prescott 2004). Their
surrounding field also contains a number of fainter galaxies whose
relationship to our targets has remained unclear.  Recently,
some new insight has been provided by observations 
from the Sloan
Digital Sky Survey (SDSS).\footnote{Funding for the creation and
distribution of the SDSS Archive has been provided by the Alfred P. Sloan
Foundation, the Participating Institutions, the National Aeronautics and Space
Administration, the National Science Foundation, the U.S. Department of Energy,
the Japanese Monbukagakusho, and the Max Planck Society. The SDSS Web site is
http://www.sdss.org/.  The SDSS is managed by the Astrophysical Research
Consortium (ARC) for the Participating Institutions.  The Participating
Institutions are The University of Chicago, Fermilab, the Institute for
Advanced Study, the Japan Participation Group, The Johns Hopkins University,
Los Alamos National Laboratory, the Max-Planck-Institute for Astronomy (MPIA), 
the Max-Planck-Institute for Astrophysics (MPA), New Mexico State University,
University of Pittsburgh, Princeton University, the United States Naval
Observatory, and the University of Washington.}  

\begin{figure}
\plotone{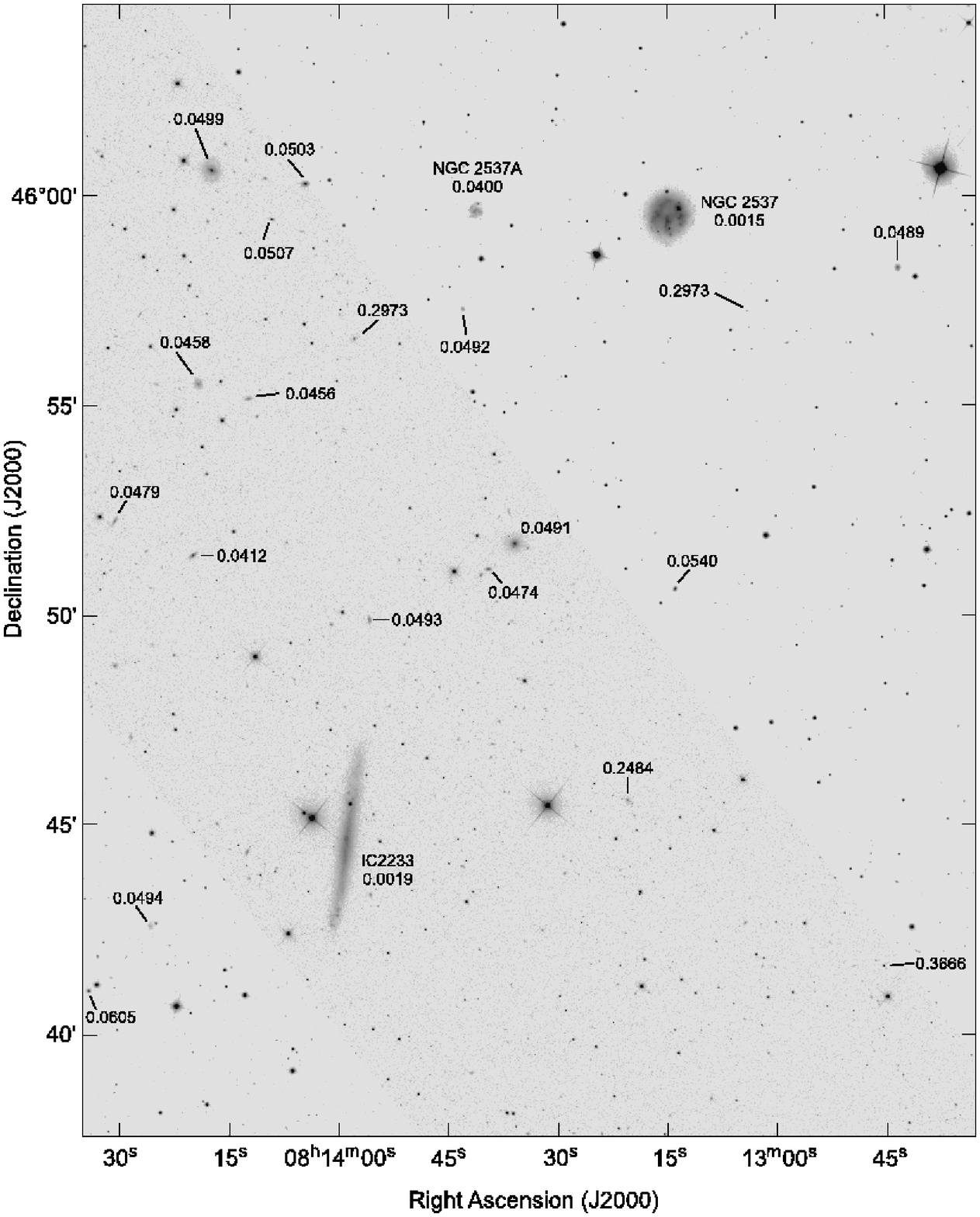}
\figcaption{Image of the IC2233/NGC2537 field obtained from a mosaic of
archival data from the SDSS.  The grey-scale representation
is based on photometric data from the $u$, $g$ and $r$ bands.  The overplotted
redshifts were also obtained from the SDSS archive.
\protect\label{fig:opticalfield}}
\end{figure}

We have obtained the publically available spectra and
five-band photometry of the galaxies observed by the SDSS
within one degree of the position of IC~2233.  Figure~\ref{fig:opticalfield} is
a grey-scale representation of a composite $u$, $g$, $r$ mosaic frame of the field
with the SDSS redshifts of the galaxies indicated.  The SDSS spectra
confirm the redshifts of IC~2233 (0.0019) and NGC~2537 (0.0015), but also reveal
a group of galaxies at $z \sim 0.04$ that includes NGC~2537A, 
whose association with
NGC~2537 had been uncertain (see Davis \& Seaquist 1983,
Gil de Paz et al. 2000a).  It now appears likely that a 
prior report of a redshift for NGC~2537A
equal to that of NGC~2537 was due to confusion of the 
two galaxies in a single-dish \HI\ measurement.
Indeed, we have not detected any \HI\ emission toward NGC~2537A with
the VLA. As described in \S\ref{discussion}, we
have also performed a systematic search 
for other gas-rich neighbors to IC~2233 and NGC~2537, but have not 
identified any candidates.

The association between IC~2233 and NGC~2537 has 
remained controversial because previous
distance estimates for both galaxies span a range of values.  
Assuming that the
galaxies are at rest with respect to the cosmic background radiation leads to
corrected recessional velocities of $\sim 710$~\kms (IC~2233) and $\sim 590$~\kms
(NGC~2537) and Hubble-flow distances of $\sim 10.0$~Mpc (IC~2233) and
$\sim 8.3$~Mpc (NGC~2537), with a formal uncertainty of 5\% (assuming a
Hubble constant of 71~\kmsm).

Estimates of the distance to IC~2233 derived from velocity flow models range from
10.4-10.9~Mpc (Tully 1988; Schwarzkopf \& Dettmar 2000; 
Garc\'\i a-Ruiz et al. 2002).
From the $B$-band Tully-Fisher relation, Garc\'\i a-Ruiz et
al. (2002) derived a distance of 8.1~Mpc, while, using the $I$-band Tully-Fisher
relation, Gil de Paz et al. (2000a) stated that IC~2233 is located at a
distance of 13~Mpc (quoting an unpublished, private communication;
these estimates all assumed a Hubble constant of 75~\kmsm).  
Based on {\it Hubble Space Telescope (HST)} 
data from Seth et al. (2005), Tikhonov et al. (2006) recently derived a
distance to IC~2233 of $10.4\pm 0.4$~Mpc using the ``tip of the red-giant branch''
method.  However, the uncertainty quoted by these latter authors appears 
to be purely statistical, without allowing for the
uncertainties in the zero-point calibration and metallicity
corrections.  Re-analyzing the same data, A. Seth (private
communication) finds a distance modulus of 30.25 using
an edge-detection algorithm and 30.11 using a Monte-Carlo approach
(see Seth et al. 2005 for details on these methods).  The
difference is indicative of the low signal-to-noise of the data.  Taking into
account the uncertainty in the zero-point calibration and metallicity corrections
therefore raises the uncertainty in the distance to IC~2233 determined with this
method to $\sim 1.5$~Mpc.

In the case of NGC~2537, Tully (1988) estimated a distance of 9~Mpc
based on a peculiar velocity flow model of the local
supercluster. More recently, Sharina et al. (1999) 
derived a distance  of 6.9$^{+1.4}_{-1.2}$~Mpc based on the magnitudes
of its brightest resolved stars. 

Given the uncertainties in the distances to both NGC~2537 and IC~2233, we cannot
exclude with absolute certainty the possibility of a true physical association
between them. However, as we argue in \S\ref{discussion}, even if
they do lie at the same distance, it appears unlikely that
either galaxy has been significantly influenced 
by a mutual interaction. For the present work we adopt 
$d=$10~Mpc for IC~2233 and $d$=7~Mpc for NGC~2537, but
throughout the paper, we indicate the distance dependence of all
physical quantities using the symbols $d_{10}$ and $d_{7}$ 
to denote the true distances
to the galaxies in units of 10~Mpc and 7~Mpc, respectively.   

\section{Optical Imaging and Photometry of IC~2233\protect\label{optical}}
\subsection{Observations and Data Reduction}
Harris $B$ and $R$ CCD imaging observations of IC~2233 were obtained on 1997
November~7 using the 3.5-m WIYN telescope\footnote{The WIYN Observatory is a
joint facility of the University of Wisconsin-Madison, Indiana
University, Yale University,
and the National Optical Astronomy Observatory.} at Kitt Peak, AZ. Deeper $B$ and $R$
images, as well as a narrow-band \HA+[\NII] image were obtained with the same observing
setup on 1999 January~19 under non-photometric conditions.  The \HA+[\NII] image was
obtained with the WIYN $W015$ filter with a bandpass of 73\ang, centered at 6570\ang.

The imaging camera employed a thinned STIS 2048$\times$2048 CCD with \as{0}{2}
pixels, giving a field of view of $\sim$\am{6}{8} per side.  The gain was
$2.8 e^{-}$ per ADU and the readout noise was $\sim 8 e^{-}$ per pixel.  Exposure
times were 550 seconds in $B$ and 180 seconds in $R$ during the 1997
observations, and 900 seconds in $B$, 750 seconds in $R$, and 1200 seconds in
\HA\ during the 1999 observations.  Seeing during the 1997 run was \as{0}{7} 
and during the 1999 run was \as{0}{5}-\as{0}{7}.

The images were processed using standard IRAF\footnote{IRAF is distributed by
the National Optical Astronomy Observatories, which is operated by the
Associated Universities for Research in Astronomy, Inc., under cooperative
agreement with the National Science Foundation.} routines.  Overscan and bias
levels were subtracted from individual frames and the data were flat-fielded
using an average of 5 dome flats taken with the appropriate filter. This yielded
images flat to $\sim 2$\% in $B$ and to better than 1\% in $R$.  A single bad column
in the CCD was corrected via interpolation from the adjacent two columns.

During the 1997 November run we derived calibrations by observing standard
fields from Landolt (1992) at three different air masses.  The data show excess scatter
with respect to the best-fit secant law, which suggests that some cirrus might have been
present during part of the night.  
These uncertainties are reflected in the rather
large errors in the photometry listed in Table~1.

Photometry of IC~2233 was performed using a single elliptical aperture.
Foreground stars, cosmic rays, and background galaxies were removed first via
background interpolation from surrounding regions.  Sky brightness levels were
determined by measuring the mean sky counts in several boxes at different
locations on the image.

Basic photometric parameters for IC~2233 are presented in Table~1.  Total errors
in the magnitude measurements were computed following Matthews \&
Gallagher (1997) and take into account sky, flat-field, and Poisson errors as
well as the formal uncertainty in the photometric solution.  They are dominated
by the calibration uncertainties described in the previous paragraph.  We have
corrected all quantities for Galactic extinction following Schlegel et
al. (1998).  Our new measurements (excluding corrections for Galactic
extinction) agree within the uncertainties with those published by de~Vaucouleurs
et al. (1991; $m_{B}$=13.48$\pm$0.18) and
Swaters \& Balcells (2002; $m_{R}$=12.53$\pm$0.02).

\begin{deluxetable}{lcc}
\tabletypesize{\scriptsize}
\tablewidth{0pc}
\tablenum{1}
\tablecaption{Optical and Infrared Properties of IC~2233}
\tablehead{\colhead{Parameter} & \colhead{Value} & \colhead{Ref.}}
\startdata

$\alpha$ (J2000.0) &  08 13 58.9  & 1\\
$\delta$ (J2000.0) &  +45 44 34.3 & 1\\
Hubble type &  Sd & 2 \\
Distance\tablenotemark{a} &  8--12 Mpc & 2\\
$A_{B}$ (mag) & 0.223 & 3 \\
$A_{R}$ (mag) & 0.138 & 3\\

\\
\tableline
\multicolumn{3}{c}{Measured Quantities} \\

\tableline

Position angle &  $172^{\circ}\pm 1^{\circ}$  & 2 \\
Inclination  & $88.5^{\circ} \pm 1.5^{\circ}$ & 2\\
$a/b$\tablenotemark{b} & 7.0  & 2\\
$D_{R,25.5}$\tablenotemark{b} & \am{5}{17} & 2\\
$m_{B}$\tablenotemark{c,}\tablenotemark{d}  & $13.05 \pm 0.14$ & 2 \\
$m_{R}$\tablenotemark{d,}\tablenotemark{e} & $12.38 \pm 0.07$ & 2 \\
$B-R$\tablenotemark{d} & $0.67 \pm 0.15$  & 2 \\
$\mu_{B}(0)$\tablenotemark{f} & 21.3 mag arcsec$^{-2}$ & 2 \\
$h_{r,B}$\tablenotemark{g}  & 32$'' \pm$~2$''$ & 2  \\
$h_{r,R}$\tablenotemark{g} & 29$'' \pm$~2$''$  & 2 \\
$h_{z,R}$\tablenotemark{h} & \as{4}{9}$\pm$~\as{0}{1} & 2 \\

\\
\tableline

\multicolumn{3}{c}{Derived Quantities} \\

\tableline

$A_{R,25.5}$\tablenotemark{b} & 15.0~$ d_{10}$ kpc & 2 \\
$M_{B}$\tablenotemark{i} & $-17.65$ & 2 \\
$L_{B}$\tablenotemark{i} & 1.8$\times10^{9}d^{2}_{10}~ L_{\odot}$ & 2 \\
$\mu_{B,i}(0)$\tablenotemark{k} & $\sim$~22.6 mag arcsec$^{-2}$ & 2 \\

$L_{H\alpha}$\tablenotemark{l} & $6.1\pm1.8\times10^{39} d^{2}_{10}$~erg~s$^{-1}$ & 2 \\
$L_{FIR}$ & $7.1\pm 0.9\times10^{7}d^{2}_{10}~L_{\odot}$ & 1,2 \\

\enddata

\tablecomments{Units of right ascension are hours, minutes, and
seconds, and units of declination are degrees, arcminutes, and
arcseconds.}

\tablenotetext{a}{We adopt a distance of 10~Mpc (see
text). Distance-dependent quantities are scaled in terms of $d_{10}$,
the actual distance in units of 10~Mpc.}

\tablenotetext{b} {Measured at an $R$-band surface brightness of 
25.5 mag arcsec$^{-2}$.}

\tablenotetext{c}{Error estimate includes $\pm 0.08$ mag formal
error and 0.11 mag zero point uncertainty.}

\tablenotetext{d}{Corrected for foreground extinction only.}

\tablenotetext{e}{Error estimate includes $\pm 0.04$ mag formal
error and 0.05 mag zero point uncertainty.}

\tablenotetext{f}{Measured central surface brightness corrected for 
foreground extinction}

\tablenotetext{g}{Disk scale length based on truncated exponential fit.}

\tablenotetext{h}{Disk scale height based on exponential fit along the minor axis.}

\tablenotetext{i}{Corrected for internal extinction (0.70 mag) 
following Tully et al. 1998.}

\tablenotetext{k}{Deprojected central surface brightness based on truncated exponential fit
including finite disk thickness and 
corrected for internal and foreground extinction.}

\tablenotetext{l}{Uncorrected for the contribution of [\NII].}

\tablerefs{(1) NED database; (2) this work; (3) Schlegel et al. 1998.}

\end{deluxetable}

\subsection{Properties of the IC~2233 Stellar Disk\protect\label{opt}}
Our $R$-band image of IC~2233 is presented in Figure~\ref{fig:WIYNR} (left
panel).  Here we draw attention to a few key features of this image.

Examining first the central regions of IC~2233, we see that the galaxy
shows no well-defined equatorial dust
lane.  Only a few small patches of extinction are evident in projection along
the inner few kiloparsecs of the galaxy, together with a narrow
dust structure inclined roughly 45$^{\circ}$ to the disk plane. This
latter feature can be traced both above and below the mid-plane 
(Figure~\ref{fig:WIYNR}), and its center
corresponds with the kinematic center of the galaxy (see
\S\ref{kincenter}).  The nature of this minor axis dust lane is
presently unclear;
it could be dusty debris recently accreted onto IC~2233, or
it may comprise material being expelled from a  starburst nucleus
or AGN, analogous to what is observed in M82 (Alton et al. 1999). However,
Seth et al. (2006) reported no evidence for a nucleus in
IC~2233. Another possibility is that it represents dense gas
streaming along a vertical density perturbation such as a spiral arm.

Because of the lack of a true equatorial dust lane in IC~2233, it is difficult to
determine precisely the inclination of the galaxy.  Using the 25.5~mag
arcsec$^{-2}$ $R$-band isophote (after correction for foreground
extinction), we measure  a semi-minor to semi-major axis ratio
$b/a$=0.14. This ratio is very close to the mean intrinsic flattening
expected for Sd galaxies [$(b/a)_{0}\sim$0.15; Yuan \& Zhu 2004], suggesting
that IC~2233 is observed no more than a degree or two from edge-on.  This is
consistent with the small scale height of the stellar disk (see below)
and the symmetry in the light
distribution about the midplane. We therefore adopt $i=$\ad{88}{5}$\pm$\ad{1}{5}.

We have estimated the radial scale length of the IC~2233 disk by fitting a
truncated exponential model to the major-axis light profiles in the $B$ and
$R$ bands.  In the $B$-band we derive
$h_{r,B}=32''\pm2''$ ($\sim$1.6$~d_{10}$~kpc) with a cutoff
radius\footnote{Following Pohlen et al. 2000, we define the cutoff radius
as the location where the radial surface brightness profile vanishes
asymptotically into the noise.} of 165$''$ and in the $R$-band we find
$h_{r,R}=29''\pm2''$ ($\sim$1.4$~d_{10}$~kpc) with a cutoff radius of
185$''$. We caution however that an exponential does not provide an ideal fit
to the major axis light profile of IC~2233; it significantly underestimates the
amount of light in the intervals $r = 0''-10''$ and $r = 90''-140''$ in~$B$
and $r = 65''-140''$ in $R$.  Our scale length estimates should therefore
be considered only as approximations to the more complex underlying
stellar light distribution.

Perpendicular to the disk, the $R$-band brightness profile along
the minor axis of IC~2233 is well described by an exponential function
with scale height \as{4}{9}$\pm$\as{0}{1} (240$~d_{10}$~pc).   This is
somewhat larger than the $K_{s}$-band scale height
$h_{z,K}$=\as{3}{77} reported by Seth et 
al. (2005)\footnote{For comparison of our exponential disk fits
with the isothermal disk fits of Seth et al. 2005, we assume
$h_{z}\approx0.549z_{0}$, where $z_{0}$ is the isothermal scale
height.}. However, this difference is consistent with other edge-on, late-type galaxies
where scale heights are found to be systematically larger in $R$-band
compared with the near-infrared (Bizyaev \& Mitronova
2002). In agreement with its identification as a superthin galaxy by
Goad \& Roberts (1981), the
global  scale height of IC~2233 is on the small end for Sd galaxies
(cf. Bizyaev \& Mitronova 2002; Seth et al. 2005), although it is notably
thicker than the more massive superthin galaxy UGC~7321
($h_{z}\approx150~d_{10}$~pc; Matthews 2000). 
In addition, we see
evidence that the stellar disk of IC~2233 flares at larger galactocentric
radii (see below).

The excellent seeing during our 1999 observing run shows the disk of
IC~2233 partially resolved into a number of bright, point-like sources.
These features are most readily identifiable
in the outer reaches of the disk, where they stand out strongly against
the comparatively faint, diffuse underlying disk light.
Our \HA\ image (discussed below) does not reveal any
compact sources at these locations, implying that the
majority of these are not compact \HII\ regions.
Using aperture photometry, we have estimated magnitudes and colors for
38 point sources throughout the galaxy
that appear to be unblended and do not correspond with
bright, underlying \HA\ emission. After correction for
foreground and internal extinction, the colors we estimate for these
features lie in the range $(B-R)_{0}\approx0.2$ to 3.5, and absolute
magnitudes
range from $M_{R}\approx-7$ to $-8.7$. These ranges of color and
luminosity are  consistent with a population of blue and red
supergiant stars (Johnson 1966; Dohm-Palmer \& Skillman 2002). While
some of the bluer sources may  be young or
intermediate-age  star
clusters or associations, 15 of the sources have $(B-R)_{0}>2$, making
them redder than even the oldest globular clusters (e.g.,
Bridges et al. 1996). This is another indication that we are resolving a
population of individual luminous stars. Using {\it HST} 
images, Seth et al. (2005) also identified a population
of red and blue supergiants in a portion of the inner disk of IC~2233.

Supergiants represent massive stars that have depleted their core
hydrogen, and those luminous enough to be identified in our
ground-based images are likely to have ages of no more than
$\sim$20~Myr (see Dohm-Palmer \& Skillman 2002).  This implies that
these stars should still lie very near their birth locations.  This makes
the distribution of supergiant candidates in the outer disk of IC~2233
noteworthy, as they show no obvious concentration along the
midplane, but rather a wide spread of $z-$heights. 
This trend is particularly striking compared
with the steep, roughly exponential vertical light profile seen near the
central parts of the galaxy and implies an overall flaring of the young
stellar disk.  We also see evidence for flaring in both the ionized and
neutral layers in IC~2233 (see \S\ref{Halpha} \& \ref{totalHI}). 
Interestingly, the scale heights
of the gaseous and stellar components appear to be roughly
comparable in the outer disk of IC~2233---a factor that may be linked
with the rather low global star-formation efficiency in this galaxy (see
also Ferguson 1998; Ferguson et al. 1998; Matthews \& Wood 2003).

\subsection{The Ionized Gas in IC~2233\protect\label{Halpha}}
\begin{figure}
\epsscale{0.75}
\plotone{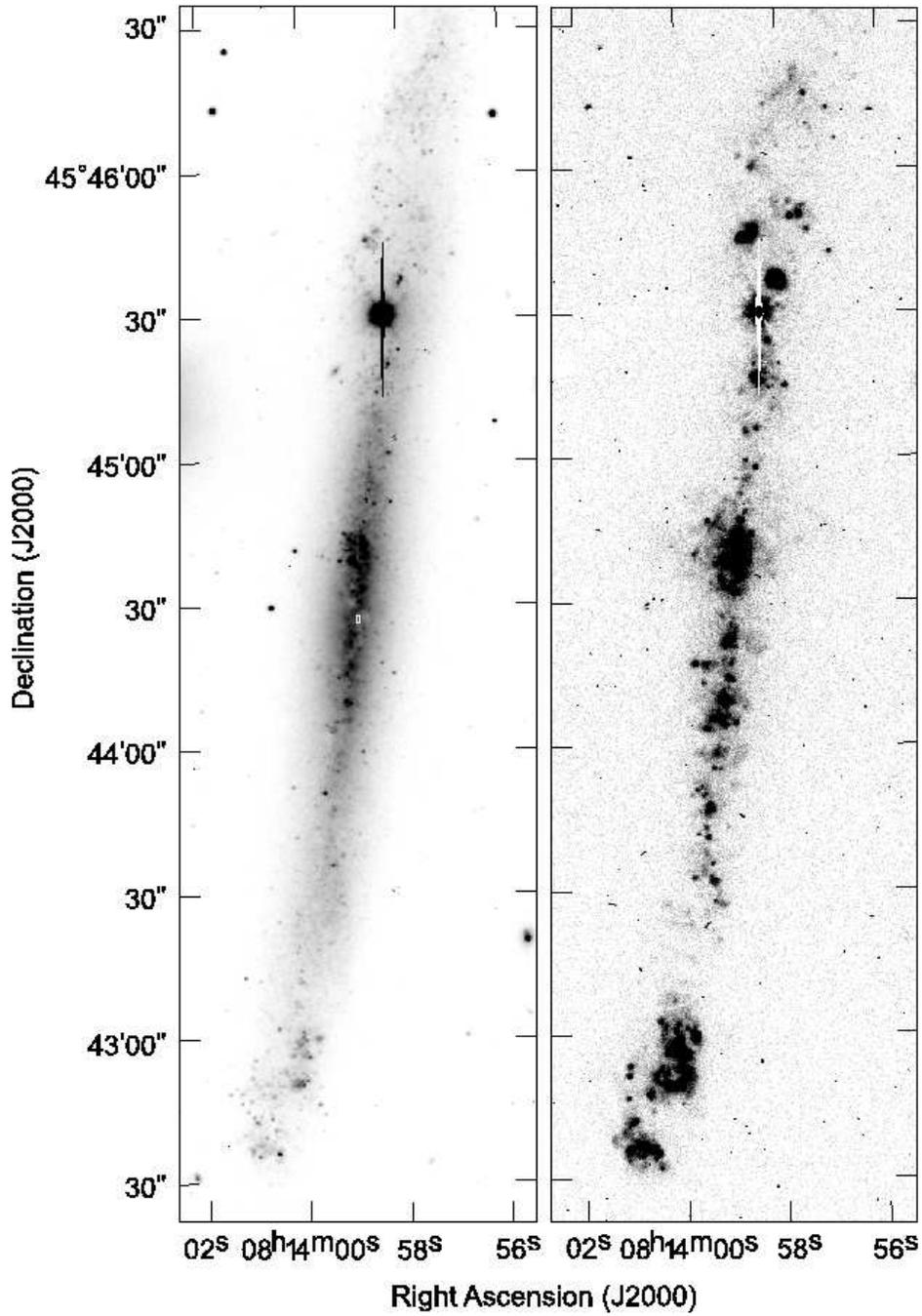}
\figcaption{WIYN images of IC~2233 in the 
$R$-band (left) and continuum-subtracted \HA+[\NII]
(right).  The small
white rectangle on the left image denotes the kinematic center of the
galaxy (see \S\ref{kincenter}).  Bleeding CCD columns from a bright
foreground star are visible along the
north-south direction near $\alpha_{\rm J2000} = 08^{h} 13^{m} 59^{s}$, 
$\delta_{\rm J2000} = +45^{\circ} 45' 30''$.
\protect\label{fig:WIYNR}}
\end{figure}

Our continuum-subtracted \HA+[\NII] image of IC~2233 is reproduced
in Figure~\ref{fig:WIYNR}.  Although our bandpass was broad enough
to encompass the [\NII] lines at 6548\AA\ and 6583\AA\ in addition to
\HA, the [\NII] emission in IC~2233 is known to be quite weak (Goad \&
Roberts 1981) and therefore should contribute minimally to this image.

The ionized gas distribution in IC~2233 has been studied previously by
Rand (1996), who drew attention to plumes of emission extending to more
than 1~$d_{10}$~kpc from the plane, near the center of IC~2233 (see also
Miller \& Veilleux 2003).  These features are also visible in our
Figure~\ref{fig:WIYNR}.  Rand (1996) suggested that this vertically-extended
emission may be due to an outflow from a central starburst.  However,
comparison with the accompanying broadband image in
Figure~\ref{fig:WIYNR} shows that the region from which the ionized plumes
originate corresponds (in projection) to a knotty, extremely blue
[$(B - R)_{0} \sim 0.3$] clump of starlight that is displaced from both the galaxy
midplane and from the isophotal and kinematic centers of the galaxy. 
This region has an extent of $\sim 380~d_{10}$~pc.  Its structure, size, and
color are consistent with a cluster of multiple OB associations (Pietrzy\'nski
et al. 2005).  The off-planar appearance of this ``superassociation'' and the
lack of any other comparable features within the galaxy suggest 
the possibility that this could be a remnant from a dwarf intruder.  This would be 
consistent with the very  different emission line ratios found by Goad \&
Roberts (1981) for the central regions of IC~2233 compared with the southern
end of the disk, although we find no kinematic evidence to support this
hypothesis (see below; see also Garrido et al. 2004).

As also noted by Rand (1996), IC~2233 does not  have a
widespread extraplanar layer of ionized gas, consistent with the low
global star formation rate of this galaxy. Nonetheless, as seen in
Figure~\ref{fig:WIYNR}, the ionized emission exhibits a great deal of
complex
structure, ranging from small-scale features (compact
\HII\ regions, filaments, and shells) to diffuse planar emission, to
bright, giant \HII\ complexes.

The largest and most luminous \HII\
complex visible in IC~2233 is situated on the southern end of the
disk. This region has a projected extension of more than $2~d_{10}$~kpc
and
contains $\sim$30\% of the total \HA\ luminosity of IC~2233.  The
location of this complex well outside the central regions of the
galaxy underscores that
recent high-mass star formation in IC~2233 has not been
centrally concentrated. Giant, off-center
\HII\ complexes are also seen in other LSB 
galaxies with similar masses to IC~2233, such as the
nearby, face-on LSB
galaxy NGC~4395 (Wang et al. 1999). These complexes may be more
prevalent in these low-mass disks with slowly rising rotation curves,
since these systems lack the shearing forces
that would tend to rapidly stretch and distort similar large-scale
density enhancements in more massive spirals
(Elmegreen \& Hunter 2000). In addition, the weaker self-gravity and 
intrinsically thicker gas disks of small spirals may allow in general 
the formation of larger bound condensations (Kim \& Ostriker 2006).
Contrasting with these giant \HII\ complexes, some lines of sight
through the IC~2233 disk show little \HA\ emission. This
implies that the recent star formation in IC~2233 has been patchy and highly
localized, perhaps occurring
along loose spiral arms analogous to those visible in NGC~4395.

Turning to smaller scales, several shell- or bubble-like features of various surface
brightnesses can be seen in the ionized ISM of IC~2233. One
example is the circular, low surface brightness feature centered at
$\alpha_{\rm J2000.0} = 08^{h}13^{m}58.5^{s}$,
$\delta_{\rm J2000.0}$ = +45$^{\circ}$44$'$\as{21}{1} (Figure~\ref{fig:WIYNR}),
which has a radius of $\sim2''$ ($r \sim 100~d_{10}$~pc) and a
projected height of $\sim100~d_{10}$~pc from the disk midplane.  Its size
and morphology are consistent with a wind-blown superbubble produced
by an association of hot, young OB stars (Mac Low \& McCray 1988).
Supporting this interpretation,
our $B$ and $R$ continuum images show a blue point source at the
apparent center of the bubble.

Another interesting aspect of the ionized gas morphology of IC~2233 is
that the various \HII\ regions and complexes in IC~2233 are
not confined to a thin layer. This effect becomes more pronounced toward
the outer parts of the galaxy,  where a number of bright \HII\ regions or
complexes appear to be displaced by $\sim0.5h_{z}$ or more from the
midplane. The brightest \HII\ complexes also seem to trace a rather
wavy pattern about the midplane, although we are unable to identify a
clear correspondence between these undulations and the more
well-defined corrugated pattern seen in the neutral hydrogen disk 
(\S\ref{corrugations}). Overall, the complex,
``frothy'' \HA\ morphology of
IC~2233 forms an interesting contrast to the much flatter
ionized gas disk of the superthin galaxy UGC~7321, which shows very
little complex structure and no sign of giant (kiloparsec-scale) \HII\
complexes (cf. Figure~7 of Matthews et al. 1999).

Although our \HA\  image is not flux-calibrated, we have estimated an \HA\
luminosity for IC~2233 by using our $R$-band photometric calibration and
the response curves for the $R$-band  (continuum)
and narrow-band (H$\alpha$+[\NII]) filters. In this manner we
derive $L_{\rm H\alpha}\approx6.1\pm1.8\times10^{39}d^{2}_{10}$~erg
s$^{-1}$. This value is corrected neither for internal extinction
nor for the contribution from [\NII] emission in the narrowband filter (likely
to be a $\sim$10\% effect; see Goad \& Roberts 1981). Our estimate for
$L_{\rm H\alpha}$ is in
good agreement with a value derived from the data of Rand (1996)
($L_{\rm H\alpha}\approx 5\times10^{39}d^{2}_{10}$~erg~s$^{-1}$, 
with an uncertainty of $\sim 10$\%; R. Rand 2003, private
communication).   Using the relations given by Kennicutt (1998), the
observed \HA\ luminosity translates to a global star formation rate of
$\sim0.05M_{\odot}$~yr$^{-1}$.  Thus the disk-averaged star formation rate
in IC~2233 is quite low despite various signatures of ongoing,
{\it localized} star formation (e.g., \HII\ complexes; a blue global
color, OB associations, and candidate populations of supergiant stars). 
For comparison, we can also compute an estimate of the star formation
rate based on the {\it IRAS} FIR emission.  Using the 60~$\mu$m and
100~$\mu$m fluxes from the NED database we find
$L_{\rm FIR}=(7.1\pm0.9)\times10^{7}d_{10}^2 ~L_{\odot}$.  Following
Kennicutt (1998), this translates to a global star formation rate of
0.02~$M_{\odot}$~yr$^{-1}$, where we have assumed that the total
IR luminosity is $\sim1.75L_{\rm FIR}$.  This estimate is more than
a factor of two smaller than the \HA\ estimate, consistent with the
suggestion of Bell (2003) that the FIR may systematically underestimate
the total star formation rate of low-luminosity galaxies.  In any case, both
estimates represent the extreme low end of observed star formation
rates for Sd spirals (Kewley et al. 2002).  This is consistent with 
the low emission line intensity ratios measured in IC~2233 by
Goad \& Roberts (1981) and Miller \& Veilleux (2003) and with 
the general classification of IC~2233 as an LSB spiral.

\section{VLA Observations}
We observed IC~2233 and NGC~2537 simultaneously in the \HI\ 21-cm
line using the VLA in its C (3~km) configuration\footnote{Since 1998 the
C~configuration has been modified from its original form by placing the
fifth antenna of the north arm at the center of the array in order to improve
sensitivity to extended emission.} on 2000 May~28 (hereafter Day~1) and
2000 May~29 (Day~2).  The observations were performed over two 8-hour
sessions, with a total of $\sim 11.6$~hours of data acquired on the target
field. Approximately 97\% of the data were obtained at elevations greater
than 40$^{\circ}$.

Our observations were carried out using the 4-IF spectral line observing mode
with a 1.5~MHz bandpass per IF pair (left and right circular polarization) and
on-line Hanning smoothing. This resulted in 63~channels of width 24.4~kHz
($\sim$5.2~\kms) per IF~pair. One IF pair was centered on the published
redshifted velocity of IC~2233 ($V_{\rm hel}$=565~\kms) while the other
IF pair was centered on that of NGC~2537 ($V_{\rm hel}$=447~\kms).
The field center was
chosen to be $5'$ north of the optical center of IC~2233 to prevent excessive
attenuation at the position of NGC~2537, with only a $\sim 7$\% loss at the
position of IC2233.  Five-minute observations of J0713+438 were interspersed
with 30-minute observations of the IC2233/NGC2537 field in order to determine
phase calibrations as well as bandpass corrections (see \S\ref{calib}).
In addition, J0137+331 (3C48) and J1331+305 (3C286) were each observed for
10~minutes each day for use as flux calibrators.  Details of our observations
are summarized in Table~2.


\begin{deluxetable}{ll}
\tabletypesize{\scriptsize}
\tablenum{2}
\tablewidth{0pc}
\tablecaption{Summary of IC~2233/NGC~2537 \HI\ Observations}
\tablehead{\colhead{Parameter} & \colhead{Value} }
\startdata

\multicolumn{2}{c}{Observing Set-Up} \\
\\
\tableline

\\

Array configuration &  C \\

Baseline range  & 30 -- 3385 m \\

Observing dates ($2 \times 8$~hours) &  2000 May 28 (Day~1), 29 (Day~2) \\

Pointing and phase center, $\alpha$ (J2000.0) &  08 13 58.8 \\

Pointing and phase center, $\delta$ (J2000.0) & +45 49 36.3 \\

Total on-source observing time & 696 minutes (683 min. at elevation $>40^\circ$)
\\

Flux calibrators &  J0137+331 (3C48), J1331+305 (3C286) \\
Phase calibrator  &  J0713+438 \\

Number of IFs &  4 [2 $\times$ (RR, LL)]\\

Channel width (after Hanning smoothing) &  24.4 kHz \\

Velocity separation of channels & $\sim$~5.2 km s$^{-1}$ \\

Number of channels per IF & 63 \\

Heliocentric central velocity (optical definition)    &  565.0 km s$^{-1}$
(IF pair 1); 447.0~km s$^{-1}$ (IF pair 2) \\

Primary beam (FWHM) &  $\sim$~\am{30}{6} \\

\tableline \\
\multicolumn{2}{c}{Deconvolved Image Characteristics} \\
\\
\tableline
\\

Robustness parameter (${\cal R}$)\tablenotemark{a} &  1 \\

Synthesized beam FWHM\tablenotemark{a} & $\sim$~\as{16}{1}~$\times$~\as{14}{8} \\

Synthesized beam position angle & $\sim -70^\circ$ \\

Linear resolution of synthesized beam\tablenotemark{b} &
$\sim 0.8~d_{10}$~kpc \\

rms noise per channel (1$\sigma$) & (0.37--0.44)~mJy~beam$^{-1}$\\

rms noise in total-intensity image (1$\sigma$) &
$\sim 5$~mJy beam$^{-1}$ km s$^{-1}$ \\

Limiting column density per channel (1$\sigma$) &
(8.8 -- 10.5)$\times10^{18}$~atoms cm$^{-2}$ \\

rms noise in \HI\ mass
channel$^{-1}$, $M_{\rm HI}$~($1 \sigma$)\tablenotemark{b}
& (4.5 -- 5.3)$ \times 10^4$~$d_{10}^2 M_\odot$ \\

\enddata

\tablenotetext{a}{We also used ${\cal R}$~=~0.7 (and a circular restoring beam with
FWHM =~\as{15}{0}) to make the continuum image (see \S\ref{imcont})}

\tablenotetext{b}{$d_{10}$ is the distance to the target galaxies expressed in
units of 10~Mpc (the distance to IC~2233, see Table~1).}

\end{deluxetable}


\subsection{Data Reduction\protect\label{reduce}}
\subsubsection{Calibration\protect\label{calib}}
A vector average of the visibilities from channels 5-59 at each time
stamp was used to identify interference and corrupted data.  Because our
observations were obtained in the daytime during solar maximum, our data contain
significant solar interference at short spacings, and noticeably poorer phase
stability compared with \HI\ observations of other targets obtained by us on
adjacent dates, but  after sunset (Paper~I).  The Sun was $\sim 52^\circ$
away from our target field during the observations and relatively  
``quiet'' on our
first day but far more active on the second day, which included a  
period of about
one hour where only baselines between the farthest two antennas of  
each arm
gave usable data.  In addition, on the second day,  a significant
fraction of the data from three antennas (located at the East-2,
East-8, and East-16 stations, respectively) had to be
discarded due to electronic problems.  In spite of these losses, we have succeeded in  
making images of excellent quality,
and comparison with past single-dish observations
suggests that we have recovered the bulk of the \HI\ emission in both  
IC~2233 and NGC~2537 (see \S\ref{totalHI} \& \ref{NGCtotalHI}).

The flux density scale was determined using the primary calibrators 3C48 and
3C286 and adopting fluxes of 15.90~Jy and 14.74~Jy, respectively, at
1417.64~MHz. These fluxes were 
scaled to our other frequency setting following the precepts of the
VLA calibration manual.  Using one of the two primary flux calibrators to solve
for the flux of the other, and comparing this flux with its known (predicted)
value, indicates that our internal flux calibration uncertainty is less than
1\%.

We determined antenna-based amplitude and phase calibrations using the
observations of J0713+438 and an initial bandpass calibration 
from the average of the data obtained on 3C48 and 3C286.  This initial bandpass
calibration resulted in some curvature in the spectra taken through strong
point sources in the IC~2233/NGC~2537 field.  We therefore evaluated and applied further
scan-based bandpass corrections by fitting 5th order Chebyshev polynomials to
channels 5-59 of each (scan) averaged spectrum of J0713+438 and corrected the
IC2233/NGC2537 data by linear interpolation (in time).   This procedure led to flat
bandpasses through the bright continuum sources in the field.  The
spectrum of the strongest source (4C+46.17) shows no significant deviations at
the level of 0.13\% (2.5~$\sigma$).

The IC~2233 field contains significant background continuum emission (totaling
$\sim 1.1$~Jy observed, i.e. before correction for the attenuation of the
primary beam) which is dominated by two bright point sources (4C+45.15 at 
$\alpha_{\rm J2000.0}$ = 08$^{h}$12$^{m}$42.0$^{s}$,
$\delta_{\rm J2000.0}$ = +45$^{\circ}$36$^{'}$\as{51}{3} 
and 4C+46.17 at $\alpha_{\rm J2000.0}$ = 08$^{h}$14$^{m}$30.4$^{s}$,
$\delta_{\rm J2000.0}$ = +45$^{\circ}$56$^{'}$\as{39}{4}). For these ``4C'' sources
we observed flux
densities of 531$\pm5$~mJy (best fit as two extended components with
formal uncertainty below the 1\% adopted here) and 1089$\pm11$~mJy (1\% calibration
uncertainty, with a formal error of 0.5~mJy). These measurements are in good agreement
with values from the NRAO VLA Sky Survey (NVSS; $528\pm16$~mJy and
$1106\pm33$~mJy, respectively; Condon et al. 1998).  The strength
of these two sources results in significant ghosts in the images of channels on both
edges of each IF pair.  These ghosts, which appear at the symmetric positions of
these sources with respect to the phase center of the observations 
were first discussed by
Bos (1984,1985) in his study of the Westerbork array.  They are always present but
rarely noticed at the VLA.  We observe them at a strength of $\sim4.3$\% in
channel~1 and can trace them up to $\sim$~channel~10 (although they would be
difficult to detect beyond channel~4 if their spatial location were unknown).  On
the other end of the band, the ghosts are clearly seen in channels~61--63.  The
cumulative effect of such ghosts is responsible for the well-known increase in
the noise of the channel images at the edges of the band in VLA spectral
observations (Uson 2007).

\HI\ emission is observed in channels 13--63 of IF1 and 1--45 of IF2,
leaving only a
small number of line-free channels.   Avoiding the ghost emission discussed
above, we used channels 5--12 of IF1 and channels 46--59 of IF2 to
improve our gain calibration through self-calibration.  We took the image
obtained from channels 5--12 of IF1 and day~1 as a reference to compute phase
corrections in 10-minute intervals for the observations of both IFs and both
days (using channels 5--12 for IF1 and channels 46-59 for IF2).  After applying the
corresponding phase corrections, we determined amplitude and phase corrections
with a 30-minute averaging time (one set per scan) restricting the baselines
used to those larger than 200$\lambda$ (the shorter baselines contained
significant amounts of corrupted data; see above).  The average of the
amplitudes was constrained to be constant in order to preserve the overall
amplitude calibration.

The frequency-averaged residual visibilities (after subtraction of the
theoretical contribution of the detected continuum emission) for each day and
each IF were used to find low-level contamination of the data.  We discarded the
corresponding data and performed one additional iteration of phase-only
self-calibration on 10-minute intervals which led to improved pseudo-continuum
images for each day and IF.

\subsubsection{Imaging of the Line Emission\protect\label{imline}}
The strong continuum present in our field precluded a fast subtraction of the
continuum emission by
fitting the real and imaginary components of each visibility 
(Cornwell et al. 1992).  Instead, we have used an alternate, two-step
continuum subtraction procedure. First, we computed a continuum model
for each IF pair on each day by imaging the available line-free channels.  We
used channels 6-12 for IF1
and channels 46-59 for IF2.  The data were gridded channel by channel to avoid
bandwidth smearing.  We subtracted these models (channel by channel) from each
respective database and concatenated the residual data for each IF pair for
each of the two days.  Next, we obtained channel images using a 3$''$~cell size and a
robustness factor ${\cal R}=1$. The resulting synthesized beam was
\as{16}{0}$\times$\as{14}{8} at a position angle of $-70^{\circ}$.
Deconvolution to a level of 0.40 mJy beam$^{-1}$ was followed by restoration of
isolated components if their absolute value inside a 6-pixel radius was less
than 1~mJy~beam$^{-1}$. The rms noise in the resulting image cube ranged from
0.40-0.46~mJy beam$^{-1}$ channel$^{-1}$ (for channels 5--59).  We found that
this procedure removed the  bulk of the continuum emission, although some
low-level sidelobes (of order $\pm$1~mJy) from the brightest continuum sources 
in the field remained.  

To improve the continuum subtraction further, we combined the data for the
two IFs (which overlap in frequency) to make a ``global cube'' containing
86~channels and a 2.1~MHz bandwidth. The corresponding channels from both IFs
were averaged with equal weight after discarding four edge channels of each IF
(60--63 of IF1 and 1--4 of IF2).  The frequency misalignment between the two
IFs was approximately one-fifth of a channel, so this resulted in a small
amount of frequency smearing ($\sim$1.1~\kms); however, this is inconsequential
for the analysis of the spectral images that follows.  We blanked regions that
contained line emission in the global cube (leaving only the residual continuum)
and fit a first-order polynomial (in frequency) to each (spatial) pixel in the
blanked cube (see Cornwell et al. 1992).  Finally, we used the average level and slope
derived from this fit to remove the residual continuum from each
pixel/channel of the global cube containing the line emission.

The rms noise in the resulting continuum-subtracted global cube was reduced to
0.37-0.44~mJy beam$^{-1}$ channel$^{-1}$, and the noise showed a Gaussian
distribution exclusive of the line emission.  Noise in the channels where the
data from the two IF pairs were averaged have $\sim$10\% lower noise.  Spectra
through the locations of the brightest continuum sources appear extremely flat,
with an rms noise of $\sim$0.42~mJy and mean flux density levels across the
bandpass that are indistinguishable from zero.  We have used this resulting
continuum-subtracted global image cube for the analysis of the line emission in
IC~2233 and NGC~2537 described hereafter.
\subsubsection{Imaging of the Continuum Emission\protect\label{imcont}}
Although the continuum images used to process the spectral line
data were adequate for the purpose described in the previous section, they
contained a number of systematic errors that became more pronounced when
we averaged the four images (2 IFs, 2 days) to obtain a higher sensitivity continuum
image for the purpose of studying the emission
of IC~2233 and NGC~2537.  
As discussed in \S\ref{calib}, the field contains two ``4C'' sources
that are located at positions where the response of the primary beam
is at
$\sim 80$\% and $\sim 35$\%, respectively, of its maximum value.  
In addition, the off-axis
geometry of the Cassegrain optics of the VLA antennas induces a 
beam-squint which separates the right and left circular polarization beams by
$104''$ on the sky (Weinreb et al. 1977).  The response at the position of these
sources is different for both polarizations and it is modulated oppositely with
parallactic angle.  Therefore, the visibilities do not obey a simple convolution
equation and the image contains a significant and serious error pattern in the
form of concentric rings around the two main continuum sources. 
This pattern extends over
the full field-of-view and affects the flux density at the locations of IC~2233
and NGC~2537.

\begin{figure}
\epsscale{0.75}
\plotone{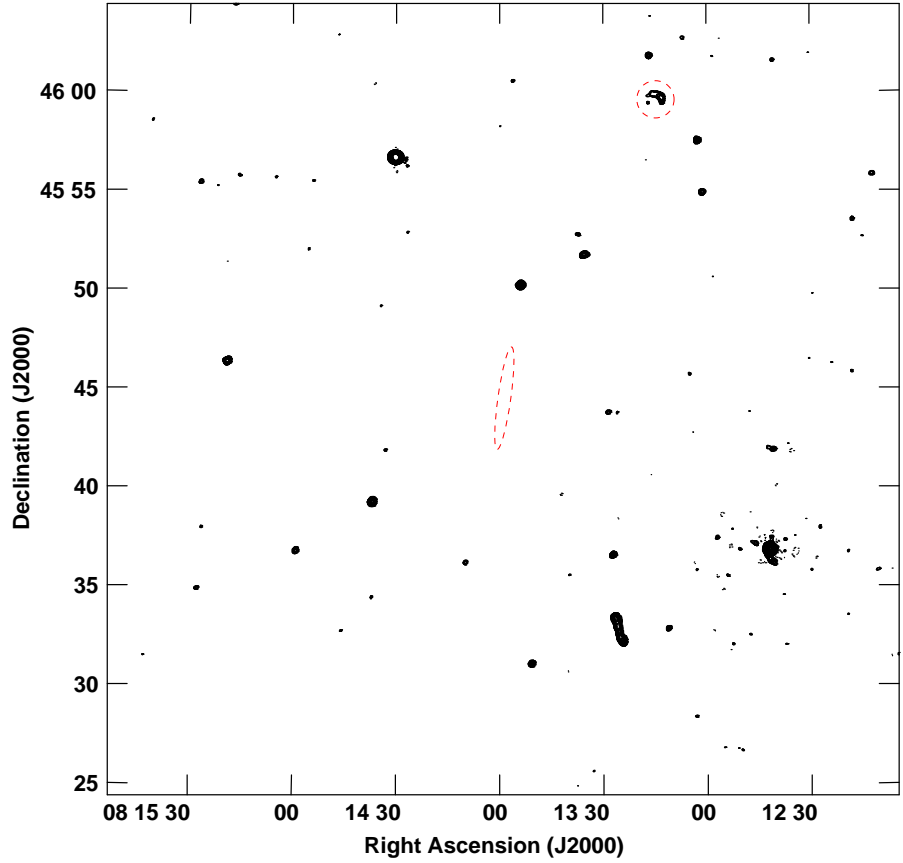}
\figcaption{Image of the IC~2233/NGC~2537 field in the 
21-cm continuum, showing the central $20' \times 20'$ region. The
position of 
IC~2233 is indicated by an ellipse and that of NGC~2537 by a circle.  
The contours are 
($-$1.4 [absent],-1,1,1.4,2,2.8,4,\dots,1024,1408)$\times
0.46$~mJy~beam$^{-1}$, where the lowest contour is 4$\sigma$. 
We detect continuum emission from NGC~2537 but no significant emission
from IC~2233.\protect\label{fig:continuum}}
\end{figure}

Although the beam-squint imposes a significant modulation on the continuum
emission in the IC~2233/NGC~2537 field, the effect is negligible provided that
the data are restricted to those visibilities for which both polarizations are
available.  In addition, it is important to ensure that amplitude self-calibration
does not ``follow the squint'' on the strongest source in the field.  Thus, we have
made an image of the continuum emission, restricting the data to those visibilities
for which both polarizations are available and pre-averaging both polarizations
prior to determining corrections to the amplitudes through self-calibration.  We
have restricted the baselines to those longer than 1600$\lambda$ to minimize
low-level cross-talk and solar contamination.  This results in a 12\%
drop in the number of
visibilities compared to the previous estimate of the continuum
(\S\ref{calib}).  Furthermore,
we have found that the ghosts discussed in \S\ref{calib} are still present in
channels~5 and 6 of IF1 and in channel~59 of IF2.  Therefore we made a new
image using channels~7--12 of IF1 as well as channels~46--58 of IF2.  We used
``3-D'' gridding and a robustness factor ${\cal R}=0.7$, which yields the optimal
compromise between noise and resolution for VLA imaging observations
obtained with full $u$-$v$ coverage (Briggs 1995).  The resulting image has an rms noise of
0.12~mJy~beam$^{-1}$ and is free of systematic errors, except at the positions of
the two brightest ``4C'' sources, where it is still somewhat dynamic-range limited.  This image,
which corresponds to an average frequency of 1418.3~MHz, is shown in
Figure~\ref{fig:continuum}.

\section{The 21-cm Continuum Emission in the IC~2233 Field\protect\label{contem}}
As previously described, the IC~2233/NGC~2537 field contains significant continuum
emission from background sources.  
The observed flux density (uncorrected by the
attenuation of the primary beam) in the central $1^\circ\times1^\circ$ is 1.129~Jy, with
a total of 1.145~Jy when  two flanking fields located outside the main
beam are included. The flux density is dominated by the two
aforementioned ``4C'' sources (see \S\ref{calib}).
 
For NGC~2537 we detect continuum emission
totaling $10.3\pm0.6$~mJy---in good agreement with
the NVSS ($10.5\pm1.7$). The peak emission of $2.4\pm0.2$~mJy lies at 
$\alpha_{\rm J2000} = 08^{h} 13^{m} 13.1^{s}$, $\delta_{\rm J2000} 
= +45^{\circ} 59' 40''$,
consistent with the location of the most prominent star forming
region in this galaxy.

We do not detect any significant 
continuum emission from IC~2233.  Within an area defined by the
stellar disk of the galaxy we measure an integrated
flux of $1.7\pm0.6$~mJy, after correction for the attenuation of the
primary beam. However,
given the dynamic range limitations and residual sidelobes discussed
in the previous section,
this should be considered non-significant and converted to
a ($2 \sigma$) upper limit of $<$3~mJy.  
The weak or absent radio continuum emission from IC~2233 is consistent
with its LSB nature and with 
the low global star formation rates derived from other tracers,
including \HA\ and FIR emission (see \S\ref{Halpha}). Using
the FIR luminosity of IC~2233 from Table~1, the
radio-FIR correlation of Condon (1992) predicts a 1.4~GHz continuum
flux of $\sim3$~mJy, consistent with our upper limit.

\section{Properties of the Neutral Hydrogen in IC~2233\protect\label{icHI}}
\subsection{Channel Images\protect\label{icchanmaps}}
\begin{figure}
\epsscale{1.}
\plotone{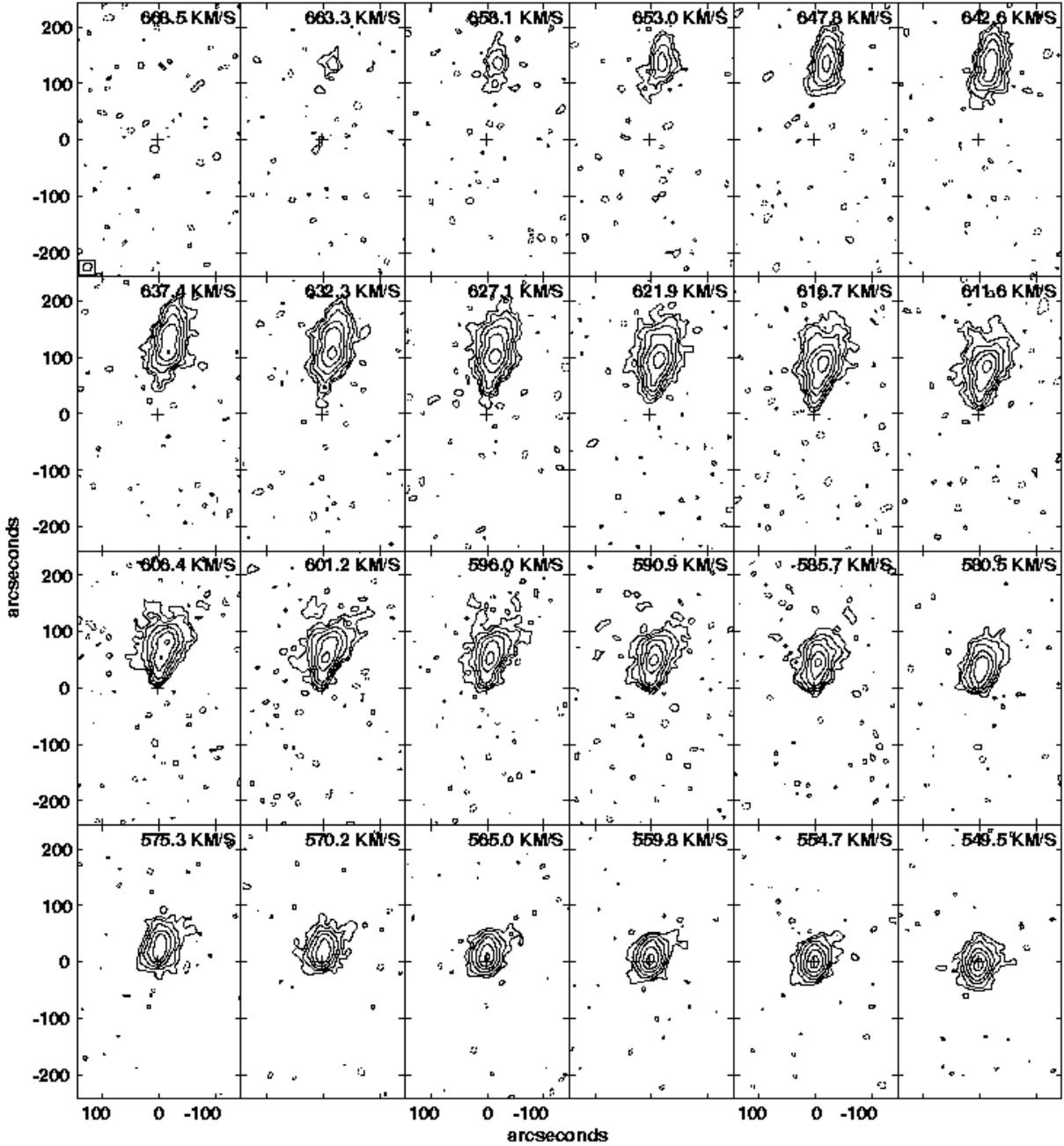}
\figcaption{\HI\ channel images for IC~2233.  The spatial resolution is
$\sim16''$.  The contours shown are
$(-2 {\rm [absent]}, -1, 1, 2, ...32 )\times1.0$~mJy~beam$^{-1}$. 
Only the channels showing line emission, plus one additional channel on either
side, are plotted.  Each panel is labeled with its corresponding 
heliocentric velocity.  The systemic velocity is $553.4\pm1.0$~\kms.
The kinematic center of the galaxy (\S\ref{kincenter}) is
indicated in each panel by a
cross.\protect\label{fig:iccmaps}}
\end{figure}

In Figure~\ref{fig:iccmaps} we show the individual channel images for all
velocities where \HI\ line emission was detected in IC~2233.  
Several key properties of the \HI\ distribution and
kinematics of IC~2233 can be gleaned from these images.

There are significant differences in the distribution and
morphology of the \HI\ emission on the approaching versus the receding
side of IC~2233.  For example, on the
receding (northern) side the emission distribution in the
individual channels appears more radially elongated;
all but the outermost few channels show emission near $x=0$ (where we
take $x$ to lie along the major axis of the galaxy, with positive $x$
toward the north), and the
channels in the velocity range 627.1--637.4~\kms\ show emission
extending along the full extent of the receding side of the
disk ($x\approx 0 - 200''$).  Within these latter channels, the
peak brightness occurs $\sim 80''$ from the galaxy center. On
the approaching (southern) side, the corresponding channels show a
more compact distribution of emission, with brightness peaking
$\sim120''$ from the galaxy center.  These differences imply an
asymmetry in the \HI\ distribution and velocity field
on the two sides of the galaxy (see also below).

The channel images in Figure~\ref{fig:iccmaps}
also provide insight into the vertical structure of
the neutral hydrogen in IC~2233.
On both the approaching and receding
sides of the galaxy---most notably in intermediate velocity
channels---the isointensity contours
broaden with increasing projected distance from
the  galaxy center, implying an overall flaring of the gas layer with
increasing galactocentric radius (see
also \S\ref{totalHI}).  Three-dimensional modeling would be necessary to
characterize the flaring in detail (see Olling 1996; Matthews \& Wood
2003). However, because IC~2233 is nearly exactly edge-on,
measurements of the vertical extent of the
emission in select channel images can provide estimates of how much
the gas layer thickness changes as a function of radius.
Channels near the velocity extrema contain
emission predominantly from the outer regions of the disk. Therefore,
the thickness of the emission along the
$z$ direction in these channels
provides an estimate of the maximum thickness
of the gas layer. At the same time (owing to the curvature of the
isovelocity contours),
emission observed near $x=0$ in channels of intermediate velocity
should arise from locations close to the true galaxy center
and be largely uncontaminated by outer
disk gas along the line-of-sight. The vertical thickness of this
emission therefore should give an estimate of the \HI\ scale height in the
inner galaxy.

To gauge the thickness of the \HI\ layer in the outer
disk of IC~2233, we have used Gaussian fits along the $z$ direction to the
emission in channels corresponding to $V=461.6$~\kms\ and
$V=647.8$~\kms.  These channels provide better
signal-to-noise than the very outermost channels, but still
sample primarily outer disk emission.
The fitted intensity profiles were extracted
through the location of the peak intensity in each of these channels.
The resulting deconvolved FWHM thickness in both cases is $\sim20.3''$
($\sim1~d_{10}$~kpc).
To estimate the thickness of the inner gas disk, we limit our
measurement to the receding side of the galaxy. We fit the emission
along $x\approx0$ in the channel corresponding to $V$=606.4~\kms\ and
find a deconvolved FWHM of \as{10}{4} ($\sim500~d_{10}$~pc).
Similar results were obtained in adjacent channels. We conclude that
the \HI\ scale height of IC~2233 increases by at least a factor of two
from the center to the outskirts of the galaxy.

In addition to the trends described above, several channels show
plumes of emission stretching to
heights of $\ga50''$ above and below the plane that cannot readily be
attributed to warping or flaring of the disk. This phenomenon 
is particularly prevalent on the receding side of the
galaxy. For example, in the channel images
corresponding to heliocentric radial velocities between
$\sim586-658$~\kms,
we see a faint envelope of ``extra'' emission lying between roughly
$x=100''-150''$,
on both the $\pm z$ sides of the disk. The high-$z$ material
shifts little in position from channel to channel; therefore
toward the highest velocity
channels, this gas lies at smaller projected radii than most 
of the emission, while in channels corresponding to $V=586-617$~\kms,
this material is seen ``trailing'' behind the bulk of the
\HI.  Although there is relatively little \HI\ mass associated with these
extraplanar 
features, the large spread of velocities observed near a single spatial location
implies a large velocity dispersion for this material. 
The characteristics of this emission are therefore consistent with those
expected for a rotationally lagging \HI\ ``halo'' (see Figure~\ref{fig:icZPV} below),
analogous to those now identified in several other spiral galaxies
(e.g., Swaters et al. 1997; Fraternali et al. 2001), including the LSB
spiral UGC~7321 (Matthews \& Wood 2003).
The possible presence of an \HI\ halo
around IC~2233 is of particular interest as a second example of this
phenomenon in a galaxy with a low global star
formation rate and minimal
extraplanar ionized gas component. However, a more rigorous characterization of this
extraplanar \HI\ emission will require  
three-dimensional kinematic modeling
and is beyond the scope of this paper.
Note that the estimates of disk flaring computed above
should be largely unaffected by a rotationally lagging halo, since
the halo gas is expected to manifest itself as non-Gaussian wings to the
vertical intensity profiles rather than a global broadening (see Matthews \&
Wood 2003).

\subsubsection{The Kinematic Center of IC~2233\protect\label{kincenter}}
The systemic velocity and kinematic center of spiral galaxies often are defined
using either the centroid of the global \HI\ profile or the location that minimizes
the asymmetry in the rotation curve. However, given the frequency of
lopsidedness and other perturbations in galaxies (see
\S\ref{discussion}), such choices are not
always physically justified. Here we propose an alternate method of defining the
systemic velocity and kinematic center of an edge-on galaxy. Our method is
most appropriate for late-type spirals, where \HI\ typically has a
high filling factor. 

For a differentially rotating disk whose velocity field is spatially
and spectrally resolved, the gas emission will be expected to have its
minimum observed spatial extent along the direction of the disk
major axis ($\Delta x_{\rm min}$) within the channel corresponding to the velocity
range $V_{\rm sys} \pm 0.5 \Delta V$, where $V_{\rm sys}$ is the true
systemic velocity and $\Delta V$ is the
spectral resolution. For all other observed velocities, $V_{i}$, emission
observed within $V_{i} \pm 0.5 \Delta V$ will span a larger $x$-extent
($\Delta x_{i}>\Delta x_{\rm min}$) owing to the projection along the $x$
direction of the curved isovelocity loci
(see e.g., Figure~6 of Sancisi \& Allen 1979).  In the absence of severe
optical depth effects, the emission is also likely to exhibit
its peak brightness in the channel encompassing the true systemic velocity
owing to the combination of its minimum projected spatial extent and
the maximum path length through the disk.

Using elliptical Gaussian fits, we derived
a position angle and FWHM major and minor axis diameters for
the \HI\ emission distribution in each of the channel images of IC~2233
(corrected for the broadening due to the synthesized beam).
A simple parabolic fit to the values of  the major-axis diameters 
($\Delta x_{i}$) as a function of channel yielded
a minimum at channel $34.25\pm0.1$ with $\chi^2\sim5$
with 5 degrees of freedom (Figure~\ref{fig:kincentfit}), corresponding to
a heliocentric velocity of $553.4\pm0.5$~\kms .  The orientation of the major
axis in channel~34 
corresponds to a PA$=172.0^\circ\pm0.9^\circ$ ,
well aligned with the disk of IC~2233.  A similar series of fits 
to the peak \HI\ brightness
as a function of channel yielded a maximum corresponding to
channel~$34.75\pm0.2$.  However, self-absorption as well as local
changes in the gas phase can affect the column density of \HI\ . In addition, the peak
brightnesses in channels 34 and 35 are essentially identical.  Therefore, we
adopt a value of $V_{\rm sys}= 553.4\pm 1.0$~\kms\ ($2 \sigma$) for the
systemic velocity of IC~2233.   We derived the location of the kinematic center
of IC~2233 by linear interpolation to ``channel''~34.25 using the
centroids of the
\HI\ emission in channels 34 and 35, which yields:
$\alpha_{\rm J2000.0} = 08^{h}13^{m}58.9^{s}$,
$\delta_{\rm J2000.0} = +45^{\circ}44' $\as{27}{0} with formal uncertainties of
$0.1^{s}$ in right ascension and \as{1}{0} in declination ($2 \sigma$).
The location of the \HI\ kinematic center coincides within the errors
with the location of the minor axis dust feature seen in our optical
images of IC~2233 (see \S\ref{opt}; Figure~\ref{fig:WIYNR}). 

\subsection{The Total \HI\ Content\protect\label{totalHI}}
To derive a global \HI\ profile for IC~2233 and measure its integrated
\HI\ flux, we have applied a ``percolation'' procedure 
to the individual channel images containing galactic
line emission. For each channel we defined an irregular blotch containing
line emission above 1~mJy beam$^{-1}$ ($\sim2.5\sigma$) and
successively expanded this region for $n$ iterations using  bands of
$2$ pixels, until the flux within the blotch converged.  
Each step increases
the noise in the total flux, as it is proportional to the square-root of the number
of synthesized-beam areas inside the blotch.  
Convergence in each channel was assumed if an
expansion band lead to no significant increase in the total flux
(i.e., the SNR of the total flux did not increase with an additional step).
To avoid biases, the algorithm begins by finding pixels with
absolute value above the 1~mJy beam$^{-1}$  threshold, with spurious noise spikes
(positive and negative) being discarded.  We corrected the
channel images for primary beam attenuation prior to determining the flux
in the regions localized with the percolation algorithm, and estimated an error
based on the rms noise in the channel and the number of beam areas in the
blotch. These uncertainties are $\la$4~mJy channel$^{-1}$. The resulting global
\HI\ profile is shown in Figure~\ref{fig:icglobal}.

The global \HI\ profile of IC~2233 exhibits a classic double-horn
shape, but with a clear lopsidedness. As illustrated in
the inset of Figure~\ref{fig:icglobal}, the slopes of
the two edges of the profile
are also different: the side having the brighter
horn shows a steeper slope than the side with the weaker and
narrower horn. We comment on the significance of this
further in \S\ref{discussion}.

From our global \HI\ profile we measure velocity widths at
20\% and 50\% of peak maximum of 195.0~\kms\ and 174.1~\kms,
respectively. These values are in good agreement with previous
measurements in the literature (e.g., Haynes et al. 1999). The
centroid of our global \HI\ profile as defined by the 20\% peak
maximum value is 557.1~\kms. This is slightly higher than the systemic
velocity of 553.4$\pm$1.0~\kms\ defined by the kinematic center of the
galaxy (\S\ref{kincenter}).

The integrated \HI\ flux that we derive for IC~2233 is $47.2\pm0.6$~\jks. The
error includes a statistical contribution from each channel (between
1~mJy and 4~mJy) 
and a calibration uncertainty of $\sim$1\%.
Our integrated flux is consistent
with the mean of various published single-dish values including:
$53.1\pm4.6$~\jks\ (Fisher \& Tully 1981); $43.1\pm2$~\jks\ (Tifft
1990); and $47.55\pm0.5\pm2.0$~\jks\ (Haynes et al. 1999).  Published values of  the
integrated \HI\ flux from aperture synthesis measurements include
$44.9\pm2.0$~\jks\ (Stil \& Israel 2002a) and $52.3\pm7.8$~\jks\ (Swaters
et al. 2002).  Assuming that the \HI\ emission in IC~2233 is optically thin, we
derive a total \HI\ mass of \mhi=(1.11$\pm0.01)\times10^{9}~d^{2}_{10}$~\msun.

\subsection{The \HI\ Intensity Distribution\protect\label{intensity}}
We derived the total \HI\ intensity distribution (zeroth moment map) for
IC~2233 by summing the emission identified in each channel by the
percolation algorithm described in the previous section.  The resulting total \HI\
intensity map is shown in Figure~\ref{fig:icmom0}, overplotted as
contours on our \HA+[\NII] image.  Our new \HI\ map of IC~2233
reveals a variety of features not seen in previous \HI\ images of this galaxy.

Along the midplane of IC~2233, the \HI\ distribution of IC~2233 appears
clumpy in spite of the line-of-sight averaging effects of our edge-on
viewing angle.
At the resolution of our data, these clumps have peak column densities of
$\sim(6-8)\times 10^{21}$ atoms~cm$^{-2}$---roughly 30\% higher than
the surrounding gas.  The clumps are not distributed symmetrically
about the kinematic center of the galaxy, nor do they lie in a flat plane (see
also \S\ref{corrugations}).  As seen in Figure~\ref{fig:icmom0},  the locations
of all but the northernmost clump correspond with some of the brightest \HII\
complexes in the galaxy.  Viewed face-on, the \HI\ disk of IC~2233 may
appear similar to that of the dwarf LSB spiral NGC~4395, where the
\HI\ shows a lopsided distribution with little central concentration and a
number of denser, sub-kiloparsec-scale clumps are visible scattered
throughout the disk  (see Swaters
et al. 1999).  Most of these latter clumps correspond  with the
locations of the brightest star-forming regions in NGC~4395.

Further from the midplane of IC~2233, the \HI\ isophotes become
increasingly complex in shape.  Various ripples and protrusions form
a network of material extending to $|z|\la 60''$ along the full radial extent
of the galaxy.  The bulk of these features cannot be attributed to noise, as
most are visible in multiple successive contours.
Figure~\ref{fig:icmom0} illustrates that the full vertical extent of
the \HI\ disk of IC~2233 is significantly larger than that of the
ionized gas disk, even after accounting for resolution effects.
While part of this effect comes from flaring and warping
of the \HI\ layer, it seems likely that at least some portion of this
vertically-extended emission comprises a distinct \HI\ ``halo'' (see also
\S\ref{icchanmaps} and \S\ref{icmajorPV}).

Along the radial direction, we measure the \HI\ extent of IC~2233 to be
\am{6}{66}$\pm$\am{0}{04} at a limiting column density of
$10^{20}$ atoms cm$^{-2}$.  This is $\sim$1.3~times the optical
diameter (defined by the 25.5~mag~arcsec$^{2}$ $R$-band isophote).
Perpendicular to the disk, the deconvolved FWHM thickness of the \HI\
layer through the location of the kinematic center is \as{20}{0}$\pm$\as{0}{2}
($\sim970~d_{10}$~pc).  Similar fits to the emission in the individual channel
images (\S\ref{icchanmaps}) show that the \HI\ layer in the inner galaxy is
intrinsically thinner, then flares at larger radii, especially on the northern side
where the (projected) thickness of the  gas layer reaches FWHM~$\sim 30''$. 
Despite projection effects, this
flaring is also evident in Figure~\ref{fig:icmom0}, where we find that
the global,
deconvolved FWHM of the \HI\ layer increases systematically with
increasing galactocentric radius.

It is interesting to contrast the vertical structure of the \HI\ disk of
IC~2233 with that of the superthin galaxy UGC~7321 studied in
Paper~I. We measured the global FWHM of the \HI\ disk of UGC~7321 to
be $\sim810~d_{10}$~pc --- roughly 15\% smaller
than we measure for IC~2233. In addition, the distance-independent
\HI\ axial ratio of UGC~7321 is significantly larger than that of IC~2233:
$(a/b)_{HI}\approx$29 versus $\approx$18, respectively.  These differences
cannot be attributed to projection effects given the similar
orientation of the two galaxies, and
imply that the mean thickness of the \HI\ disk of IC~2233 is intrinsically
larger than that of UGC~7321, consistent with its thicker stellar and ionized
gas disks.  While part of this difference could result from dynamical
heating due to a recent minor merger (e.g., Reshetnikov \&
Combes 1997) we find no direct evidence
of such an event (see \S\ref{discussion}), and 
it is quite likely that at least part of the difference in thickness stems from the lower
total mass of IC~2233 and thus the lower self-gravity of its disk. 
As the \HI\ velocity dispersion of galaxy disks seems to be nearly constant
at $\sim$6-9~\kms\ in disk galaxies of a wide variety of types and
morphologies (van der Kruit \& Shostak 1984; Dickey 1996), it follows that
the gas scale height of lower-mass disks should be proportionately larger
(see also Brinks et al. 2002). This in turn may have important implications for
the formation of molecular clouds in low-mass galaxies and their overall
regulation of star formation (Elmegreen \& Parravano 1994; Ferguson 1998; 
Dalcanton et al. 2004; Matthews et al. 2005).

Knowledge of the gas surface density of a galaxy is important to
characterizing its star-formation efficiency.
To estimate the deprojected \HI\ surface density in the disk of IC~2233,
we have used an Abel inversion technique (e.g., Binney \& Tremaine 1987). 
We extracted a $3''$-wide slice along the major axis of the galaxy, sampled
at $16''$ increments, and then computed deprojected intensity distributions
for the two sides of the galaxy independently, using a fortran program
adapted from
Fleurier \& Chapelle (1974).  The results are presented in 
Figure~\ref{fig:icHIsurf}.  We assume that the \HI\ is
optically thin and cylindrically symmetric on each of 
the two sides of the galaxy. Because the observed intensity profile fluctuates
rapidly near the galaxy center, the deprojected surface density at
small radii ($r\lsim30''$ or $ \lsim 1.5~d_{10}$~kpc) 
is not well constrained. To minimize the impact of this 
effect, we have smoothed the
measured intensity profile by a factor of three before each fit.
Despite these small-scale uncertainties,
the resulting profiles for the two sides of the galaxy
should give a reasonable estimate for the mean \HI\ surface density as
a function of radius. Indeed, the integrated \HI\ mass derived from
the deprojected profiles shown in Figure~\ref{fig:icHIsurf} is
$9.9\times10^{8}~M_{\odot}$---in excellent agreement with our observed
value (see Table~3). Modest differences are apparent on the two sides
of the disk. At all radii, the mean \HI\
surface density of IC~2233 is considerably below the mean value of
$\sim10M_{\odot}$ pc$^{-2}$ found by Cayatte et al. (1994) for high
surface brightness Sd galaxies. However, values of $\Sigma_{\rm HI}$
within the disk of IC~2233 
are comparable to those seen in other LSB spirals (van der Hulst et
al. 1993), including the superthin LSB spiral UGC~7321 (Paper~I). 

\begin{deluxetable}{lc}
\tabletypesize{\scriptsize}
\tablewidth{0pc}
\tablenum{3}
\tablecaption{21-cm Radio Properties of IC~2233}
\tablehead{\colhead{Parameter} & \colhead{Value}}
\startdata

\multicolumn{2}{c}{Measured Quantities} \\

\tableline

$\alpha$ (J2000.0)\tablenotemark{a} &  08 13 58.9 ($\pm 0.1$)  \\
$\delta$ (J2000.0)\tablenotemark{a} &  +45 44 27.0 ($\pm 1.0$) \\

Peak \HI\ column density\tablenotemark{b}  & $8\times10^{21}$~atoms~cm$^{-2}$ \\

$D_{\rm HI}$\tablenotemark{c} & \am{6}{66}$\pm$\am{0}{04} \\

$\theta_{\rm HI,b}$\tablenotemark{d} (FWHM) & \as{22}{6}$\pm$\as{0}{2}\\

$(a/b)_{\rm HI}$\tablenotemark{e} & 18 \\

$\int F_{\rm HI}{\rm d}\nu$ & $47.2 \pm 0.5$~Jy km s$^{-1}$ \\

$W_{20}$\tablenotemark{f} & 195.0 km s$^{-1}$ \\

$W_{50}$\tablenotemark{f} & 174.1 km s$^{-1}$ \\

$V_{\rm sys,HI}$\tablenotemark{g}  & 553.4$\pm$1.0~km~s$^{-1}$ \\

$V_{\rm sys,HI}$\tablenotemark{h}  & 557.1~km~s$^{-1}$ \\

$V_{\rm max}$ & $\sim$85~\kms \\

$F_{\rm cont}$ (21-cm)  & $<3$~mJy (2$\sigma$) \\

\tableline
\multicolumn{2}{c}{Derived Quantities}\\
\tableline

$M_{\rm HI}$\tablenotemark{i} & $(1.11 \pm 0.01) \times10^{9} ~d_{10}^2 ~M_{\odot}$  \\

$M_{\rm dyn}$\tablenotemark{j} & 1.55$\times10^{10} ~d_{10} ~M_{\odot}$ \\

$M_{\rm HI}/L_{B}$ & 0.62 $M_{\odot}/L_{\odot}$ \\

$M_{\rm dyn}/M_{\rm HI}$ & 14~$d_{10}^{-1}$ \\

$D_{\rm HI}/D_{25.5}$ & 1.3 \\

\enddata

\tablenotetext{a}{Kinematic center (see \S\ref{kincenter})}

\tablenotetext{b}{At 16$''$ resolution.}

\tablenotetext{c}{Measured at a column density of $10^{20}$ atoms cm$^{-2}$.}

\tablenotetext{d}{Projected thickness of \HI\ layer along minor axis,
measured at a column density of $10^{20}$~atoms~cm$^{-2}$
and corrected for the resolution of the synthesized beam.}

\tablenotetext{e}{Axial ratio of \HI\ disk measured at a column density
of $10^{20}$ atoms cm$^{-2}$ and corrected for the resolution of the
synthesized beam. }

\tablenotetext{f}{$W_{P}$ is the \HI\ profile width measured at $P$\% of
the mean peak value of the two profile horns. }

\tablenotetext{g}{From kinematic method (heliocentric frame; optical definition).}

\tablenotetext{h}{From global \HI\ profile (heliocentric frame; optical definition).}

\tablenotetext{i}{Assuming the \HI\ is optically thin and $d_{10}$
is the distance to IC~2233 expressed in units of 10~Mpc. }

\tablenotetext{j}{From $M_{\rm dyn} = 2.326\times 10^{5}rV^{2}(r)$, where
we have taken $r = 9.2$~$d_{10}$~kpc (190$''$) and $V(r)=$85~km~s$^{-1}$.}

\end{deluxetable}

\begin{deluxetable}{ccc}
\tabletypesize{\tiny}
\tablewidth{0pc}
\tablenum{4}
\tablecaption{IC~2233 Rotation Curve}
\tablehead{\colhead{Major axis distance (arcsec)} & \colhead{Circular
Velocity (\kms)} & \colhead{Uncertainty (\kms)}}
\startdata
  -189.0    &  472.8   &  4.9 \\
     -171.0  &    473.9 &     2.8\\
     -162.0  &    470.5 &     1.5\\
     -153.0  &    472.9 &    0.8\\
     -144.0  &    473.0 &    0.7\\
     -135.0  &    475.7 &    0.4\\
     -126.0  &    467.0 &    0.3\\
     -117.0  &    469.2 &    0.2\\
     -108.0 &    475.9 &    0.1\\
     -99.0  &    484.0 &   0.1\\
     -90.0  &    488.8 &   0.1\\
     -81.0  &    491.5 &   0.1\\
     -72.0  &    493.2 &   0.1\\
     -63.0  &    498.0 &   0.1\\
     -54.0  &    504.0 &   0.1\\
     -45.0  &    508.7 &   0.1\\
     -36.0  &    514.5 &   0.1\\
     -27.0  &    523.4 &   0.1\\
     -18.0  &    531.7 &   0.1\\
     -9.0  &    542.6 &   0.1\\
      0.0  &    551.5 &   0.1\\
      9.0  &    573.6 &   0.1\\
      18.0  &    589.6 &    0.1\\
      27.0  &    597.3 &    0.1\\
      36.0  &    599.4 &    0.1\\
      45.0  &    603.9 &   0.1\\
      54.0  &    608.3 &   0.1\\
      63.0  &    617.0 &   0.1\\
      72.0  &    618.2 &    0.1\\
      81.0  &    620.3 &   0.1\\
      90.0  &    623.9 &   0.1\\
      99.0  &    627.8 &   0.1\\
      108.0  &    632.7 &   0.1\\
      117.0  &    635.7 &    0.1\\
      126.0  &    639.0 &    0.1\\
      135.0  &    645.8 &    0.1\\
      144.0  &    641.3 &    0.1\\
      153.0  &    641.8 &    0.1\\
      162.0  &    642.7 &    0.2\\
      171.0  &    641.6 &    0.3\\
      180.0  &    642.8 &    0.4\\
      189.0  &    640.0 &    0.6\\
      198.0  &    641.4 &     1.1\\
      207.0  &    641.2 &     3.0\\
\enddata
\end{deluxetable}


\subsection{Warping and Undulations in the IC~2233
Disk\protect\label{corrugations}}
A careful examination of Figure~\ref{fig:icmom0} reveals that
neither the brightest \HI\ clumps nor
the underlying, smoother gas component in IC~2233 lie along a flat
plane. To illustrate this more clearly,
we have plotted in Figure~\ref{fig:warpcurve} 
the vertical displacement in the centroid of the \HI\ layer as
a function of radius. These displacements were derived from Gaussian
fits to a sequence of slices
perpendicular to the mean galactic plane.  The resulting plot highlights two
important features of the IC~2233 HI disk.

At the edge of the stellar
disk of IC~2233 (denoted by arrows on Figure~\ref{fig:warpcurve}), we see that the
\HI\ layer bends away from the plane, forming an ``integral sign''
warp.  The amplitude of the warp is asymmetric, with an amplitude $\sim$50\%
higher on one side of the disk than on the other 
($\Delta z\approx +$\as{3}{5} versus $\Delta z\approx -$\as{5}{5}). 
On the higher-amplitude side, the gas layer is observed to twist back
toward the midplane at its outermost extent.  These various
traits are common features of galactic warps (Burton 1988;
Garc\'\i a-Ruiz et al. 2002; S\'anchez-Saavedra et al. 2003;
Paper~I). 
Recently Saha \& Jog (2006) have suggested that amplitude asymmetries in
warps are a natural consequence of dynamical wave interference
between $m$=0 and $m$=1 bending modes of a disk. The process
that might excite such modes is still unclear, but one possibility is
accretion of gas---either from a minor merger or from the
intergalactic medium. We return to this issue in \S\ref{discussion}.

In addition to the warp, Figure~\ref{fig:warpcurve} reveals another
particularly intriguing feature of the neutral gas disk of IC~2233: a
pattern of positive and negative vertical displacements spanning
the full extent of the stellar disk.   These undulations appear remarkably
regular, with a wavelength of $\sim$150$''$ ($7~d_{10}$~kpc).  Their
amplitude increases with distance from the center of the galaxy, reaching
a maximum of $|\Delta z|\sim 3''$ ($\sim150~d_{10}$~pc). 

In the Milky Way, systematic deviations of the \HI\ layer from the mean
principal plane along the $z$ direction (so-called ``corrugations'') were
first noted by Gum et al. (1960).   Subsequent studies found that
corrugations are present along both radial and azimuthal directions
over a significant fraction of the disk of the Galaxy (e.g.,
Quiroga 1974,1977;  Spicker \& Feitzinger 1986). The corrugations
are reflected not only in the distribution of neutral hydrogen, but in a
variety of tracers, including molecular clouds and \HII\ regions 
(Lockman 1977; Alfaro \& Efremov 1996  and references therein).  As
measured in \HI, scales of the corrugations range from $\sim$50--350~pc
in amplitude and from one to several kpc in wavelength (Spicker \&
Feitzinger 1986), comparable in scale to the undulations seen in
IC~2233. Although it has been suggested that corrugations
should be a common phenomenon in disk galaxies, similar phenomena
have so far been observed (in optical light) in only a few galaxies (Florido
et al. 1991,1992; Alfaro et al. 2001).  {\it To our knowledge, corrugations
have not been reported previously in the \HI\ disk of an external galaxy.}

Inspection of Figure~\ref{fig:WIYNR} reveals that both the \HA\ emission
and the optical continuum of IC~2233 also show deviations from a purely planar
geometry, although these undulations are not as regular as those seen
in \HI.  Analysis of the vertical displacements via optical tracers is
complicated by the patchy distribution of bright \HII\ regions and by the
presence of a superimposed foreground star in the northern part of the
galaxy.  In the case of the $R$-band data, 
we find a displacement of the intensity-weighted centroid from the
midplane of $\Delta z=$+\as{1}{8} at $x=-140''$; the displacement
systematically decreases with increasing $x$, reaching a value of 
$\Delta z=-$\as{2}{7} at $x\approx-80''$, before increasing again
monotonically to a value of $\Delta z\approx$0 near $x=0$.  The
changes in $\Delta z$ with $x$ thus have the same sense as in the
\HI\ data, but with different amplitude and phase.  On the northern side
of the disk, analysis is hindered by the bright foreground star, and
no systematic trend in the displacement of the centroid as a function
of radius is evident.  All measured centroids between $x=50''-125''$
have positive $\Delta z$, with the exception of one point near
$x\approx100''$ with $\Delta z\approx$0.  The midplane deviations
measured from the \HA\ data show no obvious systematic trend with
radius, and using an autocorrelation analysis we were unable to
identify a periodicity in the \HA\ undulations on spatial scales
$\ge10''$.  These differences between the vertical displacements
measured from the \HI\ and optical data may result from intrinsic
differences between the vertical displacements of the stars and the
gaseous components, but projection effects and differences in the
radial extent and filling factor of the different tracers are also likely to
contribute.  Disentangling these effects in an edge-on galaxy is further
complicated by the fact that disk corrugations may occur along the
plane as well as orthogonally to it (see Spicker \& Feitzinger 1986). 

The origin of the corrugations in the disk of IC~2233 is a
puzzle. We will present a more extensive investigation of this issue
based on additional multiwavelength observations in a future paper
(L. D. Matthews \& J. M. Uson, in preparation). Here we offer a brief
discussion  of this question based on results from the present data.

In the Milky Way and in the face-on spiral galaxy NGC~5427, clear
relationships have been found
between corrugations and existing spiral arms, suggesting that the two phenomena
are linked (Quiroga 1977; Spicker \& Feitzinger 1986; Alfaro et
al. 2001; see also Nelson 1976; Martos \& Cox 1998).
While the edge-on orientation of IC~2233 makes it difficult to assess what type of
spiral arm structure it might possess, the patchy distribution of
\HA\ emission in this galaxy is suggestive of \HII\ regions organized
along loose spiral arms.  This is supported by the observed correlation
between the locations of the \HI\ density enhancements along the midplane
and the brightest \HII\ complexes (Figure~\ref{fig:icmom0}).  The \HA\
rotation
curve derived  by Goad
\& Roberts (1981; see Figure~\ref{fig:icrotcurve}, discussed below)
also shows a series of amplitude ``wiggles'' similar to those linked with
the spiral arms and velocity corrugations in 
face-on spirals (Alfaro et al. 2001 and references therein).

Using numerical simulations, Edelsohn \& Elmegreen (1997)
showed that in a Milky Way-like galaxy, tidal perturbations from a
low-mass companion will induce kpc-scale vertical oscillations
(i.e., vertical displacements coupled with velocity undulations) with
a spiral-like shape. In addition, simulations by Sellwood et al. (1998) have
shown that vertical ``buckling''---qualitatively similar in appearance to the
undulations we see in IC~2233---can be induced by the passage
of a low-mass satellite through the disk.  However, a problem with
invoking either of these scenarios is that our optical and \HI\ studies of
IC~2233 have not uncovered any direct evidence of an ongoing
tidal interaction or minor merger (see also \S\ref{thefield}, \S\ref{discussion}), and
Sellwood et al. found that if the satellite has been
completely disrupted, bending waves will not be excited.  While we
noted in \S\ref{opt} that
the unusual OB ``superassociation'' to the north of the kinematic center could
be an intruder remnant, supporting kinematic evidence is lacking. 
Moreover, it is predicted that
disk undulations would not persist  once the orbit of the
satellite has decayed into the disk (Sellwood
et al. 1998).\footnote{We note that the
simulations of Sellwood et al. 1998 strictly pertain only to 
collisionless systems, thus their direct applicability to a gas-rich
system like IC~2233 is somewhat unclear.} 

If the
midplane undulations in IC~2233 are not caused by either a recent
intruder or linked with (pre-existing) 
spiral arms, another possibility is that they reflect spontaneous
bending instabilities in the disk (e.g., Griv \& Chiueh
1998).  This might offer a natural explanation for the increase in
amplitude of the undulations with radius, as such an increase could be
a consequence of the change in the group velocity of this wave
as it reaches regions with larger gas scale-height and lower gas density 
in the outer disk (Spicker \& Feitzinger 1986; Hofner \&
Sparke 1994). If bending waves are the cause of the corrugations then
similar features should be observable in the
disks of other isolated galaxies.

One final aspect of the corrugated structure of IC~2233 that is
worth noting is that, regardless of its origin, this type of coherent, 
large-scale pattern can
only exist if the gas disk of this LSB spiral is largely self-gravitating
(e.g., Bosma 2002; Revaz \& Pfenniger 2004).  This adds to
other recent evidence for ``heavy'' disks in some LSB disk
galaxies based on the analysis of their spiral structure
(Fuchs 2002,2003; Masset \& Bureau 2003). 

\section{The \HI\ Kinematics of IC~2233\protect\label{ickin}}
\subsection{\HI\ Velocity Field\protect\label{icmom1}}
The \HI\ velocity field (first moment map) of IC~2233 is shown in
Figure~\ref{fig:icmom1}.  Near the midplane this diagram shows fairly
straight and parallel isovels, as expected from the high inclination of the galaxy and
its inner-disk kinematics, which resemble a solid-body rotator (see also
\S\ref{icmajorPV} below).  However, away from the midplane, many of the 
isovels curve to form 
{\sf S}- or {\sf L}-shaped lines.  
Such pronounced twisting of the velocity field is rarely observed
in edge-on galaxies and could be a signature of warping or of
gas on non-circular orbits (e.g., as a result of a lopsided or triaxial
potential or large-scale
streaming motions through spiral arms). Because IC~2233 is viewed so close
to edge-on, the curved portions of the isovels 
trace primarily faint, extraplanar material, hence 
warping seems to be the most likely explanation, although the presence
of some gas on non-circular orbits and/or a lopsidedness in the
overall potential cannot be excluded (see also \S\ref{discussion}). In addition, the
kinematically anomalous gas discussed in \S\ref{icchanmaps} and
\S\ref{icmajorPV} may also contribute 
to the observed complexity of the extraplanar velocity structure.

\subsection{Position-Velocity Plots\protect\label{icmajorPV}}
In Figure~\ref{fig:icPV} we show a position-velocity (P-V) diagram
along the major axis of IC~2233, together with additional P-V profiles extracted
$\pm15''$ from the midplane. The major-axis P-V diagram of IC~2233 displays
noticeable asymmetries, with the redshifted side exhibiting
a more obvious flattening or turnover at large radii, and the
blueshifted side showing an apparent ``upturn'' near $x=-140''$ and
maximum observed gas velocities $\sim$25~\kms\
larger than on the redshifted side.
As is frequently seen in lopsided galaxies, 
the ``flat'' side of the P-V profile corresponds to the
brighter horn in the global \HI\ profile (see
Figure~\ref{fig:icglobal} and \S\ref{discussion};
see also Swaters et al. 1999; Noordermeer et al. 2001).
We also draw attention
to the features seen in Figure~\ref{fig:icPV} 
near $x=\pm100''$ that indicate deviations of $\sim$+20~\kms\ and $-$20~\kms,
respectively, from the spread of ``permitted'' velocities at these locations.

It is interesting to contrast the major axis P-V plots of IC~2233 with those we
derived for UGC~7321 in Paper~I.
In spite of having only a slightly smaller
peak rotational velocity than UGC~7321, the major axis P-V plot for
IC~2233 is strikingly different: while that of UGC~7321 is
reminiscent of a ``scaled-down'' giant spiral galaxy (see Figure~10 of
Paper~I), the P-V profile of IC~2233 has a nearly solid-body shape,
characteristic of dwarf and Magellanic spiral galaxies (cf. e.g., Swaters et
al. 2002). At any given position along the major axis of IC~2233,
the spread of observed velocities is narrower than what we observed in
UGC~7321, and the peak intensities lie at
smaller values of $V_{\rm rot}$.  These differences imply less rotational
shear in IC~2233, a factor that may be important in allowing it to
form the giant \HII\ complexes that are absent in UGC~7321 (see \S\ref{Halpha}).

A sample of P-V diagrams extracted parallel to the minor axis of
IC~2233 is shown in Figure~\ref{fig:icZPV}. This figure highlights the
previously mentioned high-latitude emission, extending 
to $z\approx\pm60''$, and the tendency for the observed velocities of this
emission to lie at
smaller values compared with the emission along the plane 
(see also \S\ref{icchanmaps}). This type
of velocity structure is consistent with the 
expected signatures
of a rotationally lagging \HI\ halo (e.g., Matthews \& Wood 2003).
Kinematically anomalous extraplanar gas is observed
on both sides of the plane of IC~2233, although it is slightly
more prevalent in the northeast quadrant of the galaxy. 

\subsection{Disk Rotation Curve\protect\label{icrotcur}}
Using the data shown in Figure~\ref{fig:icPV}
we have derived a major axis 
rotation curve for IC~2233 following the method 
described in Paper~I. The result is plotted in
Figure~\ref{fig:icrotcurve} and provided in electronic form in Table~4. 
The optical (\HA) rotation curve from Goad \& Roberts (1981) is
overplotted for comparison. The generally 
excellent agreement between the \HA\
rotation curve and our new \HI\ rotation curve suggests that the
latter is not significantly affected by beam smearing. The one
notable discrepancy occurs over the interval $x \approx 50''$-100$''$,
where the \HI\ rotation curve is much smoother than the \HA\ curve.
This may result from  an irregular or patchy \HII\ region distribution
and/or  non-circular motions.
Indeed, we see evidence of a larger spread in \HI\ velocities at the
corresponding location (Figure~\ref{fig:icPV}).

A maximum rotational velocity of $\sim$85~\kms\ is reached on both sides of
the disk of IC~2233. Both sides also show a small but 
statistically significant drop in rotational
velocity of $\sim$10~\kms\ near the edge of the stellar disk 
($x\approx\pm130''$), just before the
terminal velocity is reached. The locations of these dips correspond
closely to where the edges of the \HI\ disk are 
seen to warp (Figure~\ref{fig:warpcurve}). Compared with UGC~7321, the \HI\ 
rotation curve of
IC~2233 shows a more leisurely rise relative to the stellar
disk and has a less extended flat portion. 

Despite the asymmetries visible in the major axis P-V plot
of IC~2233 (Figure~\ref{fig:icPV}) and the global \HI\ profile
(Figure~\ref{fig:icglobal}), our derived rotation curve is relatively
symmetric, both in terms of shape and in terms of peak rotational velocity on
the two sides of the disk. The difference in extent of the rotation
curve on the two sides
is also rather modest, with the receding side extending roughly $20''$
further than the approaching side. Only between 
$x=110''-135''$ is a significant difference in amplitude between the
two sides seen, with an offset of $\sim$25~\kms. 

A closer examination  reveals that the reason why the underlying
disk rotation curve appears fairly symmetric in spite of the
asymmetries visible in Figure~\ref{fig:icPV} is that the bulk
of the underlying gas in IC~2233 has a fairly smooth and
symmetric distribution (note the isophotes corresponding to
intermediate intensities), with additional
irregularities and asymmetries present in the form of the
emission traced by
the brightest, innermost contours, and the faintest, outermost
contours. The innermost contours
correspond to the bright clumps along the midplane visible in
Figure~\ref{fig:icmom0}; based on their locations in the P-V plane, it 
appears that these clumps are located at
small to intermediate galactocentric radii and may be subject to modest
non-circular motions. On the approaching side of the disk, 
the outermost contours seen in our data reveal gas
extended to velocities $\sim$25~\kms\ 
beyond the asymptotic rotational velocities shown
in Figure~\ref{fig:icrotcurve}; however, this highest-velocity 
gas is of insufficient
quantity to receive substantial weight in the determination of the rotation curve
(see Paper~I).

\section{The Blue Compact Dwarf Galaxy NGC~2537}
As described in  \S\ref{thefield},
the blue compact dwarf (BCD) galaxy NGC~2537 lies at a projected distance
of \am{16}{7} (49~$d_{10}$~kpc = 34~$d_{7}$~kpc) from IC~2233.  
Here we describe our new \HI\ measurements of
this galaxy and briefly discuss how our observations constrain
scenarios for its evolution (see also \S\ref{comparison}). 

\subsection{Optical Morphology}
NGC~2537 has been studied extensively at a variety of
wavelengths, and some of its key properties are summarized in
Table~5.  This galaxy was observed at optical and near-infrared
wavelengths by Gil de Paz et al. (2000a,2000b), who found that
its nuclear region  hosts an intermediate-age starburst 
($t\sim$30 Myr) containing numerous star-forming knots.  Surrounding
the central starburst is a diffuse, red outer disk with an underlying
stellar population of age 5-13~Gyr, implying that NGC~2537 has been
forming stars for a significant fraction of a Hubble time.  The
metallicity of NGC~2537 is somewhat uncertain, but it is
well below solar ($Z\sim0.13-0.41 Z_{\odot}$; Meier et al. 2001;
Wu et al. 2006). NGC~2537 has  been imaged previously in \HI\ by
other authors (Stil \& Israel 2002a,b; Swaters et al. 2002; 
Wilcots \& Prescott 2004), but the improved sensitivity and
resolution of our data provide additional insight into the
\HI\ properties and kinematics of this galaxy.


\begin{deluxetable}{lcc}
\tablewidth{0pc}
\tablenum{5}
\tablecaption{Optical and Infrared Properties of NGC~2537}
\tablehead{\colhead{Parameter} & \colhead{Value} & \colhead{Ref.}}

\startdata

$\alpha$ (J2000.0) &  08 13 14.7 & 1 \\
$\delta$ (J2000.0) &  +45 59 26.3 & 1 \\
Hubble type &  SB(s)m pec & 1 \\
Distance\tablenotemark{a} &  $5.5 - 9$~Mpc & 2 \\
$A_{B}$ (mag) & 0.232 & 3 \\
$A_{R}$ (mag) & 0.144 & 3 \\

\tableline
\multicolumn{3}{c}{Measured Quantities} \\
\tableline

$D_{B,25}$ & \am{1}{7}$\times$\am{1}{5} & 1 \\
$m_{B}$\tablenotemark{b} & 12.07$\pm$0.03 & 4\\
$B - R$\tablenotemark{b} & 0.58$\pm$0.11 & 4 \\
$h_{r,B}$ & \as{14}{4} $\pm$ \as{0}{4} & 4 \\
$\mu_{B,i}(0)$\tablenotemark{b} & $20.57 \pm 0.09$ mag arcsec$^{-2}$ & 4 \\

\tableline
\multicolumn{3}{c}{Derived Quantities} \\
\tableline

A$_{B,25}$ & 3.5~$d_{7}$~kpc & 1,4\\
$M_{B}$\tablenotemark{b} & $-$17.16 & 2,4 \\
$L_{B}$\tablenotemark{b} & 8.4$\times10^{9}d^{2}_{7}~L_{\odot}$ & 2,4\\
$L_{\rm FIR}$\tablenotemark{c} & 2.8$\times10^{8}~d^{2}_{7}~L_{\odot}$ & 1,2 \\

\enddata

\tablenotetext{a}{We adopt a distance of 7~Mpc (see \S\ref{thefield}).
Distance-dependent quantities are scaled in terms of $d_{7}$,
the actual distance in units of 7~Mpc.}

\tablenotetext{b}{Corrected for foreground extinction only.}

\tablenotetext{c}{Based on $IRAS$ 60$\mu$m and 100$\mu$m fluxes from the NED
database.}

\tablerefs{(1) NED database; (2) this work; (3) Schlegel et al. 1998;
(4) Gil de Paz et al. 2000a,b}

\end{deluxetable}


\subsection{Properties of the Neutral Hydrogen in NGC~2537\protect\label{bearpawHI}}

\subsubsection{Channel Images\protect\label{ngcchanmaps}}
In Figure~\ref{fig:ngccmaps} we present the \HI\ channel
images for NGC~2537.  The same data are represented both as contours
and as greyscale in order to highlight the existence of numerous
small clumps of \HI\ superposed on a lower surface density 
background.  Many of these clumps are unresolved by our beam
(i.e., they have sizes $\lsim500~d_{7}$~pc).  The background emission
underlying the clumps is itself rather diffuse and fragmented, and
the emission within the individual channel images does not appear
contiguous at the sensitivity limit of our data.

\subsubsection{The Global HI Profile of NGC~2537}
We derived a global \HI\ profile for NGC~2537 in the same manner as
for IC~2233 (\S\ref{totalHI}). The result is shown in
Figure~\ref{fig:ngcglobal}. The uncertainty in our global profile for
NGC~2537 is larger than that of IC~2233, both because
NGC~2537 was farther from the center of the primary beam,
and because the LSB,
noncontiguous nature of its \HI\ emission made
it more difficult to sum  the total flux in each channel accurately.  Nonetheless, 
consistent with the ordered velocity field seen in Figure~\ref{fig:ngcglobal},
the derived \HI\ profile confirms that 
NGC~2537 is clearly rotationally dominated (the measured profile
width at 20\% peak maximum is 121~\kms, 
compared with the peak \HI\ velocity dispersion of
$\sigma_{V,\rm HI}\sim$25~\kms; see \S\ref{ngcveldisp}). The \HI\ profile is
somewhat lopsided, and the slopes of the two edges of
the profile differ, with the blueshifted side having
a shallower slope. The global \HI\ profile we have
derived is in good agreement with the single-dish profiles published by
Thuan \& Martin (1981) and Huchtmeier \& Richter (1986) and the 
profiles derived by Swaters et
al. (2002) and Stil \& Israel (2002a) using Westerbork aperture
synthesis measurements. 

We derive an integrated \HI\ flux for NGC~2537 of
19.2$\pm$0.6~\jks. The
uncertainty is a combination of a statistical error of 0.35~\jks\  
and an overall calibration uncertainty of 1\%.
Our integrated flux lies within the range of previously published
single-dish values, including: 21.6$\pm$1.1~\jks\ (Thuan \& Martin 1981);
17.2$\pm$2.2~\jks\ (Davis \& Seaquist 1983);
20.1$\pm$1.1~\jks\ (Huchtmeier \& Richter 1986); 21.6$\pm$2.5~\jks\
(Fisher \& Tully 1981); 18.5$\pm$1.2~\jks\ (Tifft 1990); and
20.2$\pm$1.0~\jks\ (Bottinelli et al. 1990).
Previous aperture synthesis measurements include
18.8$\pm$0.9~\jks\ (Stil \& Israel 2002a) and 14.8$\pm2$~\jks\ (Swaters et
al. 2002).

\subsubsection{The Total \HI\ Intensity Distribution\protect\label{NGCtotalHI}}
An \HI\ total intensity (moment~0) image of NGC~2537 is shown in
Figure~\ref{fig:ngccomposite}.  In the upper left panel, the \HI\ intensity
contours are overplotted on a greyscale representation of the data, while
in the upper right panel, we show the \HI\ contours overplotted on a blue
image of the galaxy from the Digitized Sky Survey (DSS).  Because of the
very diffuse and patchy nature of the \HI\  emission in this galaxy,
the percolation method described in \S\ref{totalHI} picks up a large
amount of low-level noise. Therefore to compute the zeroth moment 
we have summed all positive emission over the velocity range 
$V_{r}$=368.4-502.9~\kms. 
To increase signal-to-noise for the
higher order (first and second) moments (discussed below), 
we rejected data that did not
exceed 1.5 times the rms noise per channel in a version of the data
that had been smoothed spatially with a Gaussian kernel
of width 5 pixels and in velocity with a boxcar function of width 3
channels. 

As seen in the upper panels of Figure~\ref{fig:ngccomposite}, we detect
\HI\ emission in NGC~2537 extending to $\sim$1.5--2~times the limiting
isophote on the blue DSS image, with the extent of the \HI\ relative to the
starlight varying with azimuth.  The mean ratio of the \HI\ diameter
to the $D_{25}$ optical 
diameter is $\sim$3.5. As is evident from the individual
channel images (Figure~\ref{fig:ngccmaps}), the global morphology of
the neutral gas is quite clumpy and fragmented, and several patches
within the disk appear nearly devoid of \HI\ emission. This flocculent
appearance of the \HI\ is not confined to the star-forming portion of the
galaxy, implying that it is not linked to feedback
processes or to the instabilities giving rise to star formation.
The overall \HI\ distribution and morphology of NGC~2537 are similar to
other low-mass starburst and post-starburst galaxies observed
with comparable spatial resolution (e.g., Israel \& van Driel 1990; Meurer et
al. 1996;  van Zee et al. 1998; Wilcots \& Miller 1998; Hunter et
al. 1999; Thuan et al. 2004), although frequently BCDs show global
\HI\ morphologies that are much more disorganized and irregular than
seen in NGC~2537 (cf. van Zee et al. 1998,2001; Thuan et al. 2004).

A notable feature of the outer \HI\ disk of NGC~2537 is the spiral arm-like
feature on the eastern edge, which extends nearly 180$^{\circ}$ around
the galaxy.  This feature has no optical counterpart, and as seen in
\S\ref{ngcvelfield}, the gas in this arm follows the same ordered
rotation as the rest of the disk. One possibility is that this feature
could be a
remnant from a past interaction (\S\ref{discussion}). However,
gaseous arms induced by interactions tend to be short-lived, persisting
only one or two rotation periods (e.g., Quinn 1987; Iono et al. 2004),
implying that the interloper should still be nearby.  As discussed in
\S\ref{thefield} and \S\ref{discussion}, we are unable to identify 
a likely culprit. Moreover,
the tight winding of the NGC~2537 arm, together with its regular rotation
and lack of a counter-arm all are inconsistent with classic
tidal features (see Toomre \& Toomre 1972).  Explaining the
arm through more indirect
dynamical effects, such as swing amplification (Toomre 1981), also
appears problematic given the extremely low \HI\ surface density of the
disk.  The Toomre $Q$ parameter of the outer disk is quite high
($Q\gsim$10) implying a high degree of stability against axisymmetric
perturbations; in such a case, the swing amplification mechanism will be
very inefficient, although this 
problem could be overcome if the \HI\ disk contains sufficient dark matter
to render it gravitationally unstable  (see the discussion in Masset \& Bureau 2003). 

The most striking feature of the inner \HI\  disk of NGC~2537 is the bright
``ring'' approximately 1$'$ (2.0$d_{7}$~kpc) in diameter.  Typical column
densities along the ring are $n_{\rm HI}\sim7.6\times10^{20}$~cm$^{-2}$,
although several brighter knots with $n_{\rm HI}\sim9.5\times10^{20}$~cm$^{-2}$
are superposed. In contrast, inside the ring the mean column density
is roughly a factor of two lower
($\sim4.7\times10^{20}$~cm$^{-2}$), although part of the gas inside
is likely to be molecular.  One of the regions of bright optical
continuum emission in NGC~2537 lies coincident with the center of the
\HI\ ring, but based  on the \HA\ and [\OIII] images of Gil de Paz et
al. (2000b),  there does not appear to be any significant amount of ionized
gas at this location.

Using the data from Figure~\ref{fig:ngccomposite}a, we have derived a
radial \HI\ surface density profile for NGC~2537 by taking measurements
within a series of concentric, elliptical annuli.  The center of the annuli was
taken to be the kinematic center listed in Table~6; this position approximately
corresponds to the center of the bright \HI\ ring in
Figure~\ref{fig:ngccomposite}a,b.
The ellipticity of the annuli was taken to be 0.1 and their major axis position
angle to be \ad{172}{0} (see \S\ref{ngcvelfield} below).  
The results are presented in Figure~\ref{fig:ngcHIsurf}.
We have not corrected the inferred \HI\ surface density for inclination
since this correction is small ($\sim$18\%).  We see that  the \HI\
surface density in the outer disk of NGC~2537 is extremely low, comparable
to values found in typical dwarf irregular and LSB spiral galaxies,
including IC~2233 (Figure~\ref{fig:icHIsurf}).
However, in contrast to IC~2233, NGC~2537
shows a steep rise in \HI\ density toward its center, where the \HI\ surface
density reaches a peak of $\sim11M_{\odot}$ pc$^{-2}$---comparable to values
seen in the inner parts of normal late-type 
spiral galaxies (Cayatte et al. 1994).  Similar
enhancements in the central \HI\ surface density are also observed in other BCD systems
(Meurer et al. 1996; van Zee et al. 1998).

We have overplotted as a dashed line on  Figure~\ref{fig:ngcHIsurf} the
canonical star formation threshold of Kennicutt (1989).  We have assumed
that the gas is isothermal with a constant sound speed of 7~\kms.  We see
that the computed Kennicutt threshold lies significantly above the measured
gas surface density at all radii.  However, scaling the measured \HI\ curve by
a factor of 1.34 to account for He and by an additional factor of 1.9 reconciles
it with the computed instability threshold over the portion of the galaxy where
stars are visible (delineated by the vertical bar).  The latter scaling factor is
comparable to the expected contribution to the total gas surface density from
the molecular gas: Gil de Paz et al. (2002) measured
$M_{\rm H_2}\approx(0.4-5)\times10^{7}~M_{\odot}$ within the the central
$\sim0.7d_{7}$~kpc ($r\lsim 20''$) of NGC~2537.  Therefore, in contrast to
IC~2233, the locations of star formation within NGC~2537 appear to be in
reasonable agreement with Kennicutt's star formation criterion so long
as the total gas surface density is considered.


\begin{deluxetable}{lc}
\tabletypesize{\scriptsize}
\tablenum{6}
\tablewidth{0pc}
\tablecaption{21-cm Radio Properties of NGC~2537}
\tablehead{\colhead{Parameter} & \colhead{Value}}
\startdata

\multicolumn{2}{c}{Measured Quantities} \\
\tableline

$\alpha$ (J2000.0)\tablenotemark{a} &  08 13 14.84 ($\pm 0.04$)  \\
$\delta$ (J2000.0)\tablenotemark{a} &  +45 59 30.5 ($\pm 0.4$) \\

Peak \HI\ column density\tablenotemark{b} & $9.5\times10^{20}$~atoms~cm$^{-2}$ \\

$D_{\rm HI}$ &  \am{5}{90}$\pm$\am{0}{05} \\

$\int F_{\rm HI}{\rm d}\nu$ & $19.2 \pm 0.6$~Jy km s$^{-1}$ \\

$W_{20}$ & 120.9 km s$^{-1}$ \\

$W_{50}$ & 99.2  km s$^{-1}$ \\

$V_{\rm sys,HI}$\tablenotemark{a,}\tablenotemark{c} & 445.3~$\pm0.2$~km~s$^{-1}$ \\

$V_{\rm sys,HI}$\tablenotemark{c,}\tablenotemark{d} & 443.5~km~s$^{-1}$ \\

$F_{\rm cont}$ (21-cm)& $10.3 \pm 0.6$~mJy \\

Position angle\tablenotemark{e} & 172.0$^{\circ}\pm 0.2^{\circ}$  \\

Inclination\tablenotemark{a} & 33$^{\circ} \pm 1^{\circ}$  \\

$V_{\rm max}$\tablenotemark{a} & 87 km s$^{-1}$ \\

\tableline
\multicolumn{2}{c}{Derived Quantities} \\
\tableline

$M_{\rm HI}$\tablenotemark{f} & $(2.22\pm0.07) \times10^{8}$ $d_{7}^2~M_{\odot}$  \\

$M_{\rm dyn}$\tablenotemark{g} & 9.5$\times10^{9}$ $d_{7}~M_{\odot}$ \\

$M_{\rm HI}/L_{B}$ & 2.5 $M_{\odot}/L_{\odot}$ \\

$M_{\rm dyn}/M_{HI}$ & 43 $d_{7}^{-1}$ \\

$D_{\rm HI}/D_{25}$ & 3.5 \\

\enddata

\tablenotetext{a}{From tilted-ring rotation curve fit.}

\tablenotetext{b}{At $16''$ resolution.}

\tablenotetext{c}{Heliocentric; optical definition.}

\tablenotetext{d}{From global \HI\ profile (optical definition).}

\tablenotetext{e}{Based on the kinematic major axis of the \HI\
velocity field (see \S\ref{ngcvelfield})}

\tablenotetext{f}{Assuming the \HI\ is optically thin and $d_{7}$
is the distance to NGC~2537 expressed in units of 7~Mpc. }

\tablenotetext{g}{From $M_{\rm dyn} = 2.326\times 10^{5}rV^{2}(r)$, where
we have taken $r = 5.4$~$d_{7}$~kpc (160$''$) and $V(r) = V_{\rm max}
= $87~km~s$^{-1}$.}

\end{deluxetable}

\begin{deluxetable}{ccc}
\tabletypesize{\tiny}
\tablewidth{0pc}
\tablenum{7}
\tablecaption{NGC~2537 Rotation Curve}
\tablehead{\colhead{Major axis distance (arcsec)} & \colhead{Circular
Velocity (\kms)} & \colhead{Uncertainty (\kms)}}
\startdata
     -165.0  &   497.5    &  3.2 \\
     -156.0  &    492.0   &   2.0\\
     -147.0  &    491.2   &   3.2\\
     -138.0  &    493.5   &   3.1\\
     -129.0  &    485.8   &   5.3\\
     -120.0  &    503.2   &   7.6\\
     -111.0  &    497.8   &   5.4\\
     -102.0  &    493.2   &   2.8\\
     -93.0   &   487.8    &  3.8\\
     -84.0   &   486.9    &  1.8\\
     -75.0   &   488.6    &  1.1\\
     -66.0   &   492.9    & 0.7\\
     -57.0   &   490.6    & 0.9\\
     -48.0   &   486.6    & 0.8\\
     -39.0   &   484.4    & 0.8\\
     -30.0   &  483.8     &0.9\\
     -21.0   &   479.2    & 1.0\\
     -12.0   &   471.8    &  1.4\\
     -3.0    &  461.8     & 2.4\\
      6.0    &  442.4     & 2.8\\
      15.0   &   417.4    &  2.1\\
      24.0   &  410.0     & 1.3\\
      33.0   &   411.4    &  2.8\\
      42.0   &   411.4    &  8.0\\
      51.0   &   412.5    &  5.2\\
      60.0   &   409.4    &  1.9\\
      69.0   &   400.0    &  3.0\\
      78.0   &  402.6     & 1.3\\
      87.00  &    403.1   &   1.9\\
      96.0   &   394.6    &  5.6\\
      105.0  &    384.6   &  8.0\\
      114.0  &    408.0   &   1.3\\
      123.0  &    396.0   &   5.3\\
      132.0  &   401.1    &  1.5\\
      141.0  &    404.0   &   1.7\\
      150.0  &    405.1   &   1.5\\
      159.0  &   395.1    &  1.6\\
\enddata
\end{deluxetable}

\subsubsection{The \HI\ Kinematics of NGC~2537\protect\label{ngcvelfield}}
The \HI\ velocity field shown in Figure~\ref{fig:ngccomposite} reveals that
NGC~2537 exhibits well-ordered rotation.  Note that the \HI\ arm
mentioned in the previous section participates in the regular disk
rotation (although a
series of kinks can be seen in the isovelocity contours as they cross
the region between the arm and the main disk).  While BCDs as a class are
in general rotationally-supported, only a small fraction
exhibit the type of well-ordered velocity field seen in NGC~2537  (cf. van Zee
et al. 1998,2001; Thuan et al. 2004). 

In Figure~\ref{fig:ngcPV} we show a P-V diagram extracted along the kinematic 
major axis of
NGC~2537 (lying at PA$\approx172^{\circ}$; see below). 
To improve signal-to-noise, we have averaged over a 15$''$-wide strip.
Using the data shown in Figure~\ref{fig:ngcPV}, we have also measured a
disk rotation curve for NGC~2537 by fitting intensity-weighted Gaussians to
a series of one-dimensional slices
extracted at 9$''$ intervals along the major axis. The result is
shown in Figure~\ref{fig:ngcrotcurve} and provided in electronic form
in Table~7. 
Despite the rather high central velocity dispersion observed in
NGC~2537 (see below) we have not applied corrections to the observed
rotation curve for asymmetric drift. We estimate that 
such corrections are likely to be significant ($\sim20-30$\%) only within
the centralmost regions of the galaxy ($r\lsim30''$), and the
corrections to individual points are difficult to compute accurately with only a few
resolution elements across this portion of the galaxy 
(cf. Meurer et al. 1996; Masset \&
Bureau 2003). Even without correcting for asymmetric drift, we find
the rotation curve of NGC~2537 rises rather steeply,
reaching a plateau within $\sim 30''$ ($\sim 1.0~d_{7}$~kpc, i.e., two
beam widths). Beyond this,
a more extended region is seen on both sides of the
disk. This extended portion is nearly flat, although it 
continues to rise slightly with increasing
galactocentric distance. Superposed on the outer, nearly flat portion of the
rotation curve, a series of 
``wiggles''
with amplitude $\sim$10~\kms\  is visible. The onset of these velocity undulations
corresponds with the observed edge of the stellar disk.

Based on the rotation curve in Figure~\ref{fig:ngcrotcurve}, 
the peak {\it observed} rotation velocity of NGC~2537 is
$\sim$50~\kms. Correcting this to the true rotational
velocity requires an estimate of the inclination of the galaxy.
The morphology of this BCD
makes an inclination determination from optical
images difficult, as the intrinsic flattening of this type of galaxy is
poorly known. 
To better constrain the inclination and other kinematic properties of
NGC~2537, we have used the velocity-field analysis method
of van~Moorsel \& Wells (1985) which
models the galaxy as a rotating thin disk.  Assuming an exponential
parameterization of the rotation curve leads to rapid convergence but
also shows that the bright central ring is tilted with respect to the
rest of the galaxy.  Fitting the region outside the central $40''$
(in radius) results in an inclination angle of $33\pm 1^\circ$, whereas
the region of bright \HI\ clumps is well fit by a rotating ring with an
inclination of $28\pm3^\circ$.  Both regions yield a consistent  
center of
rotation at $\alpha_{\rm J2000.0}$ = 08$^{h}$ 13$^{m}$ 14.84$^{s}$ ($\pm 0.04^{s}$);
$\delta_{\rm J2000.0}$ =  +45$^{\circ}$ 59$'$ \as{30}{5}
($\pm$\as{0}{4}), where the errors are formal
values derived from the fits.  The systemic velocities derived from  
both fits
are consistent as well, yielding an average value of
$445.3\pm0.2$~\kms.
The major axis is found to lie at PA$=172.0^\circ\pm0.2^\circ$, with both
fits again showing good agreement.
Finally, the deprojected rotational velocities are
$80\pm 7$~\kms\ for the ``ring'' and $87 \pm 2$~\kms\ for the asymptotic
(maximum) rotational velocity of the disk of  
the galaxy. The two velocity curves join smoothly at the radial distance of $40''$.
This value of $V_{\rm max}$ is surprisingly high for an
optically tiny galaxy like NGC~2537, and it exceeds most of the values
measured for other BCDs, which tend to be a factor of two smaller (van
Zee et al. 1998), although a few exceptions are known (e.g., 
Meurer et al. 1996).  Nonetheless, the
rather large $V_{\rm max}$ derived for NGC~2537 is congruous with the shape of its
rotation curve, which shows a turnover and extended
flat portion as opposed to the solid body-type rotation curve seen in
lower-mass dwarfs.

\subsubsection{HI Velocity Dispersion\protect\label{ngcveldisp}}
A map of the line-of-sight \HI\ velocity dispersion  in NGC~2537 is shown in
Figure~\ref{fig:ngccomposite}d. A radial gradient in the 
velocity dispersion is apparent, with values ranging from 
$\sigma_{V,\rm HI}\sim $4-8~\kms\ in the galaxy outskirts to as high as
$\sigma_{V,\rm HI}\sim$ 25~\kms\ near the center.
The values in the outskirts are comparable to the canonical values of
$\sim$6-9~\kms\ typical of both dwarf and spiral galaxies;
however, the values near the disk center reach a significant fraction of
the rotational velocity in the disk, implying appreciable pressure support.
The gas with the highest measured velocity dispersions coincides with the
location of the bright star forming regions in the stellar disk of NGC~2537,
although a few areas with $\sigma_{V,\rm HI}>10$~\kms\ are also seen outside
the stellar disk.  A similar pattern of \HI\  velocity dispersion has also been
observed in other BCDs, including NGC~2915 (Meurer et al. 1996). In this
latter case, Wada et al. (2002) argued that a stellar bar is responsible
for driving the high central turbulence.  However, NGC~2537 does not
show any sign of a stellar bar, implying that another mechanism (most likely
feedback from star formation) is responsible for the elevated velocity
dispersion.

\section{Discussion}
\subsection{The Origin of the Structural and Kinematic
Features of the IC~2233 Disk\protect\label{discussion}}
In earlier sections we have drawn attention to a variety of interesting
features of the \HI\ distribution and kinematics 
of IC~2233, including warping and flaring of the \HI\ layer; 
a corrugated vertical structure; lopsidedness in the \HI\ intensity
distribution, global
\HI\ profile and P-V diagrams; and evidence for vertically-extended,
rotationally anomalous gas. Here we discuss possible
origins for these various features and note the likelihood that some
or all of these phenomena may share a common origin.

Some of the features of IC~2233 are known to be
quite common in disk galaxies. For example,
the study of Garc\'\i a-Ruiz et al. (2002) established that warping is
nearly ubiquitous in disk galaxies and affects nearly 100\% of galaxies where the \HI\
disk extends beyond the stellar distribution. Similarly,
lopsidedness in the stellar and gas distributions is now well known to
affect a significant fraction of galaxies (Richter \& Sancisi 1994;
Zaritsky \& Rix 1997; Haynes
et al. 1998), and asymmetries seem to
be even more common among extreme late-type disks, affecting more than
half of such systems (Matthews et al. 1998). To date, rotationally anomalous 
extraplanar \HI\ and disk
corrugations have been confirmed in only a handful of galaxies (see
\S\ref{corrugations},\S\ref{icmajorPV}), but relatively few
investigators have searched systematically for these phenomena,
thus their frequency of occurrence is not yet known. 

Tidal interactions and minor mergers frequently have been invoked to explain
a variety of structural and kinematic phenomena in galaxies. 
For example, tidal interactions have been suspected as a key
trigger of galaxy lopsidedness (e.g., Zaritsky \& Rix 1997) and have
been proposed as a possible driver of warps (e.g., Weinberg 1995) and of
disk corrugations (e.g., Edelsohn \& Elmegreen 1997). This raises
the possibility that this process could be shaping IC~2233. 
Indeed, we chose to observe IC~2233 as a suspected
``victim'' of a tidal encounter (see \S\ref{intro}).
However, numerical simulations predict that most tidally-induced
disturbances 
will be rather short-lived, and prevalent only during the interval when the
minor merger is ongoing (Walker et al. 1996; Sellwood et al. 1998; 
Noguchi 2001; Bournaud
et al. 2005). Our observations do not support this scenario, as we
find no neighbor with sufficient mass to induce tidal effects in IC~2233, and
we have not identified any stellar or gaseous remnant of a possible intruder.

In \S\ref{thefield} we noted the lack of  any candidates for
{\it bona fide} companions to IC~2233 based on optical data and
argued that its previously suspected neighbor, NGC~2537,
most likely lies at a different distance. Furthermore, our \HI\ observations of
both galaxies show that even if they lay at similar distances,
NGC~2537 and IC~2233 would not constitute a bound pair. Assuming
a minimum separation (equal to the
projected separation at 10~Mpc), a relative velocity equal to the difference in
the recessional velocities, and total galaxy masses equal to the
dynamical masses given in
Tables~3 and 6, then the
kinetic energy of the system ($\sim7.2\times10^{56}$~erg) 
would exceed the potential energy ($\sim2.6\times10^{56}$~erg) by
roughly a factor of 3. Furthermore, the velocity separation of
the centers of mass of the galaxies exceeds their ``flat'' rotational
velocities, again indicating that the galaxies are not gravitationally bound.
Finally, a comparison of
Figures~\ref{fig:icmom1} and \ref{fig:ngccomposite} shows that
any interaction would have been retrograde, thus
minimizing its dynamical effects (e.g., Velazquez \& White 1999).
Consistent with
these arguments, our deep \HI\ observations of IC~2233 and NGC~2537
have  failed to uncover any 
overt signatures of a recent or ongoing interaction, such as tidal debris
or counter-rotating gas.

To investigate the environment of IC~2233 further, we have also
performed a systematic search
for gas-rich  neighbors using the matched-filter 
method described in Paper~I. We explored a
$30'\times30'$ region over the velocity interval covered by our
observations (see Table~2) and searched for signals with velocity
widths of up to $\sim$40~\kms. This search uncovered seven features
with amplitudes
$>4\sigma$, but none greater than $5\sigma$. Four of the $4\sigma$
features are positive (emission) features, while three are negative,
suggesting these simply represent the expected Gaussian tail of the
noise distribution. Moreover, none of the positive features have optical
counterparts in the SDSS data (see \S\ref{thefield}). 
We therefore do not consider any of these
to be likely candidates for companions to IC~2233. Based on
these results, we derive 
an (8$\sigma$) upper limit to the \HI\ mass of any 
companions to IC~2233 
of \mhi$<3\times10^{6}d^{2}_{10}~M_{\odot}$
within a radius of 30$'$ ($\sim90~d_{10}$~kpc). Similar
limits for NGC~2537 are \mhi$<1.5\times10^{6}d^{2}_{7}~M_{\odot}$
within $\sim60~d_{7}$~kpc.

If the properties of IC~2233 cannot readily be accounted for by a recent
interaction, it is necessary to consider alternative explanations.
For example, Levine \& Sparke (1998) and Noordermeer et al. (2001)
have proposed that galaxy lopsidedness may be linked to
the disk lying
offset from the center of its global potential. Noordermeer
et al. (2001) presented some predictions of this model that can be 
compared with observations. The details depend on whether the disk
rotates prograde or retrograde relative to the halo. However, in
general the models predict: (1) modest asymmetries in shape and
extent of the two sides of the major axis P-V curve, 
with the more extended
side showing a slight velocity decrease and the less extended side
showing a slight ``upturn'' near the last measured point; (2) a global \HI\
profile with a higher peak on the side 
corresponding to the more extended side of the
rotation curve and a lower, sloping horn on the opposite side; (3)
various degrees of twisting or distortion of the velocity field (which
depend on the orientation of the axis of symmetry of the disk relative
to the observer).
All of these signatures are observed in IC~2233 (Figures~\ref{fig:icglobal},
\ref{fig:icmom1}, \ref{fig:icPV}). Unfortunately,
Noordermeer et al. (2001) did not offer specific suggestions as to what mechanism
might lead to an offset between the disk and the overall
galactic potential.

There are intriguing similarities between the velocity field,
rotation curve, and global \HI\ profile of IC~2233 and those of the dwarf LSB spiral
NGC~4395 (see Swaters et al. 1999; Noordermeer et al. 2001), a galaxy we
have suggested  might be an analog of IC~2233 observed face-on (\S6.3).
NGC~4395 is not known to be tidally interacting or undergoing
a minor merger, hence these similarities suggest the possibility of a more
universal and continuous driver for these traits in low-mass, late-type disks
(see also Wilcots \& Prescott 2004).  Spontaneously arising asymmetries
(e.g., those linked with spiral arm formation) are one possibility,
although Bournaud et al. (2005) argue that in most instances these are likely
to be too weak; these authors suggest instead that a more promising explanation is 
the continuing cosmological accretion of gas. Cosmological models
predict that such infall should continue to the present day,
particularly in low-density field environments such as those inhabited
by IC~2233 and NGC~4395 (e.g., Kere\v{s} et al. 2005). 

The possibility that IC~2233 might be continuing to accrete material
from the IGM is intriguing in light of several other recent theoretical
studies that predict that cosmological accretion can account for a
variety of properties observed in isolated galaxies, including the presence of anomalously
rotating extraplanar gas (Fraternali et
al. 2007) and warping of the disk (Jiang \& Binney
1999; Shen \& Sellwood 2006). While none of these studies specifically
investigated whether disk corrugations might also be triggered
by slow gas accretion, previous studies linking
corrugations with warps (Nelson 1976; Sparke 1995; Masset \& Tagger
1997) and velocity undulations
with disk lopsidedness (Schoenmakers et al. 1997) suggest this would
be an interesting topic for further research.
Additional high-resolution, high-sensitivity \HI\ 
studies of apparently isolated, low-mass spirals would also help 
to establish how ubiquitous the simultaneous appearance of these
various phenomena is in the absence of close neighbors 
and permit more rigorous comparisons with the
morphological and kinematic signatures of gaseous infall
predicted by models.

\subsection{A Comparison between NGC~2537 and IC~2233: Clues to their
Evolutionary Histories\protect\label{comparison}}
As emphasized by Matthews \& Gallagher (1997), one 
intriguing aspect of extreme late-type disk galaxies is the existence
of such a diversity of disk morphologies within a relatively narrow
range of mass and luminosity (see also Noguchi 2001). 
IC~2233 and NGC~2537 are two excellent
illustrations of this contrast; both are rotationally-dominated
galaxies having similar peak rotational
velocities, dynamical masses, and blue luminosities (see Tables~3 \& 6),
yet they exhibit a variety of differences in terms of their optical and
\HI\ morphologies. We now briefly comment on the possible origin and
implications of these differences.

One key difference between IC~2233 and NGC~2537 becomes apparent from 
a comparison of their respective disk rotation curves 
(Figures~\ref{fig:icrotcurve} \& \ref{fig:ngcrotcurve}). Compared with
IC~2233, the rotation curve of NGC~2537 rises far more steeply in the
central regions, implying a much higher central mass
concentration. Indeed, such
enhanced central mass concentrations appear to be a hallmark of BCDs
(e.g., van Zee et al. 1998). 

Previous authors have suggested that BCDs might represent a
brief starbursting state of ordinary dwarf galaxies (e.g., Taylor et
al. 1993; Papaderos et al. 1996). While we
cannot rule out such a scenario for some low-mass BCDs,
our present observations suggest that it is unlikely that 
a galaxy like NGC~2537 will
ever evolve into a ``normal'' LSB disk galaxy (e.g., similar to IC~2233), or vice
versa. 

If a tidal interaction or minor merger were invoked to
explain a transition from LSB disk to starbursting BCD (e.g., Taylor 1997), 
one difficulty would be to account for the
properties of the outer gas disk of NGC~2537, which appears quite
dynamically cold with no signs of recent dynamical heating or perturbations. Moreover, 
in contrast to typical gas-rich LSB spirals---which tend
to have blue outer disks,
shallow light profiles, and little central light concentration---the
underlying stellar disk of NGC~2537 is old, red, and 
compact; thus it is unclear that the required pre-merger progenitor would
resemble any ordinary class of disk galaxy. Finally, we note that NGC~2537
seems to join an increasing number of BCDs that are found to
reside in rather isolated environments (Telles \& Terlevich 1995; van
Zee et al. 1998).

As an alternative means of linking BCD and LSB/dwarf disk galaxies,
Papaderos et al. (1996) proposed that the reverse transition (from
BCD to LSB) might occur
via secular evolution processes. However, this would
necessitate a large-scale redistribution of mass, presumably via starburst-driven
winds followed by cooling and infall of this gas from the halo. 
While this might be possible for extremely low-mass dwarfs, 
this requirement is particularly problematic for NGC~2537, since its
gravitational potential is too steep to permit substantial
mass-loss via mechanical energy from stellar winds and
supernovae (see e.g., Ferrara \& Tolstoy 2000). 

Given the above difficulties, the
current body of observations of NGC~2537 seems to be most
consistent with the suggestion 
that at least some BCDs are a special subset of low-mass 
disk galaxies whose central mass concentrations lie at the extreme
end of the present-day distribution (Salzer \& Norton 1999; 
Ferrara \& Tolstoy 2000; Noguchi 2001). Models by
Noguchi (2001) predict that viscous
evolution would be particularly efficient in such
centrally-concentrated galaxies, thus giving a
natural explanation for their ability to funnel gas inwards to fuel
episodic central starbursts; in contrast, his models predict that 
the low central mass densities of LSB galaxies like IC~2233 will 
help them to maintain a fairly flat
and largely subcritical \HI\ surface density across much of their
disk. We conclude that IC~2233 and
NGC~2537 are therefore examples of small disk galaxies
that have evolved in relative isolation, and whose inherent
structural differences have allowed them to preserve
important differences in their morphologies and star-forming properties.

\section{Summary\protect\label{summary}}
We have presented VLA \HI\ imaging of the edge-on Sd `superthin' spiral galaxy
IC~2233 as well as the blue compact dwarf (BCD) NGC~2537. We have also described
results from
new optical $B$, $R$, and \HA\ imaging and
photometry for IC~2233.

We confirm that IC~2233 is an intrinsically LSB galaxy, having
a deprojected central surface brightness 
$\mu_{B,i}(0)\approx 22.6$~mag~arcsec$^{-2}$ and an extremely low \HI\
surface density over its 
entire disk ($\Sigma_{\rm HI}\lsim3 M_{\odot}$ pc$^{-2}$). Similar to many other
late-type, LSB galaxies, the \HI\ component of IC~2233 comprises 
a significant fraction of
its observed baryons ($M_{\rm
HI}/L_{B}\approx$0.83~$M_{\odot}/L_{\odot}$). 
Despite evidence for localized ongoing star
formation in IC~2233 in the form of prominent \HII\ complexes and shells,
red and blue supergiant stars, and a very blue
integrated color [$(B-R)_{0}=0.67\pm0.15$],  
we detect no significant 21-cm radio continuum emission from the galaxy, 
and the global star formation rate inferred from the \HA\
emission and from the {\it IRAS} FIR emission both imply an extremely
low globally-averaged star formation rate
($\lsim0.05~M_{\odot}$~yr$^{-1}$). 

Both the \HI\ and ionized gas disks of IC~2233 are clumpy and
vertically distended, exhibiting scale heights comparable to that of 
the young stellar disk. 
The thickness of both the gas and the stars flares with increasing
galactocentric radius. We have quantified the flaring of the
\HI\ disk and estimate an increase in scale height of a
factor of two across the galaxy. We
find evidence that IC~2233 also contains a component of ``anomalous'' extraplanar \HI\
emission whose rotation does not follow that of the material in the
midplane. Future three-dimensional kinematic 
modeling should help to confirm whether this material comprises a
part of a rotationally lagging \HI\ halo similar to those now confirmed
in a number of other spiral galaxies. 

The \HI\ disk
of IC~2233 exhibits a mild lopsidedness as evidenced by 
differences in the \HI\ intensity distribution on the two sides of the
disk, differences in shape and
extent of the major axis P-V curves on the two sides of the galaxy,
and by the asymmetric shape of the
global  \HI\ profile.  The origin of this lopsidedness is unclear,
although its kinematic signatures 
are consistent with model predictions for a disk lying offset from the center of its
overall halo potential.

A particularly intriguing feature of the \HI\ disk of IC~2233 is the
presence of a global corrugation pattern
with a period of  $\sim 7~d_{10}$~kpc and an 
amplitude of $\sim 150~d_{10}$~pc.
These  undulations may represent bending instabilities or 
be linked to underlying spiral structure.  Their presence
suggests that the optically diffuse disk of IC~2233 is largely
self-gravitating. Outside of the stellar disk, a mild warp of the \HI\
disk is also observed. 

The properties of IC~2233 form an interesting contrast to those of the
LSB superthin galaxy UGC~7321 (Paper~I). Although IC~2233 is only a
factor of two
less massive than UGC~7321, its kinematic characteristics resemble
those of a solid-body
rotator (\S\ref{icmajorPV}), while those of UGC~7321 are more 
similar to those of a giant spiral. The stellar
disk as well as the neutral and ionized gas layers of IC~2233 are
intrinsically 
thicker than those of UGC~7321 and exhibit far more complex
structure. These differences may reflect in part the weaker
self-gravity of the IC~2233 disk and the decreasing importance of rotation
shear. The respective properties of IC~2233 and UGC~7321 are 
consistent with previous suggestions that important changes occur in the
structural and ISM properties of galaxies  over the mass interval
corresponding to $V_{\rm rot}\sim$100~\kms\ (Dalcanton et al. 2004;
Matthews et al. 2005).

We have confirmed that the BCD NGC~2537 is a rotationally-dominated galaxy
and estimate its maximum (deprojected) rotational velocity to be
$V_{\rm max}\approx$87~\kms. This is at the high end of values observed 
for BCDs. The steep inner rise of
its rotation curve indicates a  high central matter density. 

The \HI\ velocity field of NGC~2537 is quite regular and shows no obvious
signs of recent perturbations. An arm-like feature in the outer \HI\
disk shows regular rotation and does not appear to be tidal in origin.
The \HI\ disk of NGC~2537 extends to $\sim$3.5~times the ($D_{25}$) diameter of
its stellar disk, and its
outer regions appear dynamically cold and fragmented. In
the inner \HI\ disk we measure
\HI\ velocity dispersions as high as $\sigma_{V,\rm HI}\sim25$~\kms, 
indicative of significant turbulent motions.  The inner 
disk also exhibits a tilted \HI\ ring with a diameter of
$\sim 2.0~d_{7}$~kpc and peak column density of $\sim10^{21}$~atoms~cm$^{-2}$---
several times higher than that of the surrounding gas. 

Several of the properties of of IC~2233, including its
lopsidedness, warped \HI\ disk, disk corrugations, and the presence of
anomalously rotating, vertically extended \HI\
emission, might seem readily
explained by a recent interaction or minor merger, but we have not
uncovered any direct evidence of such an event. We also argue that
NGC~2537 is unlikely to be the required perpetrator. Indeed, both
IC~2233 and NGC~2537 each appear to be evolving in relative
isolation. Although IC~2233
and NGC~2537 have sometimes been described as a ``binary pair'' in
the past literature, recent  distance estimates suggest they probably
lie at different distances. Moreover, even if they lay at the same
distance, our data show they would not form a bound pair and any
encounter would
have been retrograde, implying that any mutual interaction would have had
minimal impact
on the properties of the two galaxies. Our search for other companions to
both galaxies using
imaging and spectroscopic data from the SDSS as well as our new \HI\
data have failed to identify any additional candidates for close neighbors.
Slow, continuing accretion of
intergalactic gas may be one means of accounting for a number of the
observed properties of IC~2233 in the absence of a companion.
In the case of NGC~2537, the intrinsically high central matter
density of the galaxy coupled with efficient viscous evolution may
account for its bursting state without the need for a recent interaction.

\acknowledgements{We thank 
J. Gallagher for assistance with the WIYN observations and A. Seth
for computing the distance modulus to IC~2233 from his {\it HST}
data. We also acknowledge
valuable discussions with a number of colleagues, including 
D. Hogg, J. Hibbard, M. Roberts, L. Sparke, T. Cornwell, J. Condon,
B. Cotton, and E. Greisen. LDM was 
supported by a Jansky Fellowship from NRAO and a
Clay Fellowship from the Harvard-Smithsonian CfA. The VLA observations
described here were part of program AM649.
This research has made use of: the NASA/IPAC Infrared Science Archive
and the NASA/IPAC Extragalactic Database (NED)
which are operated by the Jet Propulsion
Laboratory, California Institute of Technology, 
under contract with the National Aeronautics and Space
Administration, and the Digitized Sky
Surveys (DSS), which were produced at the Space Telescope Science
Institute under U.S. Government grant NAG W-2166. Finally, we thank
the referee for a thoughtful and thorough reading of our paper.
}

\clearpage

%

%

%

%

%
\begin{figure}
\figurenum{4}
\epsscale{1.}
\plotone{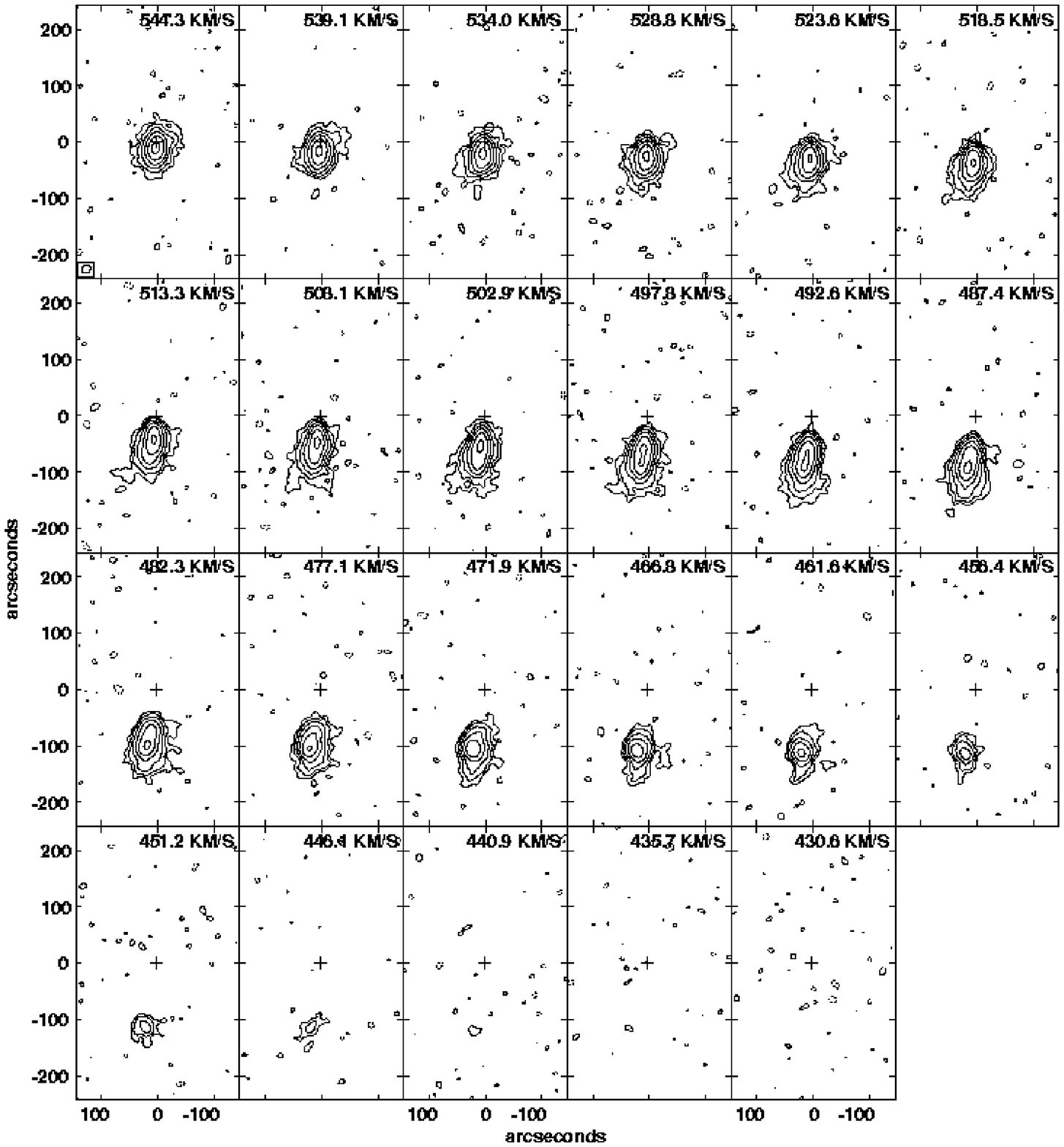}
\figcaption{continued.}
\end{figure}

\suppressfloats
\clearpage

%
\begin{figure}
\epsscale{1.}
\plotone{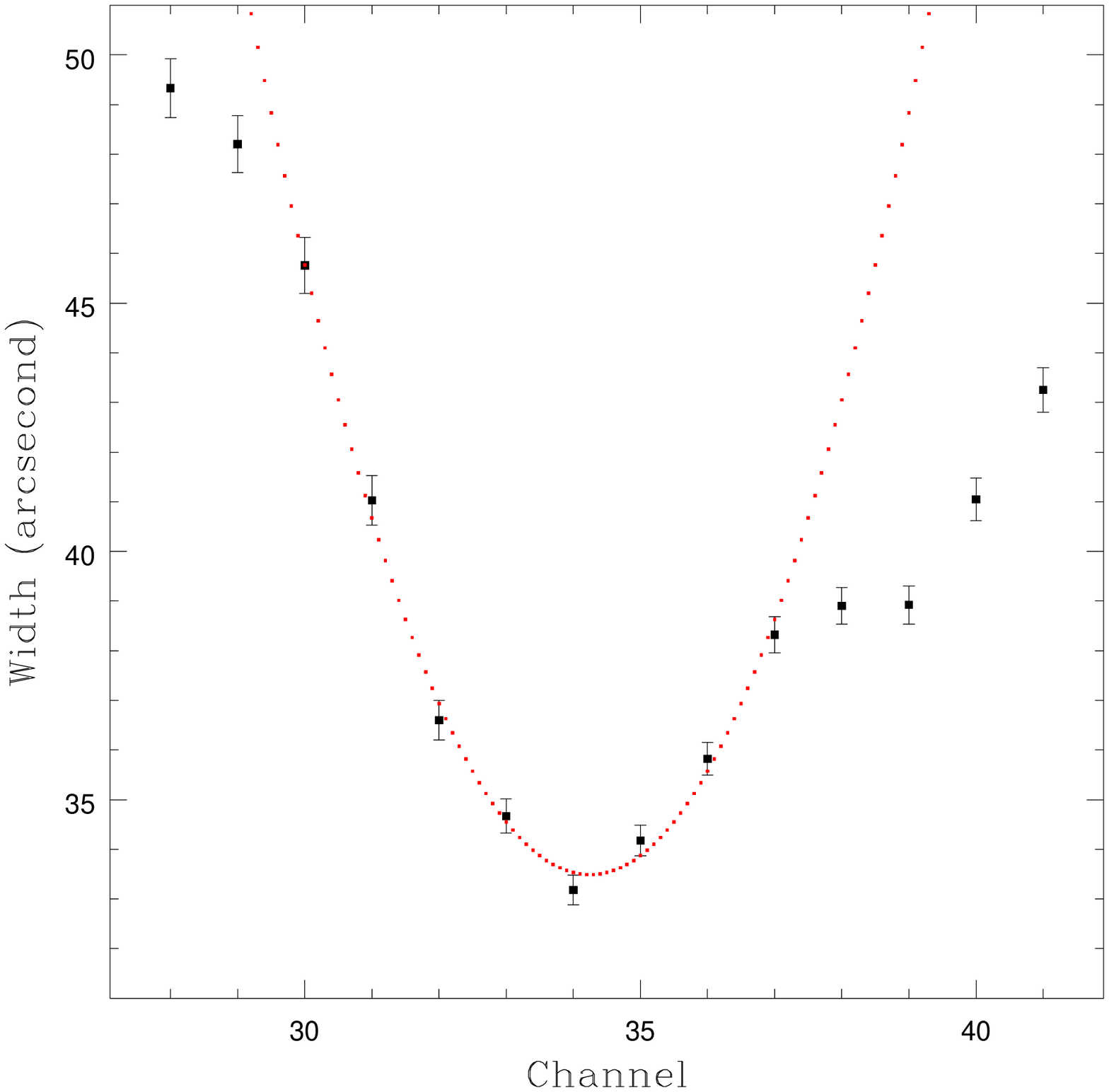}
\figcaption{Results of two-dimensional elliptical Gaussian fits to the \HI\
emission of IC~2233 in several channels. Full-width at half-maximum of
the major axis (after deconvolution of the
synthesized beam) is plotted as a function of channel.  
The  best fit parabola to the results for channels 30 -- 37 is overplotted.
The fit yields a minimum of the width for ``channel'' 34.25, corresponding to a
systemic velocity of $553.4\pm1.0$~\kms\ (see \S\ref{totalHI} for details).
\protect\label{fig:kincentfit}}
\end{figure}

%
\begin{figure}
\epsscale{1.0}
\plotone{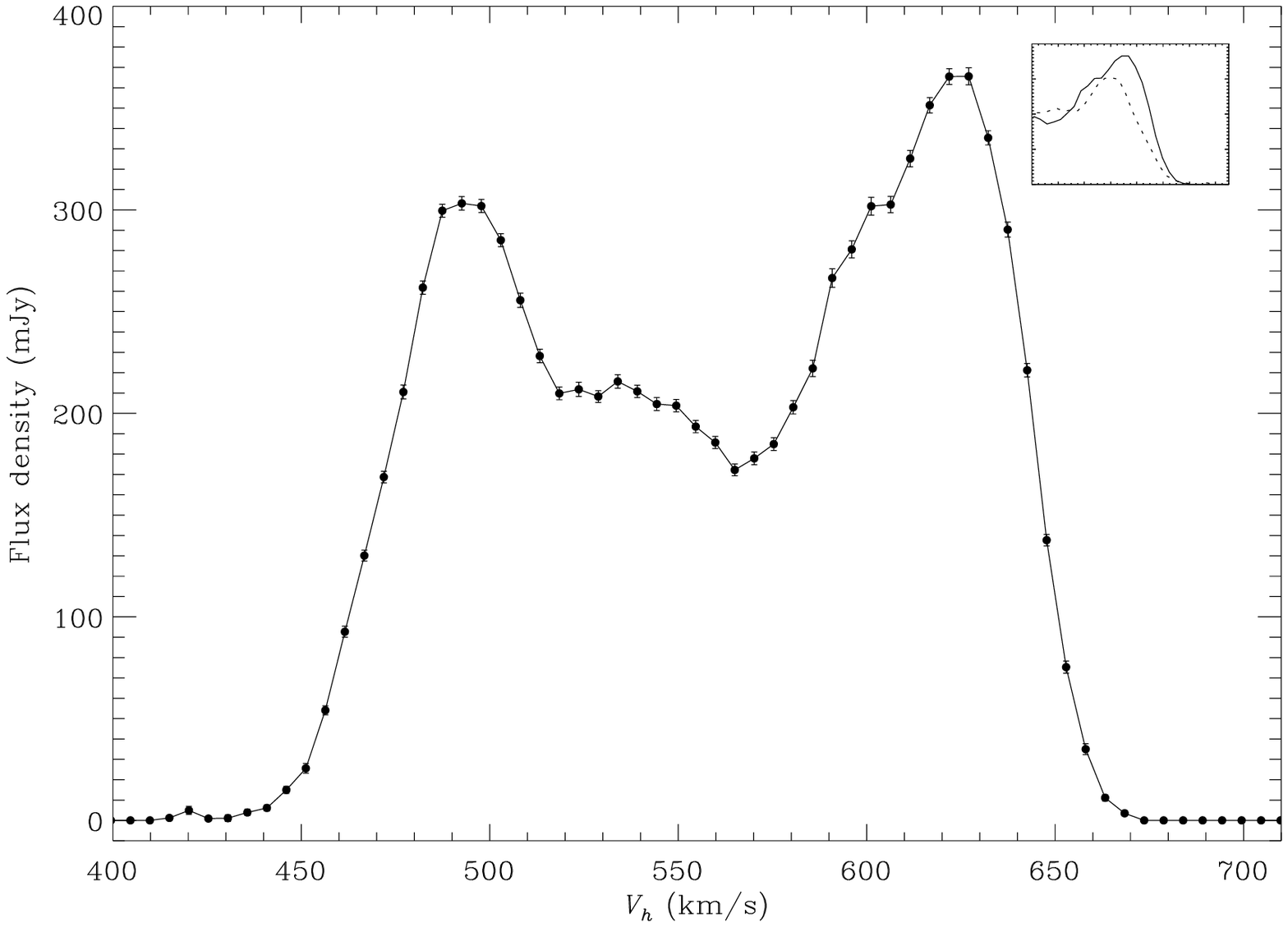}
\figcaption{Global (spatially integrated) 
\HI\ profile of IC~2233 derived from our VLA data.
The 1$\sigma$ error bars account for the statistical noise in each
channel but 
do not include the 1\% global calibration uncertainty. 
The inset shows the profile folded about the systemic velocity in
order to highlight the difference in slope of the two edges.
\protect\label{fig:icglobal}}
\end{figure}

\suppressfloats

%
\begin{figure}
\epsscale{0.6}
\plotone{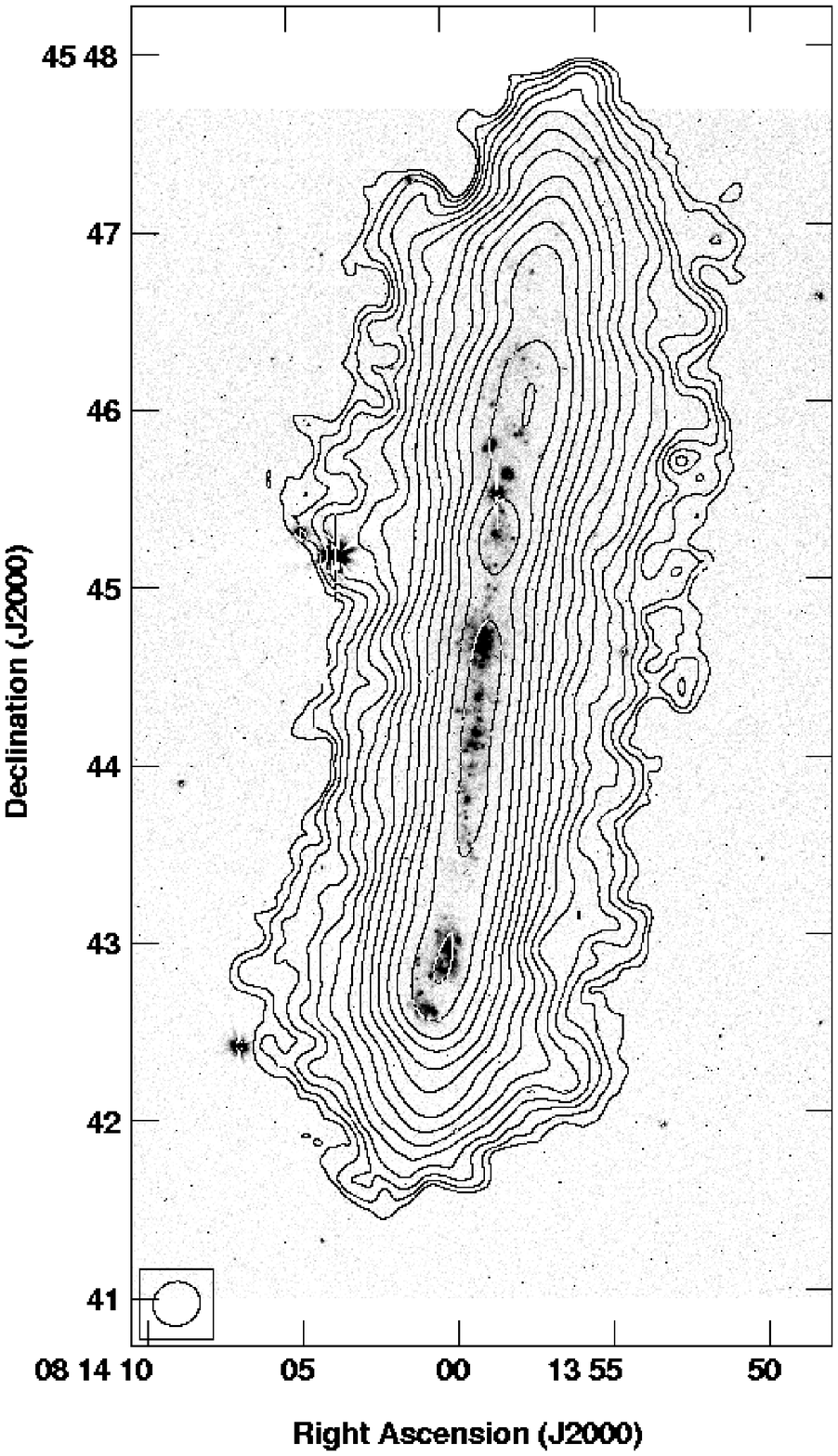}
\figcaption{\HI\ total intensity contours of IC~2233 overlaid on the
H$\alpha$+[\NII] image of the galaxy from Figure~\ref{fig:WIYNR}. 
\HI\ contour levels are $(1,1.4,2,2.8,4 ... 64,78)\times18$~mJy
beam$^{-1}$ m~s$^{-1}$. The highest contour departs from the
``$\sqrt{2}$''
sequence in order to highlight the correspondence between local maxima
in the \HI\ column density and the location of \HA\ emission features.
The synthesized beam is nearly circular with a
FWHM~$\sim16''$. 
\protect\label{fig:icmom0}}
\end{figure}

\suppressfloats
\clearpage

%
\begin{figure}
\epsscale{0.9}
\plotone{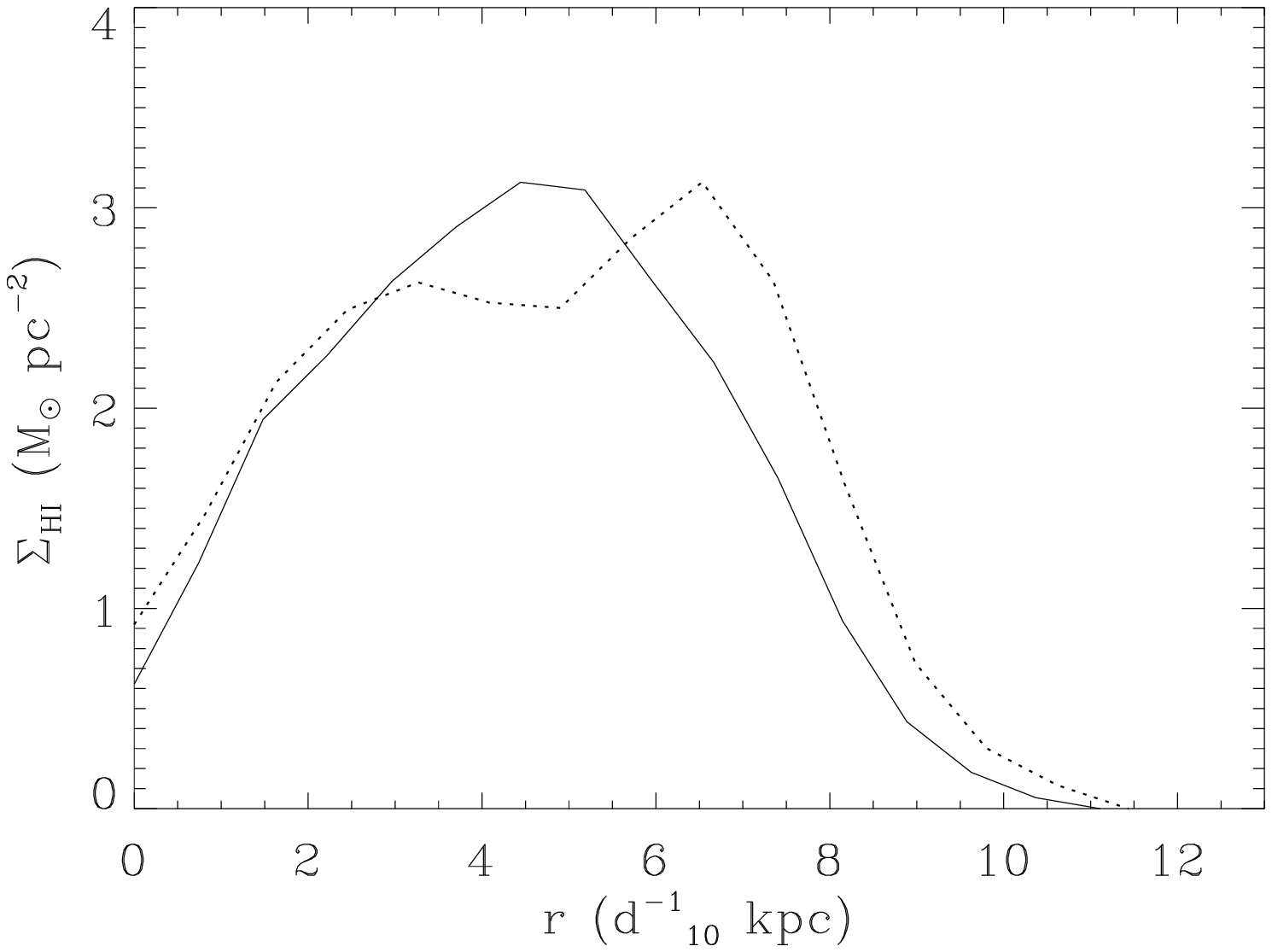}
\figcaption{Deprojected \HI\ surface density profile for IC~2233, computed
using an Abel inversion technique. \HI\ surface density
in units of solar masses per square parsec is plotted as a function
of projected radius in kpc. The southern (approaching)
side of the galaxy is shown as a dotted line and the northern (receding)
side is shown as solid line. Values at very small radii ($r\lsim$1.5~kpc) are highly
uncertain (see \S\ref{intensity}).
\protect\label{fig:icHIsurf}}
\end{figure}

%
\begin{figure}
\epsscale{0.9}
\plotone{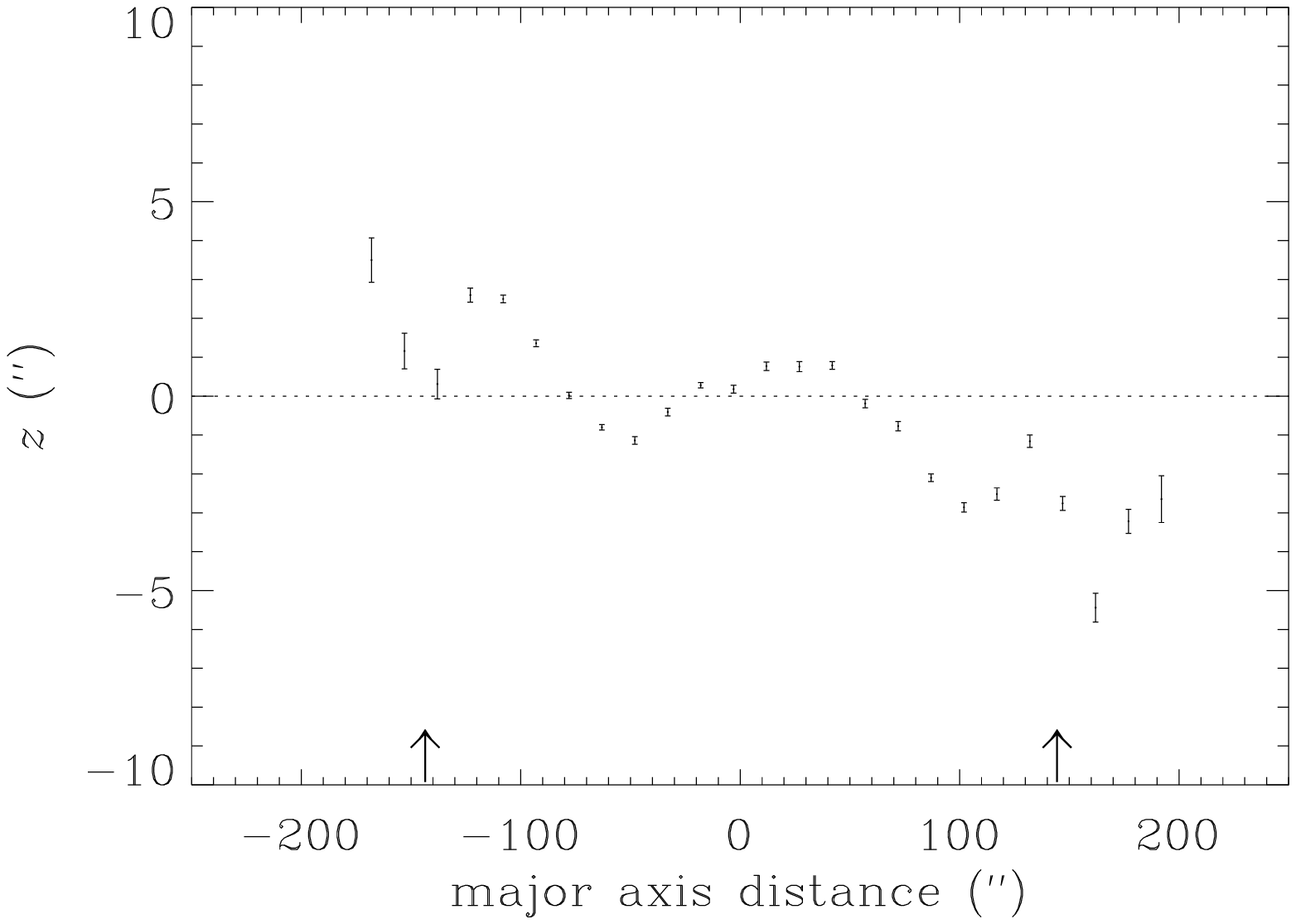}
\figcaption{\HI\ warp curve for IC~2233 derived by fitting Gaussians to
a series of \HI\ intensity profiles perpendicular to the major axis of
the disk. 
A major axis positional angle of $172^{\circ}$ was
used. The displacement of the fitted centroid from the mean disk
plane in arcseconds is plotted as a function of projected radius in
arcseconds. The 1$\sigma$
error bars indicate the formal uncertainty in the fit. Arrows along
the bottom of the figure denote
the observed extent of the stellar disk. 
\protect\label{fig:warpcurve}}
\end{figure}

%
\begin{figure}
\epsscale{0.6}
\plotone{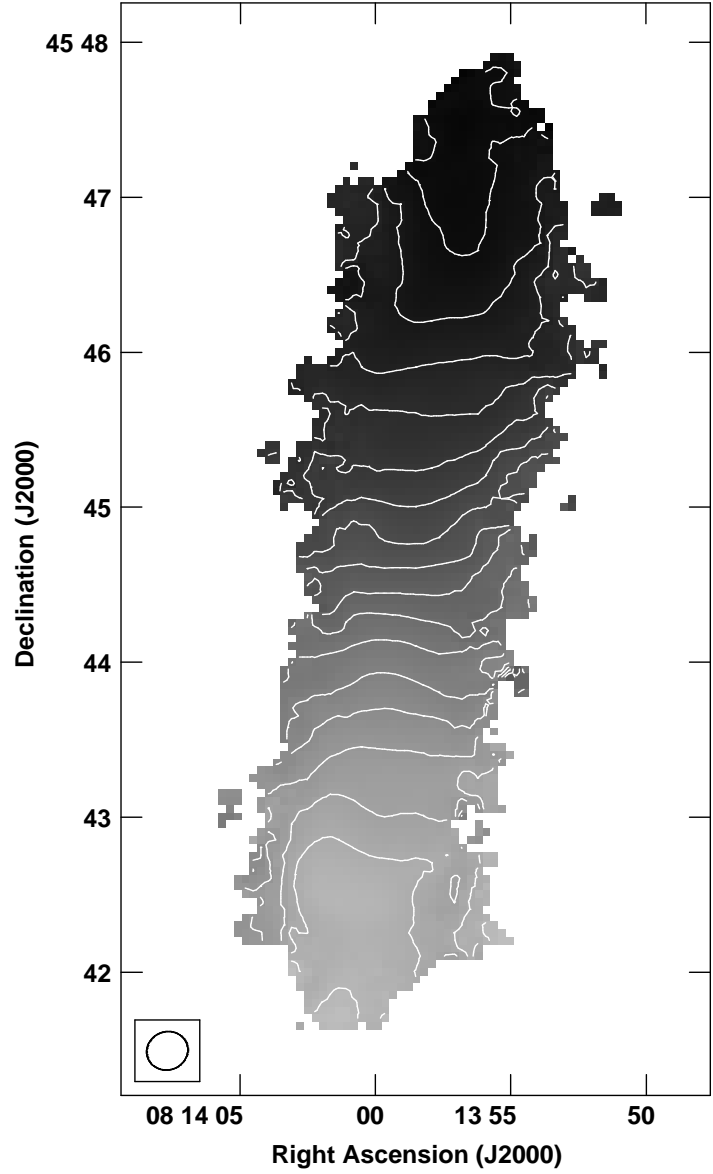}
\figcaption{\HI\ velocity field for IC~2233. Isovelocity contours
ranging from 441--668~\kms\ are shown at 10.4~\kms\ increments (i.e., every
other channel), overplotted on a linear greyscale
representation of the same map. The greyscale range is 400--700~\kms,
with darker colors corresponding to higher velocities.
\protect\label{fig:icmom1}}
\end{figure}

%
\begin{figure}
\epsscale{0.6}
\plotone{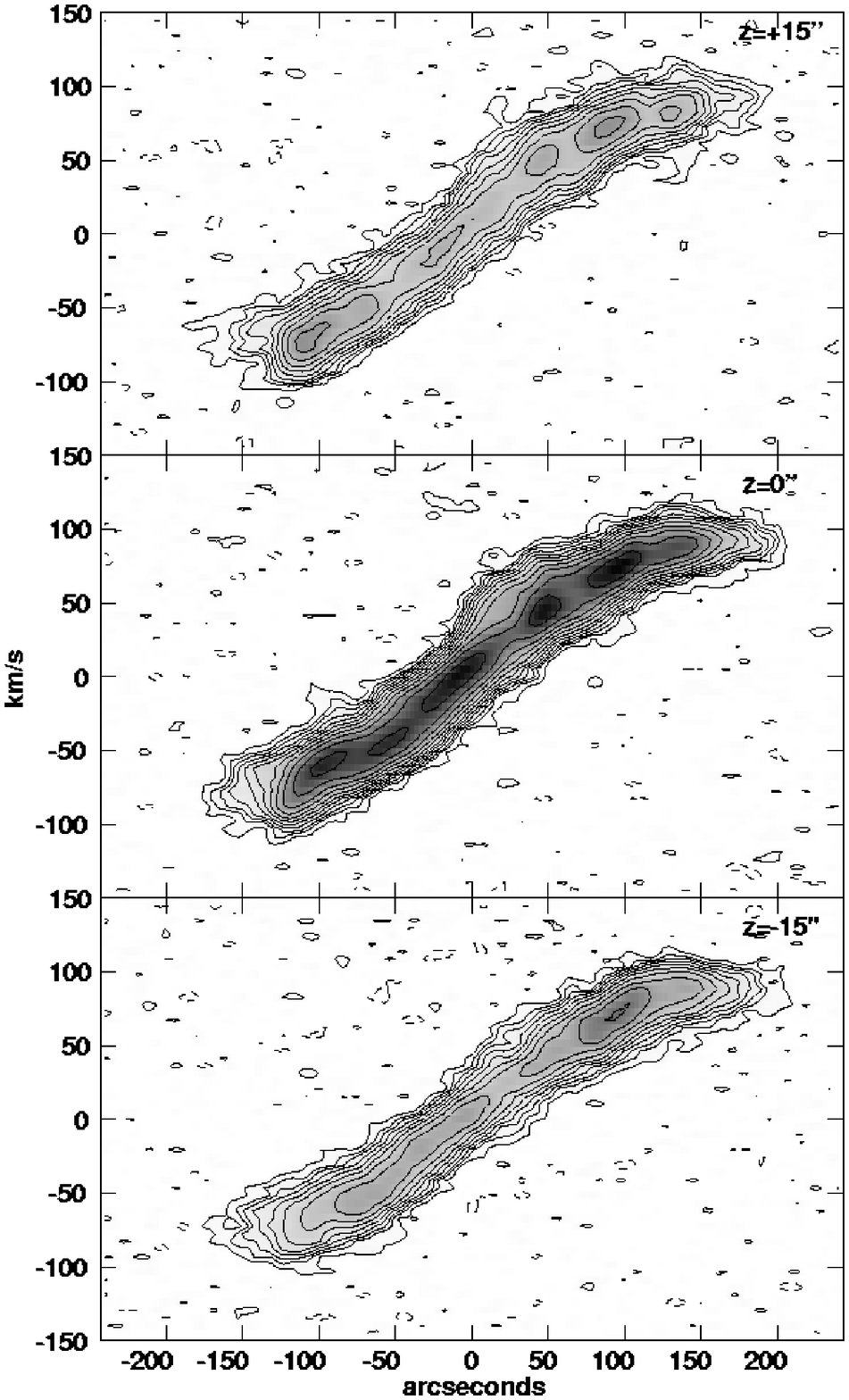}
\figcaption{\HI\ position-velocity plots parallel to the major axis
of IC~2233. The cuts were extracted in 3$''$ strips along the disk major
axis ($z=0$), and at $z=\pm15''$.  The contours are
$(-1.4 {\rm [absent]},-1,1,1.4,2,2.8,4,5.6,8,11,16,22,32,44)\times0.8$~mJy beam$^{-1}$.
The greyscale range is 0-45~mJy beam$^{-1}$.  \protect\label{fig:icPV}}
\end{figure}

%
\begin{figure}
\epsscale{1.0}
\vspace{-8.0cm}
\plotone{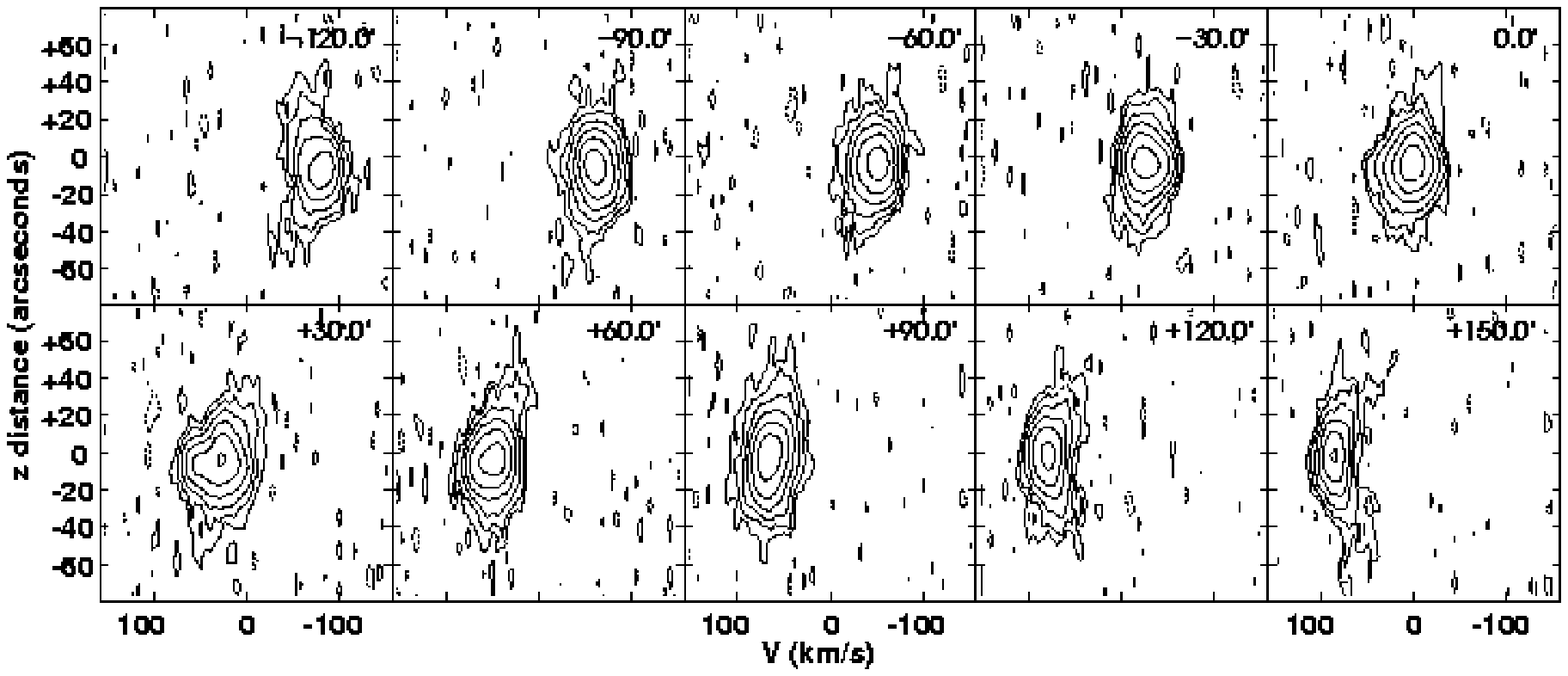}
\vspace{-0.5cm}
\figcaption{Sample minor axis position-velocity plots extracted at $30''$ intervals
along the major axis of IC~2233. The offset along the major axis relative to the galaxy
center is indicated in the upper right of each panel. 
The horizontal axes show radial velocity
relative to the systemic velocity, and the vertical axes show distance from
the midplane.  Positive values of $z$ correspond to the east
side of the galaxy. Contour levels are (-2[absent],-1,1,2,4,8,16,32)$\times$0.8
mJy beam$^{-1}$.  \protect\label{fig:icZPV}}
\end{figure}

%
\begin{figure}
\epsscale{0.85}
\plotone{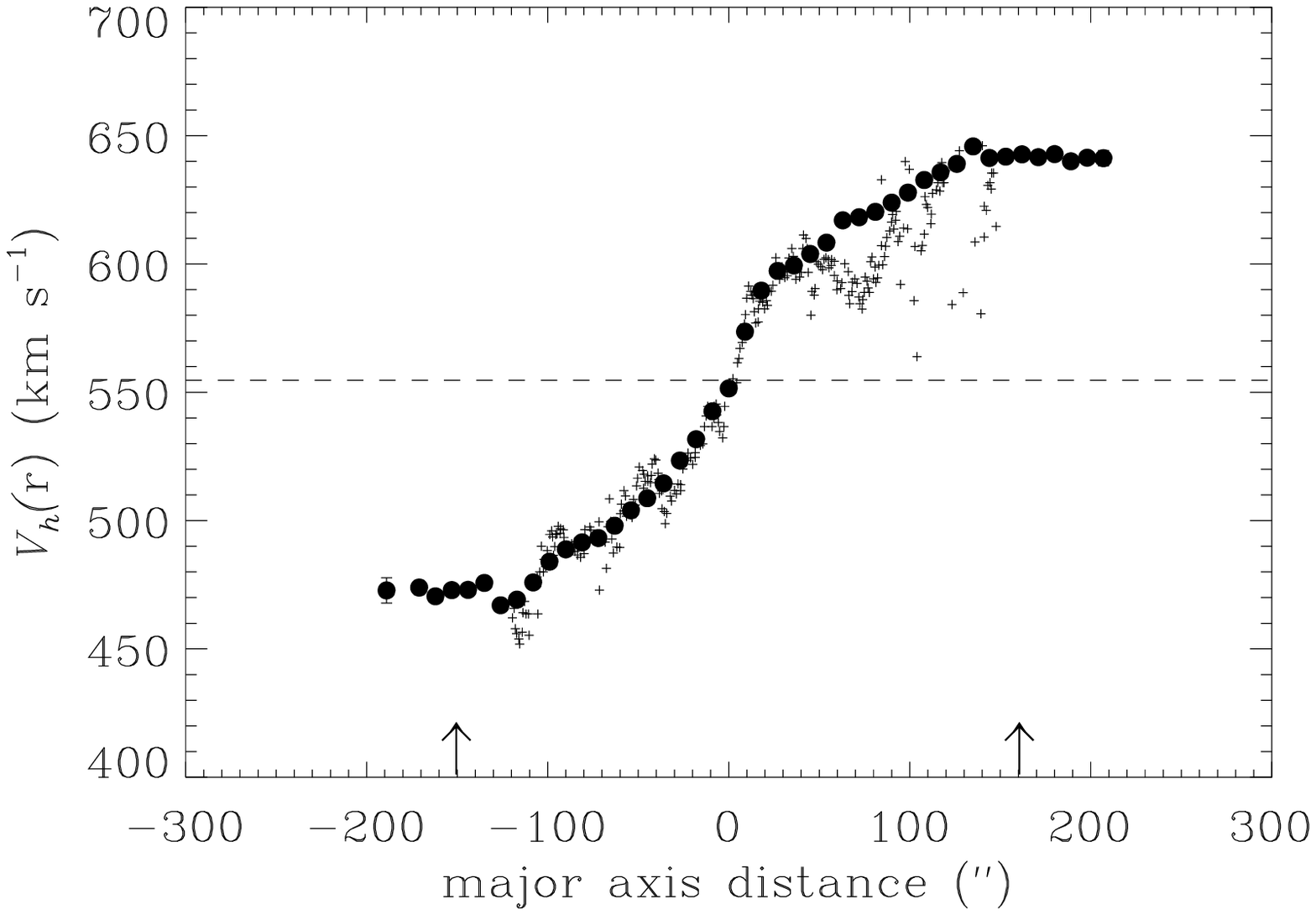}
\figcaption{\HI\ rotation curve for IC~2233 (filled circles).  
The overplotted crosses are data points derived by Goad \& Roberts 1981
using H$\alpha$ emission-line spectroscopy.  The axes are projected
distance along the major axis 
in arcseconds and rotational velocity in \kms. The dashed line
indicates the systemic velocity of the galaxy. The location of the
observed edges of the stellar disk are indicated with arrows.
\protect\label{fig:icrotcurve}}
\end{figure}

%
\begin{figure}
\epsscale{0.95}
\plotone{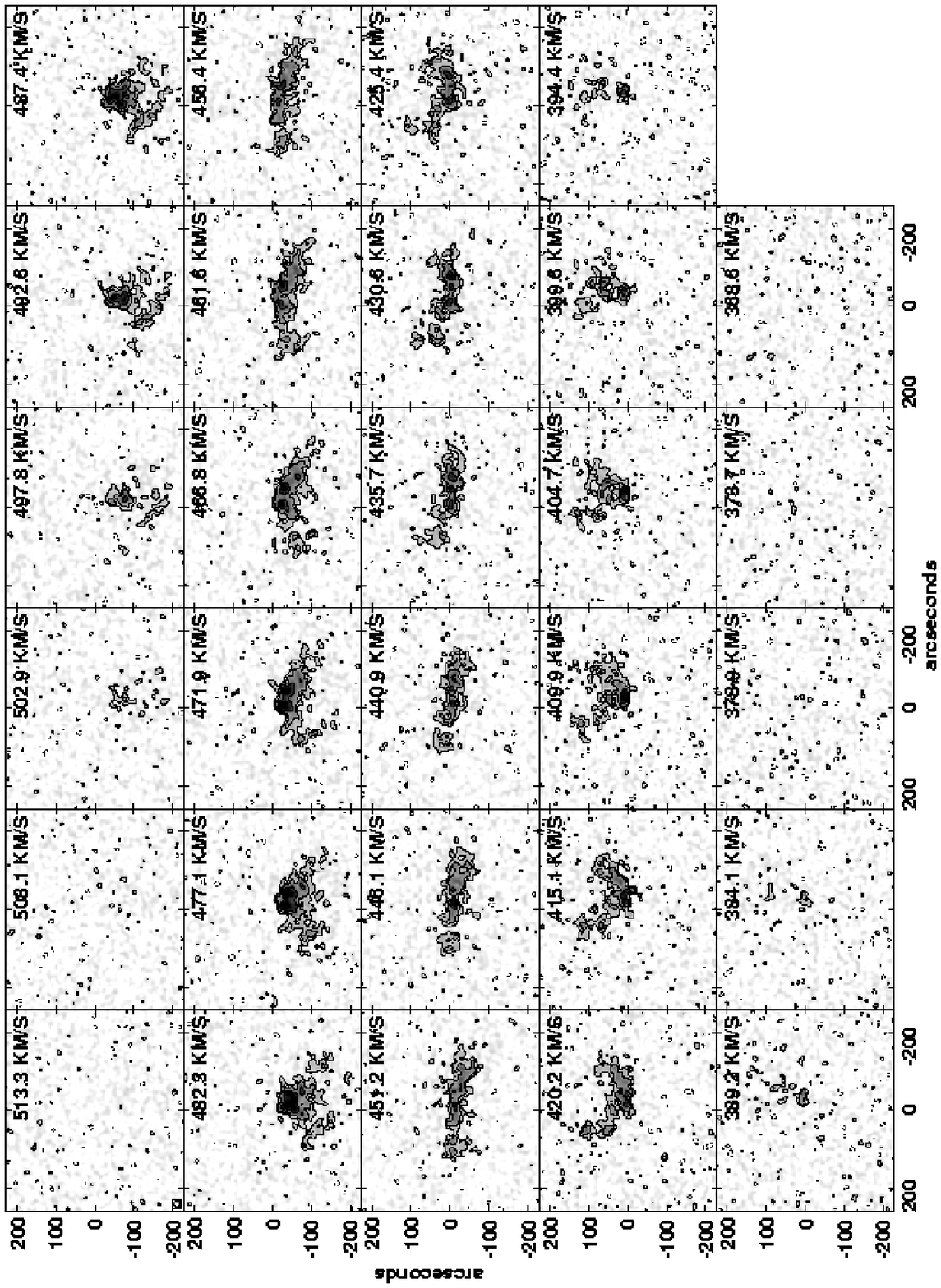}
\figcaption{\HI\ channel images for NGC~2537.  Contours are overplotted on a
greyscale representation of the same data.  Contour levels are 
($-2$[absent],-1,1,2,4)$\times$0.9~mJy beam$^{-1}$. The greyscale range is
0---5.0~mJy beam$^{-1}$.  Only the channels showing line emission, plus one
additional channel on either side, are plotted.  Each panel is labeled with its
corresponding heliocentric velocity.\protect\label{fig:ngccmaps}}
\end{figure}

%
\begin{figure}
\epsscale{1.0}
\plotone{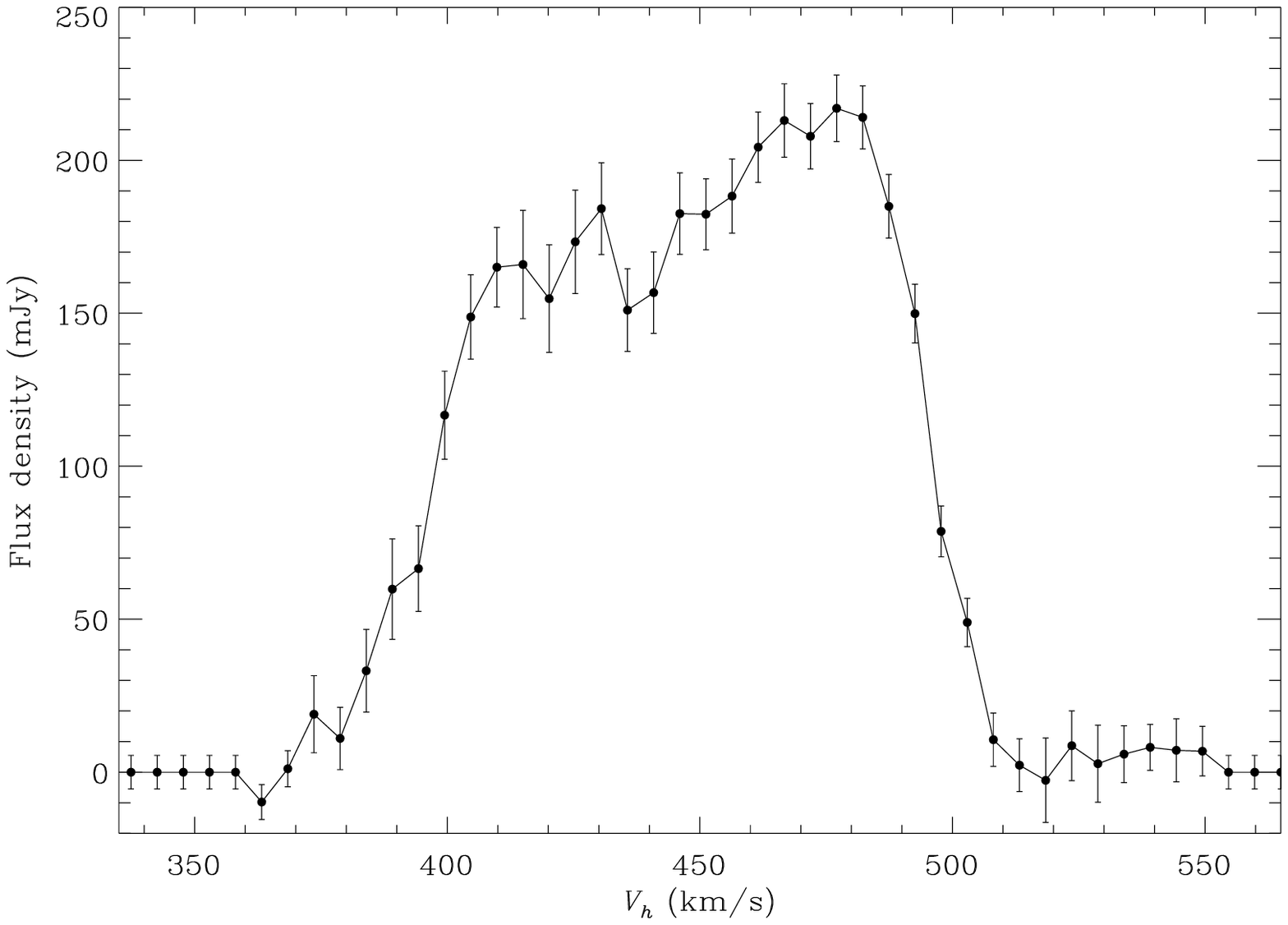}
\figcaption{Global (spatially integrated) 
\HI\ profile for NGC~2537 derived from VLA data. 
The 1$\sigma$ error bars correspond to the statistical noise in each
channel but do not include the $\sim$1\% global calibration uncertainty.
\protect\label{fig:ngcglobal}}
\end{figure}

%
\begin{figure}
\epsscale{1.0}
\vspace{-4.0cm}
\plotone{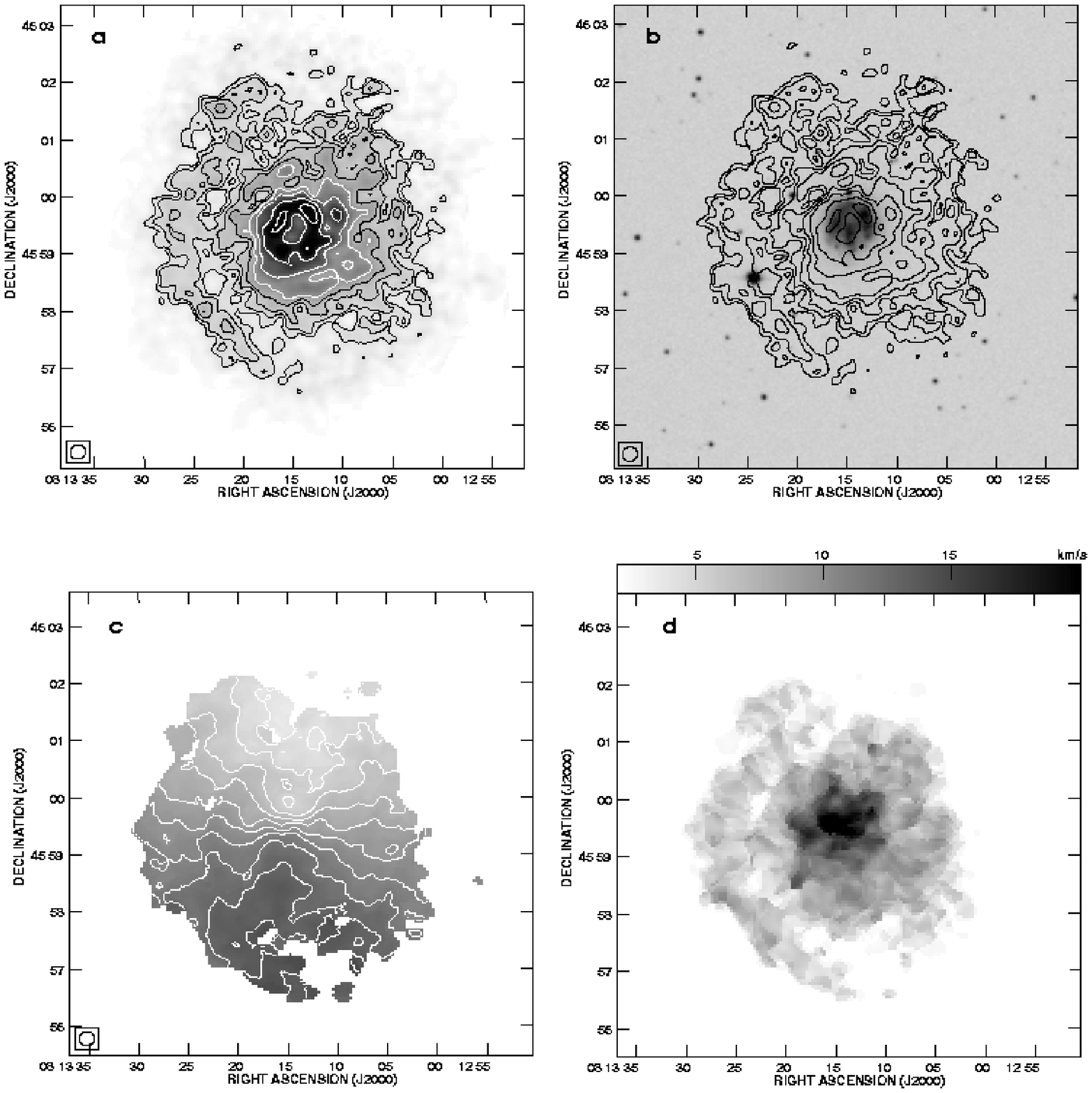}
\vspace{-1.0cm}
\figcaption{\HI\ moment maps of NGC~2537. For all data shown, 
the synthesized beam is roughly circular with a FWHM~$\sim16''$. 
(a) \HI\ total intensity contours overplotted on a greyscale
representation of the same data. Contour levels are 
(0.24,0.34,0.48,0.68,0.96,1.36,1.92)$\times$100~Jy
beam$^{-1}$~m~s$^{-1}$. The greyscale is 
linear from 0-200 Jy beam$^{-1}$~m~s$^{-1}$.
(b) Same as (a), but the \HI\ contours are overlaid on a blue image of the
galaxy from the DSS.   
(c) \HI\ velocity field.  Isovelocity contours ranging from
368--502~\kms\ are shown at 10.4~\kms\ increments (i.e., every other channel),
overplotted on a linear greyscale representation of the same map. The greyscale
range is 380--550~\kms, with darker colors corresponding to higher velocities.
(d) \HI\ velocity dispersion.  The greyscale range is
2--20~\kms.  Maximum observed values are $\sim$25~\kms.
\protect\label{fig:ngccomposite}}
\end{figure}

\suppressfloats
\clearpage

%
\begin{figure}
\epsscale{1.0}
\plotone{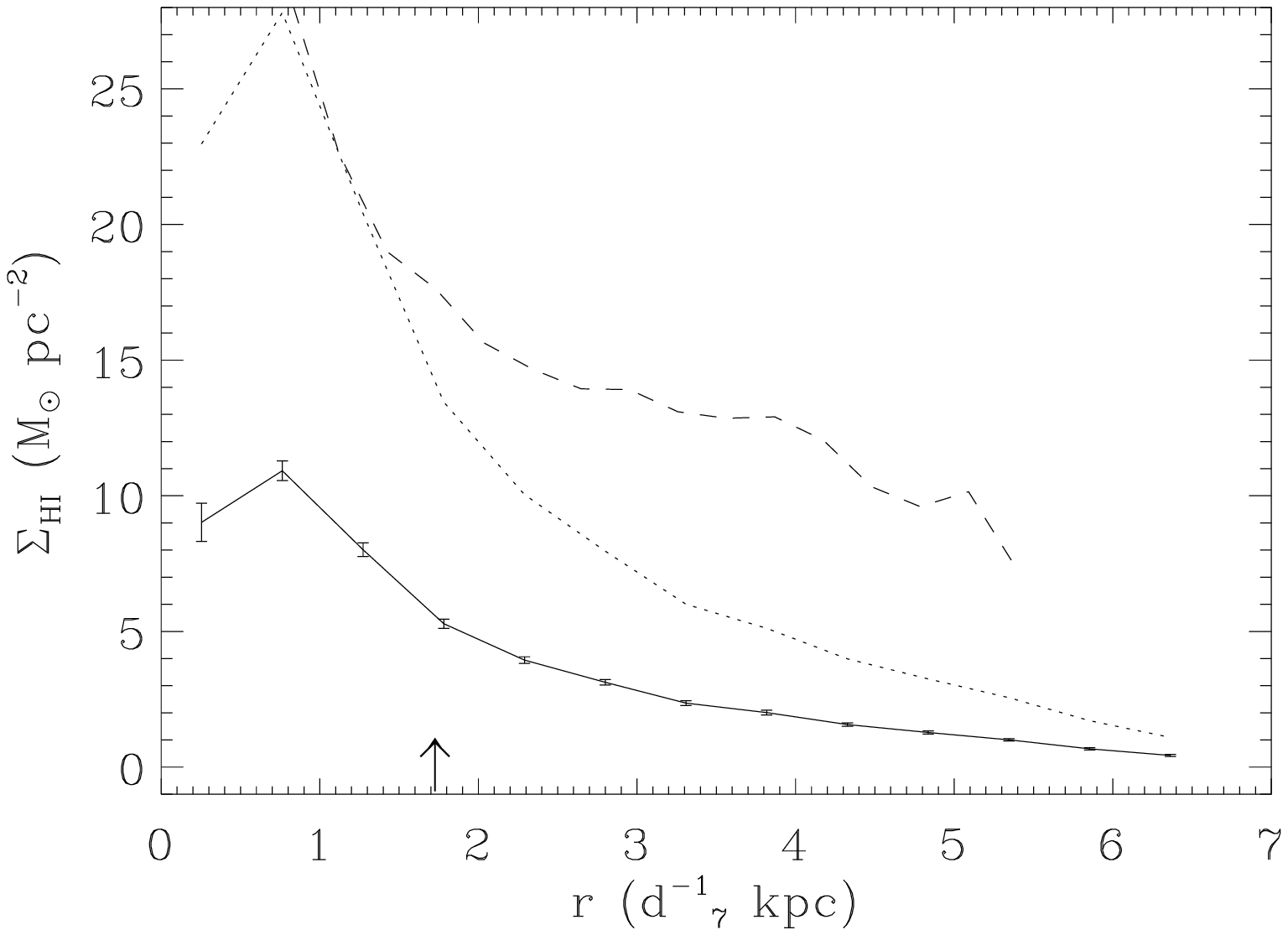}
\figcaption{Azimuthally-averaged \HI\ surface density profile for NGC~2537. 
Mean \HI\ surface density
measured in a series of concentric, elliptical annuli (in units of
solar masses per square parsec) is plotted 
as a function
of radius (in kpc). The error bars are statistical only and take into
account the number  
of synthesized beams in each annulus as well as the number of  
channels that contribute to the \HI\  signal at each position. Because
of the nearly face-on orientation of the galaxy, no correction has been
applied for inclination. The dashed line shows the critical density
for star formation according to the criterion of Kennicutt (1989; see text).
The dotted line shows the observed density scaled to
account for the mass of helium and molecular gas. An arrow denotes
the edge of the stellar disk.
\protect\label{fig:ngcHIsurf}}
\end{figure}

\suppressfloats

%
\begin{figure}
\epsscale{0.6}
\plotone{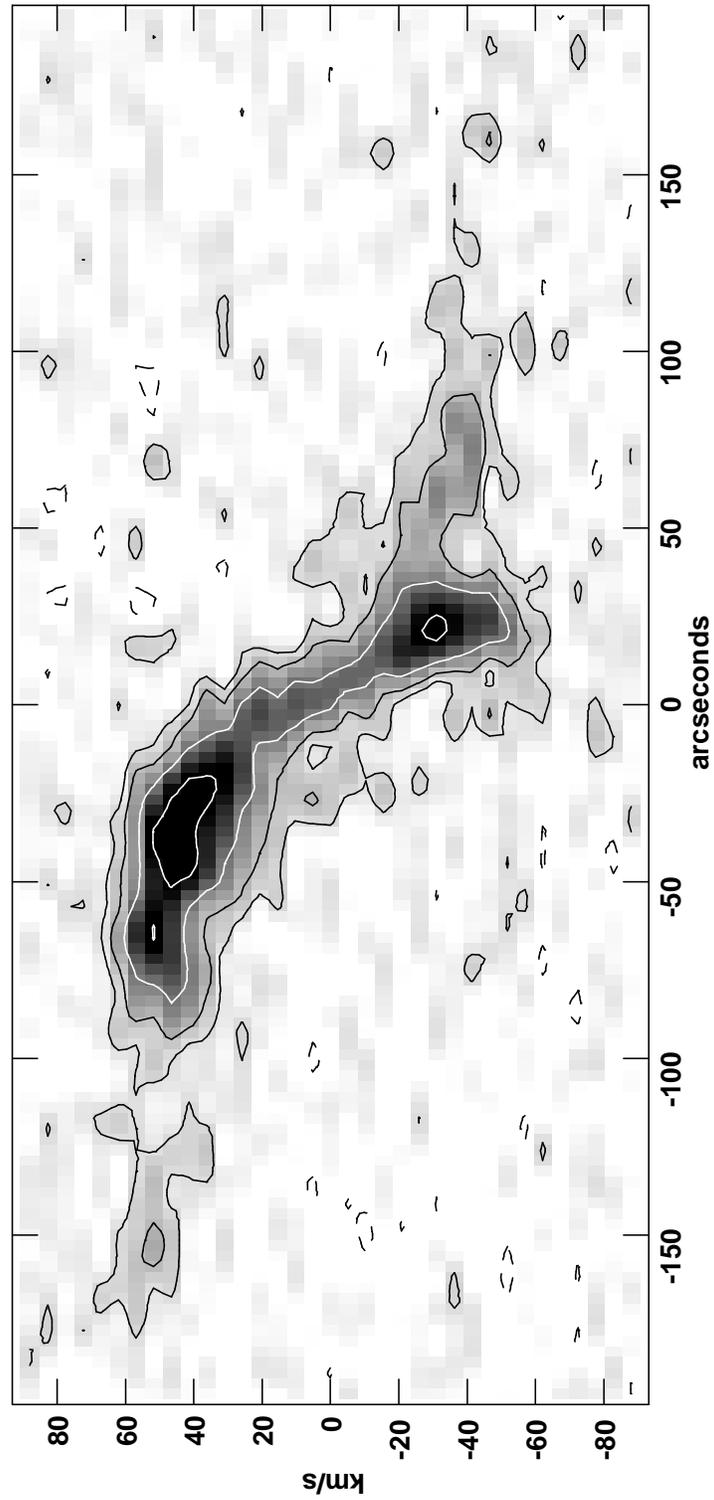}
\figcaption{\HI\ P-V plot along the major axis of NGC~2537.  The
data were averaged along a 15$''$-wide strip.  Contours are
($-$2[absent],$-$1,1,2,4,6,8)$\times$0.6~mJy beam$^{-1}$.
The greyscale range is 0$-$4.5~mJy beam$^{-1}$.   
\protect\label{fig:ngcPV}}
\end{figure}

%
\begin{figure}
\epsscale{1.0}
\plotone{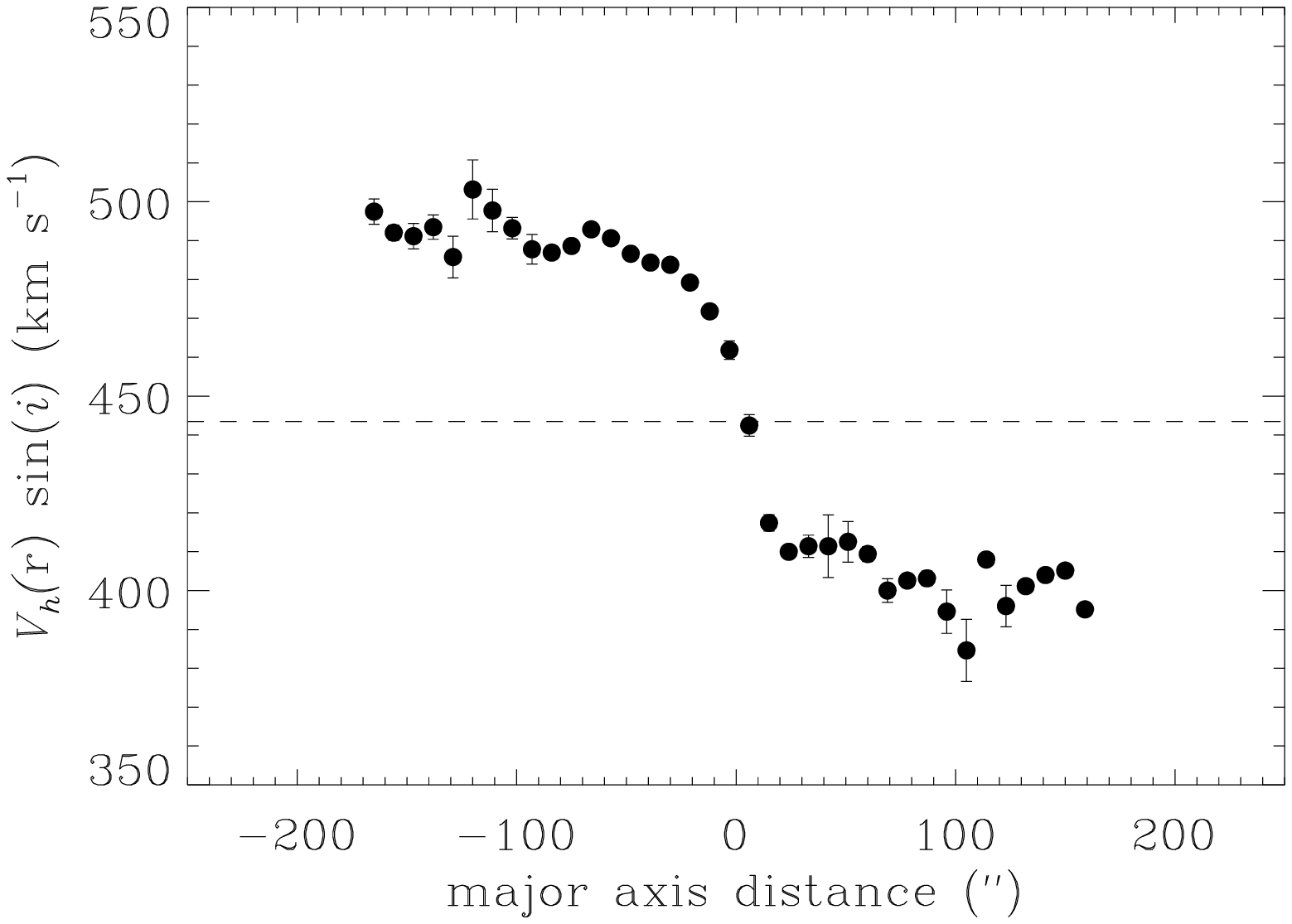}
\figcaption{\HI\ rotation curve for NGC~2537. No corrections for
inclination or asymmetric drift 
have been applied to the observed circular velocities. The dashed line
indicates the galaxy systemic velocity derived from a tilted ring fit (\S8.2.4).
\protect\label{fig:ngcrotcurve}}
\end{figure}

\end{document}